\documentclass[twocolumn, preprint2]{aastex631}

\def\kms{km~s$^{-1}$}

\usepackage{color}
\usepackage{multirow}
\usepackage{rotating}
\usepackage{amsmath} 
\usepackage{amsmath}
\usepackage{natbib}
\usepackage{graphicx}
\usepackage{mathrsfs}
\usepackage{color}
\usepackage{booktabs}
\usepackage{lipsum}
\usepackage{epstopdf}

\def\hi{H{\textsc{i}}}

\begin{document}

%\title{Template \aastex Article with Examples: 
%v6.3.1\footnote{Released on March, 1st, 2021}}
%\title{Multi-epoch OH Absorption and Stimulated Emission Measurements towards PSR J1644$-$4559}
%\title{Tiny-Scale Properties within the Interstellar Medium towards PSR~J1644$-$4559: I. Investigating HI Absorption Variabilities}
\title{Tiny-Scale Properties within the Interstellar Medium towards PSR~J1644$-$4559: I. Observational Evidence of Turbulence-induced Tiny Scale Atomic Structures}

\correspondingauthor{Mengting Liu, Di Li, Donghui Quan}
\email{liumengting@nao.cas.cn}
\email{dili@tsinghua.edu.cn}
\email{donghui.quan@zhejianglab.org}

\author[0000-0001-7790-5498]{Mengting Liu}
\affil{Research Center for Astronomical Computing, Zhejiang Laboratory, Hangzhou 311100, China}

\author[0000-0003-3010-7661]{Di Li}
\affil{Department of Astronomy, Tsinghua University, Beijing 100084, China}
\affiliation{National Astronomical Observatories, Chinese Academy of Sciences, 20A Datun Road, Chaoyang District, Beijing 100101, China}
%\affil{University of Chinese Academy of Sciences, Beijing 100049, People's Republic of China}
%\affil{Research Center for Astronomical Computing, Zhejiang Laboratory, Hangzhou 311100, China}

\author[0000-0003-0235-3347]{J.\ R.\ Dawson}
\affil{CSIRO Astronomy $\&$ Space Science, Australia Telescope National Facility, P.O. Box 76, Epping, NSW 1710, Australia}
\affiliation{School of Mathematical and Physical Sciences and MQ Research Centre in Astronomy, Astrophysics, and Astrotechnology, Macquarie University, 2109, NSW, Australia}

\author[0000-0001-9096-6543]{Joel M.\ Weisberg}
\affil{Department of Physics and Astronomy, Carleton College, Northfield, MN 55057}

\author[0000-0002-3418-7817]{Sne\v{z}ana Stanimirovi\'c}
\affil{Department of Astronomy, University of Wisconsin–Madison, Madison, WI 53706, USA}

\author[0000-0003-1502-100X]{George Hobbs}
\affil{CSIRO Astronomy $\&$ Space Science, Australia Telescope National Facility, P.O. Box 76, Epping, NSW 1710, Australia}

\author[0000-0002-7122-4963]{Simon Johnston}
\affiliation{CSIRO Astronomy $\&$ Space Science, Australia Telescope National Facility, P.O. Box 76, Epping, NSW 1710, Australia}

\author[0000-0003-3186-3266]{Lawrence Toomey}
\affil{CSIRO Astronomy $\&$ Space Science, Australia Telescope National Facility, P.O. Box 76, Epping, NSW 1710, Australia}

\author[0000-0002-0458-7828]{Siyao Xu}
\affil{Department of Physics, University of Florida, 2001 Museum Rd., Gainesville, FL 32611, USA}

\author[0000-0002-9390-9672]{Chao-Wei Tsai}
\affiliation{National Astronomical Observatories, Chinese Academy of Sciences, 20A Datun Road, Chaoyang District, Beijing 100101, China}
\affiliation{Institute for Frontiers in Astronomy and Astrophysics, Beijing Normal University,  Beijing 102206, China}

\author[0000-0003-4811-2581]{Donghui Quan}
\affil{Research Center for Astronomical Computing, Zhejiang Laboratory, Hangzhou 311100, China}

\author[0000-0002-9332-8616]{Stacy Mader}
\affil{CSIRO Astronomy $\&$ Space Science, Parkes Observatory, P.O. Box 276, Parkes NSW 2870, Australia}

\author[0000-0002-2670-188X]{James A. Green}
\affil{SKAO, SKA-Low Science Operations Centre, 26 Dick Perry Avenue, Kensington, WA, 6151, Australia}

\author[0000-0001-8539-4237]{Lei Zhang}
\affiliation{National Astronomical Observatories, Chinese Academy of Sciences, 20A Datun Road, Chaoyang District, Beijing 100101, China}
\affil{Centre for Astrophysics and Supercomputing, Swinburne University of Technology, P.O. Box 218, Hawthorn, VIC 3122, Australia}

\author[0000-0002-2169-0472]{Ningyu Tang}
\affil{Department of Physics, Anhui Normal University, Wuhu, Anhui 241002, People's Republic of China}

\author[0000-0002-3386-7159]{Pei Wang}
\affiliation{National Astronomical Observatories, Chinese Academy of Sciences, 20A Datun Road, Chaoyang District, Beijing 100101, China}
\affiliation{Institute for Frontiers in Astronomy and Astrophysics, Beijing Normal University, Beijing 102206, China}

\author[0000-0003-0985-6166]{Kai Zhang}
\affiliation{National Astronomical Observatories, Chinese Academy of Sciences, 20A Datun Road, Chaoyang District, Beijing 100101, China}

\author[0000-0003-3948-9192]{Pei Zuo}
\affiliation{National Astronomical Observatories, Chinese Academy of Sciences, 20A Datun Road, Chaoyang District, Beijing 100101, China}

\author[0000-0002-1583-8514]{Gan Luo}
\affil{Institut de Radioastronomie Millimetrique, 300 rue de la Piscine, 38400, Saint-Martin d\'Heres, France}

\author[0000-0002-0475-7479]{Yi Feng}
\affil{Research Center for Astronomical Computing, Zhejiang Laboratory, Hangzhou 311100, China}
\affil{Institute for Astronomy, School of Physics, Zhejiang University, Hangzhou 310027, China}

\author[0000-0002-9618-2499]{Shi Dai}
\affil{CSIRO Astronomy $\&$ Space Science, Australia Telescope National Facility, P.O. Box 76, Epping, NSW 1710, Australia}

\author{Aditi Kaushik}
\affil{School of Mathematical and Physical Sciences and Astrophysics and Space Technologies Research Centre, Macquarie University, 2109, NSW, Australia}
\affil{CSIRO Astronomy $\&$ Space Science, Australia Telescope National Facility, P.O. Box 76, Epping, NSW 1710, Australia}

\author[0000-0001-8018-1830]{Mengyao Xue}
\affiliation{National Astronomical Observatories, Chinese Academy of Sciences, 20A Datun Road, Chaoyang District, Beijing 100101, China}

\author[0000-0002-9441-2190]{Chenchen Miao}
\affil{Research Center for Astronomical Computing, Zhejiang Laboratory, Hangzhou 311100, China}



\begin{abstract}

We investigated \hi\ absorption toward a single pulsar, PSR J1644$-$4559, and its variability over timescales from days to years, using Murriyang, CSIRO's Parkes Radio Telescope.
Our 19 epochs of spectral observations, spanning 1.2 years with intervals as short as 1 day, provide the most comprehensive cadence coverage for monitoring \hi\ absorption to date. 
We identified two significant detections of tiny-scale atomic structure (TSAS) with spatial scales ranging from a lower limit of $\sim$11 au to an upper limit of 165 au, both exhibiting integrated signal-to-noise ratios exceeding 5.0. 
We find a relationship between linear size and optical depth variation in the cold neutral medium (CNM) component hosting the TSAS, 
described by a power-law relationship, $\Delta\tau_{\rm int} = \Delta\tau_0 (\Delta L)^{(\alpha-2)/2}$, with $\alpha = 4.1 \pm 0.4$. This is the first observational evidence explicitly connecting TSAS to turbulence in CNM. 
This power-law index is significantly steeper than previously reported values for the CNM, where $\alpha$ ranges from 2.3 to 2.9, but similar to those observed in the warm ionized gas. 
Additionally, we observe no significant variation in $\alpha$ across the entire range of spatial scales traced in our study, indicating that turbulence may be cascading down and dissipating at smaller scales. 
While there is no precise proper motion measurement for this pulsar, our estimates for the turbulence dissipation in the CNM place the lower and upper limits at less than 0.03 au and 0.4 au, respectively. 
\end{abstract}



\section{Introduction} \label{sec:int}

The neutral atomic interstellar medium (ISM) is observed to have a self-similar hierarchy of structures, ranging from scales of thousands of au to kiloparsecs \citep{2023ARA&A..61...19M,2007ARA&A..45..565M,2000prpl.conf...97W,1990ARA&A..28..215D}. 
For the atomic gas in the diffuse ISM, turbulence plays a crucial role in regulating structure \citep{2013ApJ...770..141B}. 
Turbulent energy is injected at large scales and cascades down towards smaller scales. 
If the energy transfer rate across scales remains constant, the turbulence spectrum follows a Kolmogorov-type scaling, with a 3D power-law index of 11/3 \citep{1941DoSSR..30..301K}. 
The detection of tiny-scale atomic structure (TSAS) on spatial scales of tens to thousands of astronomical units (au) through temporal and spatial variations in \hi\ absorption features observed towards extragalactic compact sources and pulsars \citep[e.g.,][]{2010ApJ...720..415S, 2012ApJ...749..144R,2020ApJ...893..152R,2021ApJ...911L..13L} has the potential to reveal the extension of the turbulence density and velocity spectrum down to au scales, or even probe the dissipation scale in the cold neutral medium (CNM).

In the multi-phase ISM model, the neutral medium consists of two steady phases in thermal and pressure equilibrium: the CNM and the warm neutral medium (WNM), with kinematic temperatures and densities of 60$-$260 K and 7$-$70 cm$^{-3}$, and 5000$-$8300 K and 0.2$-$0.9 cm$^{-3}$, respectively \citep{2003ApJ...587..278W}. 
These phases are maintained by a balance between cooling and heating processes. 
Additionally, there is a thermally unstable neutral medium (UNM) that is not in equilibrium and will eventually either cool or heat with temperatures ranging from 250 to 4000 K, transitioning into one of the two stable phases \citep{2023ARA&A..61...19M}.
However, within this picture, the origin of TSAS remains an unresolved question. This is partly because if one assumes that the line-of-sight spatial scale of a TSAS feature is similar to its width in the plane of the sky, then TSAS are characterized by significantly higher densities ($\sim10^4$ cm$^{-3}$) and pressures ($\sim10^6$ cm$^{-3}$ K) in atomic gas, compared to the typical conditions of the CNM. 

These issues may be resolved if TSAS does not originate from tiny-scale \hi\ cloudlets but is instead the manifestation of interstellar turbulence. 
\citet{2000MNRAS.317..199D} first suggested that if TSAS, observed through \hi\ opacity fluctuations, is driven by interstellar turbulence, 
a power-law relationship should exist between the opacity fluctuations and spatial scales. 
Turbulence cascading down to au scales in the CNM has been detected using the traditional statistical tools of \hi\ optical depth power spectra and structure functions. 
Interferometric \hi\ absorption mapping towards Cas A, Cygnus A, and 3C138 have estimated consistent power-law slopes of the \hi\ optical depth power spectrum of the CNM ranging from 2.7 to 2.9, from spatial scales of 5 au to 50 pc, suggesting that the turbulent dissipation scale in the CNM is less than 5 au 
\citep[e.g.,][]{2000ApJ...543..227D,2010MNRAS.404L..45R,2014MNRAS.442..647D}. 
However, these studies -- while they have measured optical depths to au-scales -- have not generally detected the \textit{individual statistically significant fluctuations} in \hi\ absorption that have been traditionally classified as TSAS.  

\citet{2018ARA&A..56..489S} and \citet{2010ApJ...720..415S} compiled all radio TSAS detections and found no clear correlation between the observed optical depth variations and spatial scales in different directions, although this analysis may be biased due to variations in turbulence properties along different lines of sight and differences in background sources. 
\citet{2005ApJ...631..376M} performed 18-epoch \hi\ absorption observations over 1.3 years toward PSR~B0329$+$54, also finding no evidence for any significant turbulent \hi\ fluctuations. To-date we are aware of only one instance where TSAS detections are present alongside a confirmed measurement of an au-scale power law slope: observations against 3C138 have shown TSAS with a 25 au spatial scale \citep{1998AJ....116.2916F,2005AJ....130..698B}, with separate analysis of the \hi\ opacity structure function on scales from 5 to 100 au \citep{2012ApJ...749..144R,2014MNRAS.442..647D}. 
However, currently, no study has simultaneously successfully constructed an \hi\ optical depth power spectrum/structure function and detected significant optical depth fluctuations. There remains a lack of direct observations to confirm whether turbulence can indeed induce such large optical depth variations on these au scales.

Pulsars, with their relatively high transverse velocities, tiny angular sizes, and rapidly varying on/off states, are ideal background sources for tracing intervening tiny-scale structures along the line of sight (LOS) via multi-epoch \hi\ absorption line measurements. 
PSR~J1644$-$4559, located in the Galactic plane ($l = 339.193$, $b = -0.195$) and exhibiting a relatively high flux density (\(\sim 0.3\) Jy at L-band), has already been observed to produce \hi\ absorption spectra of several absorption components over several epochs \citep{1976A&A....50..177A, 1991ApJ...382..168F, 2002MNRAS.337..409O, 2003MNRAS.341..941J}. 
No significant variations in \hi\ absorption had been detected towards this pulsar, possibly due to the limited number of observation epochs in each study. 
\citet{2005Sci...309..106W} detected complex interstellar hydroxyl (OH) absorption and pulsar-pulse-induced stimulated emission towards PSR~J1644$-$4559 in observations made with Murriyang, CSIRO's Parkes Radio Telescope. 
The richness of atomic and molecular structures and the environmental conditions within the J1644$-$4559 beam present a unique opportunity to investigate tiny scale structure variations.

In this work, we present the most comprehensive cadence coverage for
monitoring \hi\ absorption, with 19-epoch observations over 1.2 years using Murriyang, CSIRO's Parkes Radio Telescope towards PSR~J1644$-$4559. 
The aim was to search for TSAS and simultaneously measure the slope of the turbulent power spectrum across spatial scales, ranging from 0.03 to 11 au at the lower limit and 0.4 to 175 au at the upper limit in the CNM, thus investigating the origin of TSAS from interstellar turbulence.

In Section \ref{sec:obs}, we describe our observations and data reduction strategies. 
The \hi\ absorption Gaussian decomposition, excitation temperature estimation, and column density calculation are presented in Section \ref{sec:HI_analysis}. 
Section \ref{sec:var_HI_abs} illustrates the \hi\ absorption variations. 
In Section \ref{sec:disc}, we discuss the variability results and their relation to turbulence. 
Our conclusions are presented in Section \ref{sec:concl}. 
This paper primarily focuses on discussing the variations of \hi\ absorption features over 19 epochs. 
We recorded OH at 1612, 1665, 1667, and 1720 MHz simultaneously with \hi\ data in 19 epochs along the exact same line of sight towards PSR J1644$-$4559. 
Detailed analysis of OH absorption and stimulated emission variations over these 19 epochs will be presented in Papers II and III.

\section{Observations and data reduction} \label{sec:obs}

\begin{deluxetable}{cc|cccc}
\setlength{\tabcolsep}{0.2in} 
\tablecolumns{3} 
\tabletypesize{\scriptsize}
\tablewidth{0pt}
\tablecaption{Observing Dates and Integration Times (after RFI removal) for PSR J1644$-$4559 Observations}
\label{tab:observations_old}
\tablehead{\colhead{Epoch} & \colhead{Start Date}  & \colhead{$\rm T^{int}_{1420}$}  \\ %[1mm]
\colhead{(No.)}   &   \colhead{(UTC)}  & \colhead{(minutes)}  }
\startdata
1	&	2022-02-18	&	269	\\
2	&	2022-03-29	&	302	\\
3	&	2022-05-08	&	289	\\
4	&	2022-06-15	&	273	\\
5	&	2022-07-10	&	297	\\
6	&	2022-12-16	&	432	\\
7	&	2023-01-01	&	365	\\
8	&	2023-01-26	&	58	\\
9	&	2023-02-05	&	474	\\
10	&	2023-02-17	&	156	\\
11	&	2023-02-20	&	94	\\
12	&	2023-03-04	&	98	\\
13	&	2023-03-05	&	107	\\
14	&	2023-03-08	&	121	\\
15	&	2023-03-10	&	99	\\
16	&	2023-03-12	&	94	\\
17	&	2023-03-13	&	386	\\
18	&	2023-03-31	&	203	\\
19	&	2023-04-05	&	362		
\enddata
\tablecomments{Column 1: number of observing epochs. Column 2: observing dates. Columns 3: integration time (minutes) for 1420 MHz observations after RFI removal. 
}
\end{deluxetable}

We conducted a total of 19 epochs of \hi\ observations of PSR J1644$-$4559 using Murriyang, CSIRO's Parkes Radio Telescope. 
The observing dates and integration times for each epoch are presented in Table~\ref{tab:observations_old}. 
All observations were carried out using the ultra-wide-bandwidth, low-frequency receiver (`UWL') \citep{2020PASA...37...12H} between February 2022 and April 2023 using the zoom baseband mode with a frequency centered at 1420.0 MHz and 8 MHz bandwidth.

For the source observations, the baseband data were coherently dedispersed and folded synchronously with the pulsar period, for an integration time of 10 s with 128 phase bins and 16384 (0.49-kHz) frequency channels, using the {\tt\string DSPSR} package \citep{2011PASA...28....1V}. 
To ensure accurate flux and polarization calibration, a calibration signal (CAL) was injected after each hour of observations, with the telescope pointed at a position approximately 1 degree from the pulsar. 
For the CAL observations, the baseband data were folded synchronously with the CAL period, utilizing an integration time of 10 s and 32 phase bins with data processing performed using the {\tt\string DSPSR} package. 
We performed flux and polarization calibration for each pulsar dataset using the {\tt\string pac} command in PSRCHIVE \citep{2012AR&T....9..237V}. 
The flux calibration was executed by applying the observatory's calibration solutions obtained from observations of the flux calibrators 0407$-$658 and 1934$-$638, which were taken closest in time to our own observations. 
The frequency resolution of the flux calibration data is 1 MHz,  
which limited the precision of our spectral line bandpass calibration. 
While this is not important for derived optical depth spectra, which subtract pulsar-on and pulsar-off spectra with the same baseline structure (see in the 5th paragraph of Section~\ref{sec:obs}), it is important in cases where pulsar-off spectra are used alone, e.g. in computing \hi\ spin temperature. Baseline subtraction is discussed further in the 
appendix \ref{sec:app_blrm}. 

The units of the flux calibrated data $S$ are Jy. 
We converted this into main beam brightness temperature $T_{\rm b}$ using 
\begin{equation}
T_{\rm b}= 0.85 S,
\label{e:tb_S_parkes}
\end{equation}
where 0.85 is the factor obtained by comparing our flux-calibrated pulsar-off spectra with \hi\ emission data from the Parkes Galactic All-Sky Survey (GASS; \citealt{2015A&A...578A..78K}) at the closest possible location on the sky. 
This factor is also consistent with the expected theoretical conversion factor of $\sim 0.81$, assuming a Gaussian main beam width of 0.24 degrees at 1420 MHz \citep{2020PASA...37...12H}.

Sub-integrations strongly affected by radio frequency interference (RFI)  were identified using the \texttt{pazi} command in PSRCHIVE, and were subsequently excised from the dataset. 
To remove narrow-band time-variable RFI contamination from the pulsar-off spectra, we flag spectra that contain frequency channels with values that deviate from the mean by more than 10$\sigma_0$, where $\sigma_0$ is the standard deviation of each frequency channel over 10-minute observations during pulsar-off. (No such
flagging was possible for the pulsar-on spectra since the pulsar itself generates time-variable signals that are
difficult to distinguish from RFI.) 
Less than 0.1$\%$ of the time samples were removed in this process.

Pulsar-on spectra were computed by averaging the spectra collected from the on-pulse rotational phases of each subintegration, weighted by the square of pulsar brightness temperature at the given phase bin. 
Pulsar-off spectra were similarly computed, except with uniform weighting. 
The radiative transfer equations for the pulsar-on and pulsar-off spectra are given by
\begin{equation}
\begin{split}
T^{\rm on}_{\rm b}(v)&= (T_{\rm bg}+T_{\rm psr})e^{-\tau(v)}+T_{\rm s}(1-e^{-\tau(v)}),\\
T^{\rm off}_{\rm b}(v)&= T_{\rm bg}e^{-\tau(v)}+T_{\rm s}(1-e^{-\tau(v)}),
\end{split}
\label{e:on}
\end{equation}
\noindent where $T^{\rm on}_{\rm b} (v)$ and $T^{\rm off}_{\rm b} (v)$ are the main beam  temperatures of the pulsar-on and pulsar-off line profiles, 
$T_{\rm bg}$ is the background main beam temperature, $\tau(v)$ is the \hi\ optical depth spectrum, $T_{\rm s}$ is the \hi\ spin temperature, and $T_{\rm psr}$ is the pulsar continuum main beam temperature. The absorption spectra can then be obtained from the difference between the pulsar-on and pulsar-off spectra, scaled by the mean value of the off-line region of this difference spectrum, which is simply the pulsar main beam temperature, $T_{\rm psr}$:
\begin{equation}
e^{-\tau(v)}= \frac{T_{\rm b}^{\rm on} (v)-T_{\rm b}^{\rm off} (v)}{T_{\rm psr}}.
\label{e:tau}
\end{equation}

Baseline subtraction was carried out by constructing median filters with a width of 413 \kms\ 
and subtracting these from the scaled absorption spectra. 
A detailed description of the baseline removal process for the absorption spectra and the corresponding impact on the difference spectra can be found in appendix \ref{sec:app_blrm}.

The distances of \hi\ absorption components are determined kinematically, using a Galactic rotation curve. 
We utilized the linear fit of \citet{2019ApJ...870L..10M}, which matches observations into a Galactocentric radius $R \sim 6$ kpc (See Appendix \ref{sec:tan_vel} for details of the \citet{2019ApJ...870L..10M} fit).
The uncertainties in the distance are estimated by adding and subtracting 7 \kms\ to the LSR velocities to account for streaming and random gas motions within the Galaxy \citep{2008ApJ...674..286W}. 

\begin{figure*}
\centering
\includegraphics[width=0.6\linewidth]{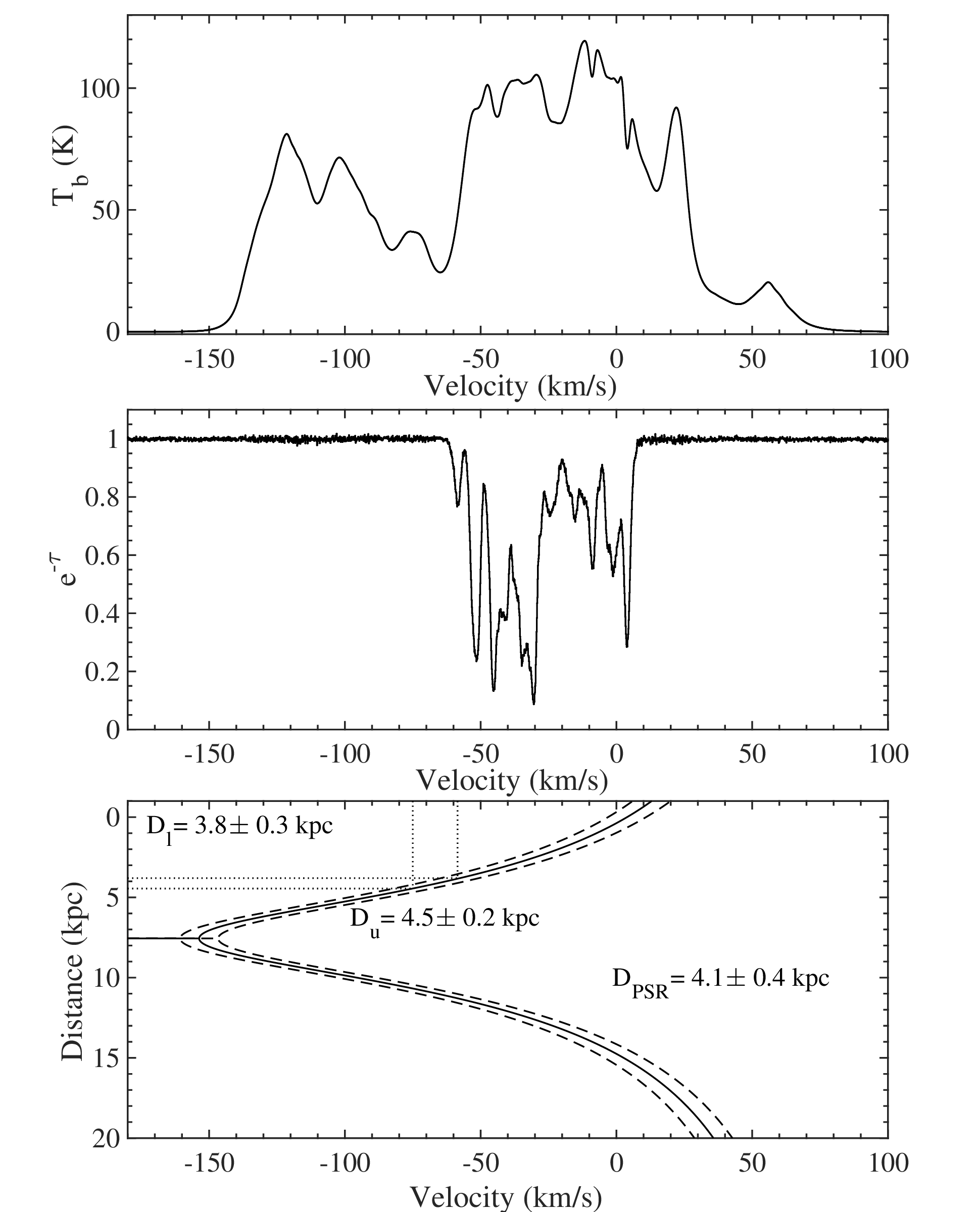}
\caption{\textit{Top panel}: \hi\ emission observed towards PSR~J1644$-$4559 with Parkes during the pulsar-off phase. 
\textit{Middle panel}: \hi\ absorption profile toward PSR~J1644$-$4559, derived using Equation~\ref{e:tau}. 
\textit{Bottom panel}: Velocity-distance relation (solid line) calculated from the \citet{2019ApJ...870L..10M} linear fit to their Galactic rotation curve. The dotted lines mark the lower and upper limits on the pulsar kinematic distance, $D_{\rm l}$ and $D_{\rm u}$. 
} 
\vspace{0.2cm}
\label{fig:HI_dist} 
\end{figure*}

\begin{figure*}
\centering
\includegraphics[width=0.6\linewidth]{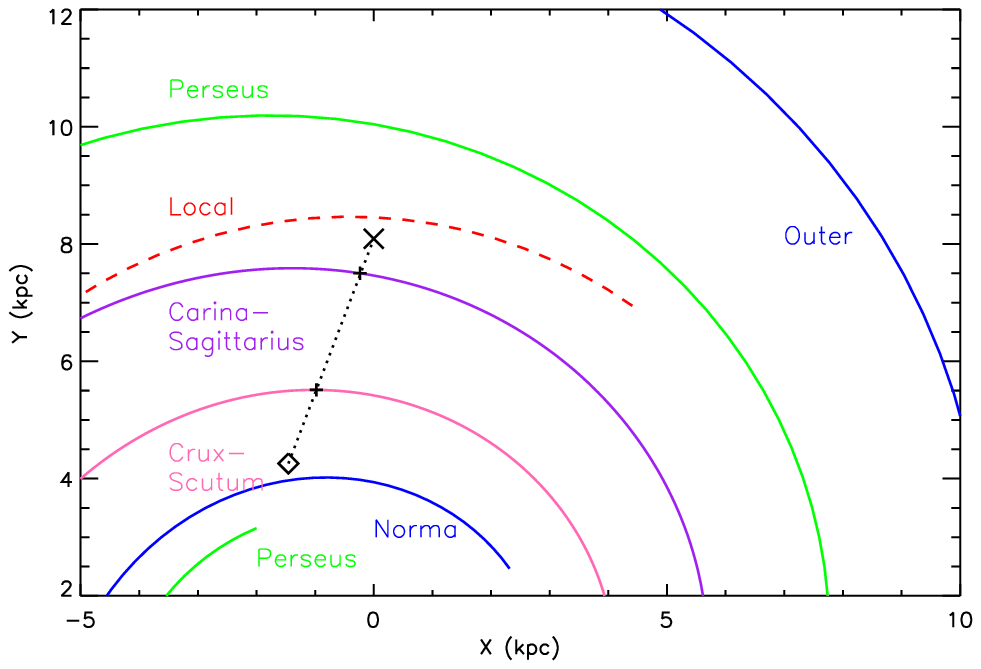}
\caption{The Earth - pulsar line of sight is displayed atop a face-on plot of the Galaxy.  The line of sight is drawn between the Earth (shown as an $\times$) and the pulsar (depicted by a diamond), 4.1 kpc away, as derived in Section \ref{sec:obs}. Spiral arm centers,   calculated from the \citet{2017ApJ...835...29Y} model, rescaled with a value of $R_0 = 8.09$ kpc, are displayed as colored arcs.  The Local Arm is
dashed to indicate that many recent analyses \citep[e.g.;][]{2018A&A...616L..15X,2019ApJ...885..131R} indicate that this ``Arm'' is more nearly parallel to adjoining arms than illustrated here.}
\vspace{0.2cm}
\label{fig:gal_arm} 
\end{figure*}

We estimated the kinematic distance $D_{\rm PSR}$ of this pulsar by comparing our \hi\ emission and absorption spectra (See Figure~\ref{fig:HI_dist}). 
Following \citet{1979A&A....77..204W}, the lower kinematic distance limit of the pulsar, $D_{\rm l}$, is set by the velocity of the most distant absorption component, while the upper limit, $D_{\rm u}$, corresponds to the velocity of the nearest emission component with a brightness temperature above 35 K. 
The final updated kinematic distance, $D_{\rm PSR}=4.1\pm0.4$ kpc, transformed from the lower and upper limits, is estimated using the likelihood analysis method from \citet{2012ApJ...755...39V}. 
Figure~\ref{fig:gal_arm} depicts the locations of the Earth, pulsar, and the line of sight between them atop a face-on plot of the Galaxy's spiral arms. 

Knowledge of the proper motion and hence transverse velocity of the pulsar is necessary in order to convert the time interval between observations into a physical length scale. In the absence of a significant measured proper motion for PSR~J1644$-$4559 \citep{2021MNRAS.508.3251L}, we resort to alternative arguments to estimate approximate lower and upper limits of the pulsar's transverse velocity and of its {\it{relative}} transverse velocity with respect to absorbing \hi\ clouds.

For the {\it{lower}} limit on the relative transverse velocity $v^{\rm relative}_{ \rm Low, Transverse},$ 
we start from the above-noted fact that there are no significant proper motion measurements available for PSR~J1644$-$4559, so we assume that  the pulsar is at rest with respect to the Solar System frame,  and hence that its velocity in that frame is zero; while all \hi\ components move according to the Galactic rotation curve at their own Galactocentric radii. 
Under these assumptions, 
the relative transverse velocity at the pulsar kinematic distance $\rm D_{\rm PSR}=4.1$ kpc is determined to be $v^{\rm relative}_{ \rm Low, Transverse}$ $\sim$ 50.5 \kms. (See Appendix~\ref{sec:tan_vel}.)

The {\it{upper}} limit on the relative transverse velocity, $v^{\rm relative}_{ \rm Up, Transverse},$ is calculated as follows.  The upper limit on the magnitude of the pulsar's transverse velocity is  $\sim787$ \kms\ in the Solar System frame, based on \citet{2021MNRAS.508.3251L} $\sim$41 mas yr$^{-1}$ upper limit on the pulsar's total proper motion. These values are so large as to render negligible any frame corrections, so we adopt 787 \kms\ as our upper limit on relative transverse velocity, $v^{\rm relative}_{ \rm Up, Transverse}$.  
We also note that this upper limit is large compared to the median proper motion of $\sim$260 \kms\ \citep{2020MNRAS.494.3663I}. 
The lower and upper limits of the spatial scale, $\Delta L^{\rm Low}$ and $\Delta L^{\rm Up}$, for a time interval $\Delta t$ are then computed using $\Delta L^{\rm Low} =v^{\rm relative}_{ \rm Low, Transverse} \times d_{\rm comp}\times \Delta t  /D_{\rm PSR}$ and $\Delta L^{\rm Up} = v^{\rm relative}_{ \rm Up, Transverse} \times d_{\rm comp} \times \Delta t /D_{\rm PSR}$, respectively, where $d_{\rm comp}$ denotes the distance to a specific absorption component.
Uncertainties in $\Delta L^{\rm Low}$ and $\Delta L^{\rm Up}$ themselves can be derived from the distance uncertainties using the same equations, but the principal source of uncertainty in the calculation of $\Delta L$ stems from the current inaccuracy in
the measurement of $v^{\rm relative}_{ \rm Transverse}$, which leads directly to the large spread in the values of $\Delta L^{\rm Low}$ and $\Delta L^{\rm Up}$ and in the quantities derived from them.  
 
The pulsar's emission is broadened by diffractive interstellar scattering. 
For PSR~J1644-4559, the corresponding  diffractive scattering disk size at 1420 MHz is 0.03", as calculated from the dual-frequency interstellar pulse-broadening timescales of \citet{2013MNRAS.434...69L}. 
It is important to note that the scattering disk is not a physical object in space, in much the 
same fashion that a gravitationally-lensed image of a background source is not “real”. The
size of the region instantaneously sampled on an absorbing HI cloud is instead a complicated
function of the cloud’s distance and details of electron density fluctuations along the line
of sight, and is beyond the scope of this paper.

While  {\it{diffractive}} interstellar scattering can lead to image broadening as discussed above, it does not cause image ``wandering.''
In contrast, {\it{refractive}} interstellar
scattering can do so, which would further complicate our analyses
of spatial scales.
However, \citet{1992ApJ...392..530K} showed that if {\it{refractive}} scattering and its attendant image steering are operative, then their timescales are $\geq$1824 d for this pulsar; i.e., significantly longer than the duration of our experiment. Hence we can safely ignore the issue of refractive interstellar scintillations in this experiment.
Regardless, the above expressions for lower
and upper limits on the spatial scale of HI sampled by the moving pulsar in two temporally separate observations, remain unchanged,
despite the possible presence of any of these scattering
phenomena.

\section{\hi\ Analysis} \label{sec:HI_analysis}

Using the time intervals ($\Delta t$) between our 19 epochs and the estimated lower and upper transverse velocities ($v^{\rm relative}_{ \rm Low, Transverse}$ and $v^{\rm relative}_{ \rm Up, Transverse}$), we trace structures with spatial scales ranging from 0.03 to 11 au (lower limit $\Delta L^{\rm Low}$) and from 0.4 to 175 au (upper limit $\Delta L^{\rm Up}$). 

\subsection{Comparison with \citet{2003MNRAS.341..941J}} 
\citet{2003MNRAS.341..941J} conducted two-epoch observations of PSR~J1644$-$4559 using the Parkes telescope, achieving a velocity resolution of  $\sim$0.25~\kms. 
Their analysis revealed no significant variations in the absorption profiles of 12 individual \hi\ clouds over a 2.5-year period. 
However, they identified a potential difference in the absorption component at 4~\kms\ when comparing it to absorption spectra obtained in 1975 and 1988 \citep{1976A&A....50..177A,1991ApJ...382..168F}.

Our observations have a higher velocity resolution of $\sim$0.1 \kms\ and span 19 observing epochs over a 1.2-year interval, enabling us to trace variations over periods ranging from days to months. 
The absorption line profiles obtained by \citet{2003MNRAS.341..941J} are broadly consistent with our findings. 
With an upgraded Gaussian decomposition algorithm \citep{2019A&A...628A..78R}, we have decomposed the \hi\ absorption spectrum into 13 individual components, compared to the 12 components identified by \citet{2003MNRAS.341..941J}.

\subsection{Gaussian decomposition, spin temperature, and column density} 
\label{sec:fitting}

We followed the method outlined in \citet{2021ApJ...911L..13L} to decompose our \hi\ optical depth spectra into 13 Gaussian features observed across 19 epochs and to estimate the spin temperature for each component. 
The optical depth spectrum $\tau(v)$ along the line of sight can be decomposed into $N$ Gaussian components
\begin{equation}
\tau (v)= -\ln{\frac{T_{\rm b}^{\rm on} (v)-T_{\rm b}^{\rm off} (v)}{T_{\rm psr}}} = \sum_{n=0}^{N-1} \tau_{0,n} \cdot e^{- 4 \ln{2} \left(v-v_{0,n}\right)^2/ \Delta v_{0,n}^2},
\label{e:tau_abs}
\end{equation}
\noindent where $\tau_{0,n}$, $v_{0,n}$, $\Delta v_{0,n}$ are the amplitude, central velocity, and FWHM of the $n$th Gaussian component, respectively.
The pulsar-off \hi\ emission contains contributions from both the CNM ($T_{\rm CNM}$) and WNM ($T_{\rm WNM}$), which is given by 
\begin{equation}
T_{\rm b}^{\rm off} (v) = T_{\rm CNM} (v) + T_{\rm WNM}(v).
\label{e:tb}
\end{equation}

\noindent $T_{\rm CNM}$ %traced by \hi\ absorption 
is given by
\begin{equation}
T_{\rm CNM}(v) = \sum_{n=0}^{N-1} T_{s,n} \left(1-e^{-\tau_n(v)}\right) e^{-\sum_{m=0}^{M-1} \tau_m(v)},
\label{e:em_CNM}
\end{equation}
\noindent where $T_{s,n}$ is the spin temperature for the $n$th CNM component, and there are $m$ clouds lying in front of the $n$th cloud. $T_{\rm WNM}(v)$ with $K$ Gaussian components can be written as
\begin{equation}
T_{\rm WNM}(v) = \sum_{k=0}^{K-1} \left[ \mathscr{F}_k + \left(1- \mathscr{F}_k\right)e^{-\tau(v)} \right] \cdot T_{0,k} e^{\frac{- 4 \ln{2} \left(v-v_{0,k}\right)^2}{ \Delta v_k^2}},
\label{e:em_WNM}
\end{equation}
where $T_{0,k}$, $\Delta v_k$, and $v_{0,k}$ represent the brightness temperature, FWHM, and central velocity for the $k$th WNM component, respectively. 
There is a fraction $\mathscr{F}_k$ of the WNM located in front of all CNM components for each $k$th WNM component. 
Following \citet{2003ApJS..145..329H}, we adopted values of 0, 0.5, and 1 to each $\mathscr{F}_k$. 
The final estimated $T_{\rm s}$ and corresponding uncertainties are the weighted mean and standard deviation over the three cases. 
The CNM components in our absorption spectra are well-separated with corresponding derived radial distances. 
We therefore did not consider different ordering of the CNM Gaussian features along the line of sight when performing the fitting.

The results of the Gaussian decomposition for our 19 epochs of observations are presented in Tables~\ref{tab:fit_tau_1420_amp_compB}-- \ref{tab:fit_tau_1420_N_compB}, and shown in Figures~\ref{fig:compB_gaussian_Tex_1420_fit0} through \ref{fig:compB_gaussian_Tex_1420_fit3}. %
The estimated spin temperatures, ranging from 7.4$\pm$2.8 to 437.4$\pm$1.3 K, are relatively stable across the 19 observing epochs. 
Among these features, we detected two UNMs components exhibiting spin temperatures of 305.7$\pm$3.1 K and 437.4$\pm$1.3 K, centered at velocities of $-23.9$ and $-14.9$ \kms, respectively. 
The remaining 11 Gaussian features are CNM, with averaged spin temperatures ranging from 7.4$\pm$2.8 to 220.5$\pm$1.7 K. 
We derived the column density for each component 
using $N{(\rm HI)} = 1.064 \cdot C_0 \cdot \tau_0 \cdot \Delta v_0 \cdot T_s$, where $C_0=1.823\times10^{18}$ cm$^{-2}/( $ km s$^{-1}$ K). 
The column densities of the two UNM features are $(6.4 \pm 0.3) \times 10^{20}$ and $(17.7 \pm 0.8) \times 10^{20}$ $\rm cm^{-2}$. 
The remaining 11 CNM components have column densities ranging from $3.8 \times 10^{18}$ to $1.4 \times 10^{21}~\rm{cm^{-2}}$, consistent with the typical range observed in CNM clouds across various directions by \citet{2003ApJS..145..329H}.

\section{Temporal Stability of \hi\ absorption} \label{sec:var_HI_abs}

We searched for potential variations in \hi\ absorption for each Gaussian component over time in the $(1-e^{-\tau})$ spectra. 
We assume that the frequency channels and all 19 epoch spectra are independent.
The process for detecting potential variations and calculating the signal-to-noise ratio (SNR) from the epoch-differences between two $\left[1-e^{-\tau(v)}\right]$ spectra involved several steps:
\begin{itemize}
   \item[1)] Calculate the noise envelope $\sigma_{\left(1-e^{-\tau}\right)}(v)$ for each $\left[1-e^{-\tau(v)}\right]$ spectrum.  
    The $i$th epoch spectrum has a noise envelope of $\sigma_{\left(1-e^{-\tau}\right), i}(v)$, which can be expressed as 
    \begin{equation}
    \sigma_{\left(1-e^{-\tau}\right), i}(v) = \frac{\sqrt{\sigma_{\mathrm{psr\text{-}on,} i}(v)^2+ \sigma_{\mathrm{psr\text{-}off,} i}(v)^2}}{T_{\mathrm{psr,}i}},
    \end{equation}
  where $T_{\mathrm{psr,}i}$ is the pulsar main beam temperature for the $i$th epoch as defined in Equation~\ref{e:tau}, and $\sigma_{\mathrm{psr\text{-}on,} i}(v)$ and $\sigma_{\mathrm{psr\text{-}off,} i}(v)$ represent the noise envelopes of the $i$th epoch pulsar-on and pulsar-off spectra, respectively. The quantity $\sigma_{\mathrm{psr\text{-}on,} i}(v)$ is obtained by  
     \begin{equation}
     \begin{split} 
   & \sigma_{\mathrm{psr\text{-}on,} i}(v)\\
    &  = \sigma_{\mathrm{psr\text{-}on,}\mathrm{off\text{-}line},i} \times \frac{T^{\mathrm{on}}_{\mathrm{b},i}(v)}{T_{\mathrm{sys,off\text{-}line,psr\text{-}on},i} + T_{\mathrm{psr},i}},
     \end{split}
    \end{equation}
    where $T^{\mathrm{on}}_{\mathrm{b},i}(v) = T_{\mathrm{\hi,psr\text{-}on},i}(v)+ T_{\mathrm{sys,off\text{-}line,psr\text{-}on},i} + T_{\mathrm{psr},i}$, 
    $T_{\mathrm{sys,off\text{-}line,psr\text{-}on},i}$ is the mean system temperature of the off-line region (i.e. with no \hi\ emission) during pulsar-on for Epoch $i$ including the receiver temperature and the Galactic synchrotron background at the pulsar position,  $T_{\mathrm{\hi,psr\text{-}on},i}(v)$ represents the \hi\ emission during pulsar-on for the $i$th epoch, and $\sigma_{\mathrm{psr\text{-}on,} \mathrm{off\text{-}line},i}$ represents the standard deviation in the off-line region of the pulsar-on spectrum for $i$th epoch. 
    Similarly, $\sigma_{\mathrm{psr\text{-}off,} i}(v)$ is obtained from:
     \begin{equation}
     \begin{split} 
   & \sigma_{\mathrm{psr\text{-}off,} i}(v)\\
    & = \sigma_{\mathrm{psr\text{-}off,} \mathrm{off\text{-}line},i} \times \frac{T^{\rm off}_{\mathrm{b},i}(v)}{T_{\mathrm{sys,off\text{-}line,psr\text{-}off},i}},
     \end{split}
    \end{equation}
    where $T^{\rm off}_{\mathrm{b},i}(v) = T_{\mathrm{\hi,psr\text{-}off},i}(v)+ T_{\mathrm{sys,off\text{-}line,psr\text{-}off},i}$, $T_{\mathrm{sys,off\text{-}line,psr\text{-}off,}i}$ is the mean system temperature of the off-line region during pulsar-off for $i$th epoch, including the receiver temperature and the Galactic synchrotron background at the pulsar position,  $T_{\mathrm{\hi,psr\text{-}off},i}(v)$ represents the \hi\ emission during pulsar-off for the $i$th epoch, and  
    $\sigma_{\mathrm{psr\text{-}off,} \mathrm{off\text{-}line},i}$ represents the standard deviation in the off-line region of the pulsar-off spectrum for the $i$th epoch.

    \item[2)] Calculate the difference spectra $\Delta\left[1-e^{-\tau(v)}\right]_{ij}$ for the $i$th and $j$th $\left[1-e^{-\tau(v)}\right]$ epochs. For our 19-epoch observations, there are 171 $\Delta\left[1-e^{-\tau(v)}\right]$ spectra. 

   \item[3)] Calculate the noise envelope for the $\Delta\left[1-e^{-\tau(v)}\right]_{ij}$ spectrum. This can be expressed as 
    \begin{equation}
    \sigma_{\Delta (1-e^{-\tau}),ij}(v) = \sqrt{\left[\sigma_{(1-e^{-\tau}),i}(v)\right]^2+\left[\sigma_{(1-e^{-\tau}),j}(v)\right]^2},
    \end{equation}
    where $\sigma_{(1-e^{-\tau}),i}(v)$ and $\sigma_{(1-e^{-\tau}),j}(v)$ are the noise envelopes in the $i$th and $j$th $\left[1-e^{-\tau(v)}\right]$ spectrum, respectively. 

  \item[4)] For the $n$th Gaussian absorption component, define an initial velocity range of interest as the mean fitted FWHM: 
    $\left[v_{0,n}^{\rm mean}-\frac{1}{2}\Delta v_{0,n}^{\rm mean}, v_{0,n}^{\rm mean}+\frac{1}{2}\Delta v_{0,n}^{\rm mean}\right]$, where $v_{0,n}^{\rm mean}$ and $\Delta v_{0,n}^{\rm mean}$ represent the 19 epoch-averaged central velocity and FWHM for the $n$th component. 

    \item[5)] 
    Define the signal within the FWHM velocity range of each Gaussian component range as the sum over the $N$ channels in that range, $\sum ^{N}_{k=1}\left[ \Delta \left( 1-e^{-\tau}\right)_{ij}(v_k)\right]$, where $k$ represents the $k$th channel within the online region.(Note: This approach is designed to capture variations with the same sign within the line profile. )

    \item[6)] Calculate uncertainty of the signal region in the difference spectrum between epochs $i$ and $j$ on that sum as $\sqrt{\sum ^{N}_{k=1}\left[ \sigma_{\Delta \left( 1-e^{-\tau}\right),ij}(v_k)\right] ^{2}}$.

    \item[7)] Calculate the signal-to-noise ratio (SNR) by dividing the summed signal by its corresponding uncertainty.

    \item[8)] For any samples with an SNR greater than 3, adjust the velocity range of the sum until the SNR is maximised.
\end{itemize}

\begin{figure*}
\centering
\includegraphics[width=0.495\linewidth]{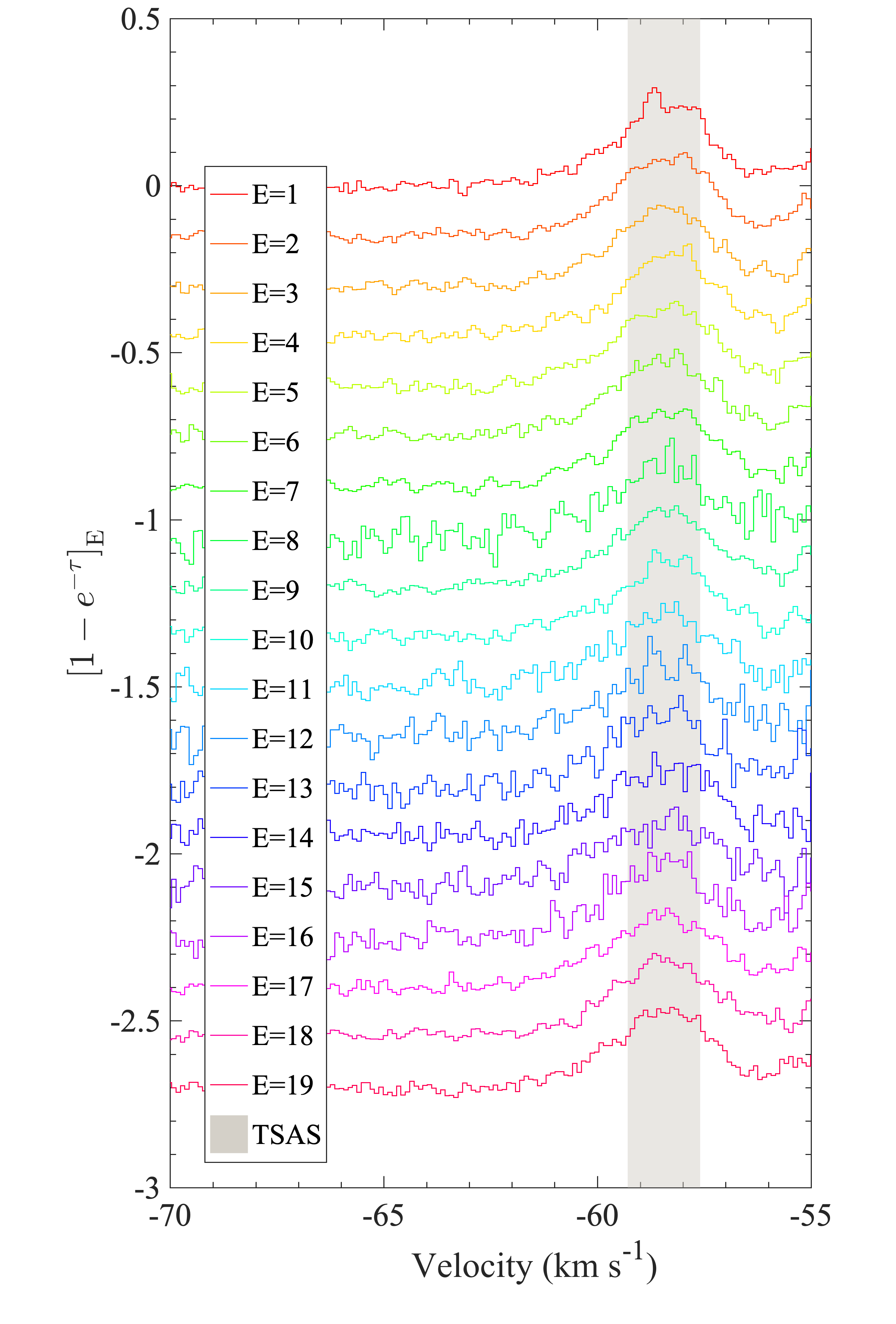}
\includegraphics[width=0.495\linewidth]{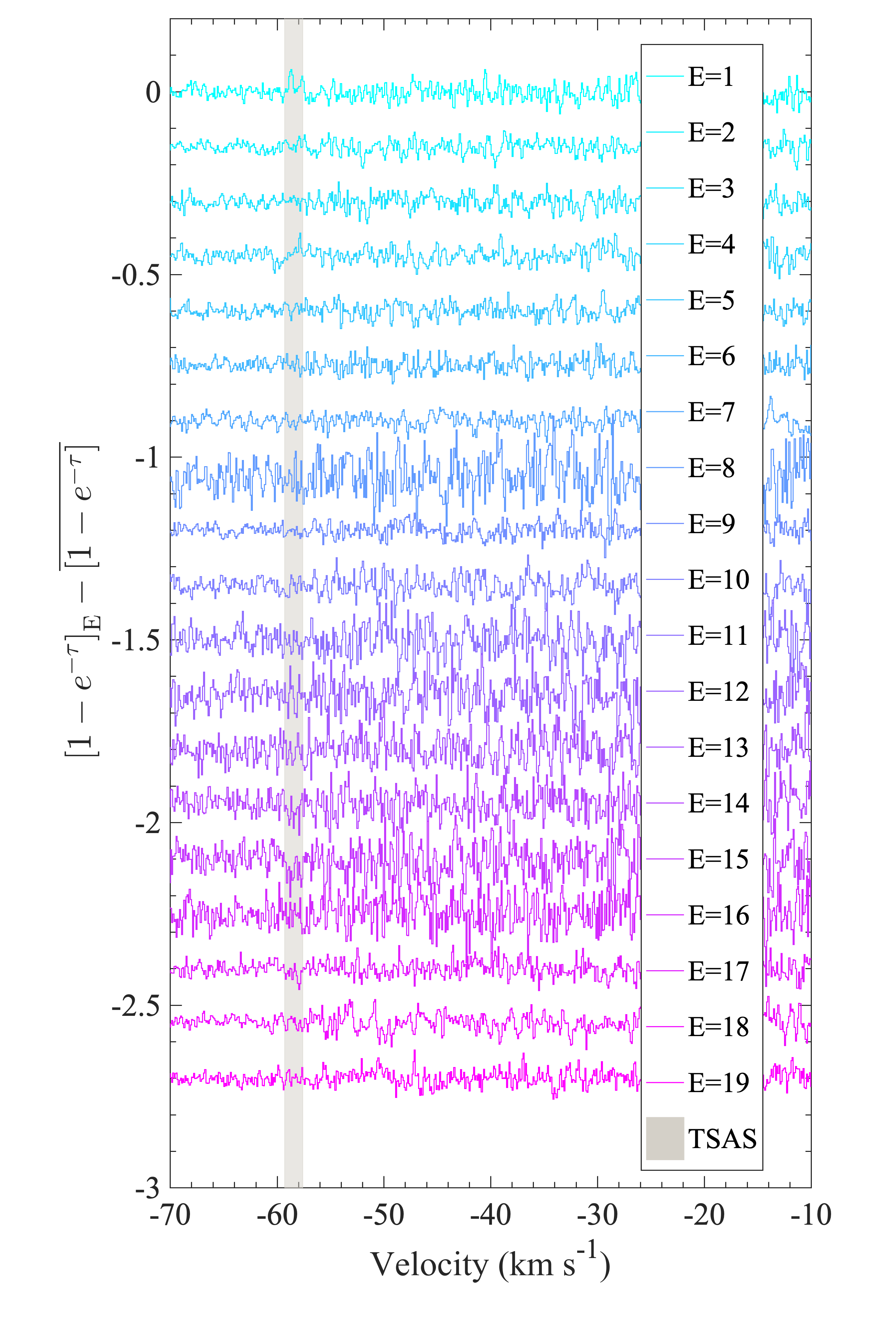}
\caption{\textit{Left panel:} The $\left[1-e^{-\tau}\right]_{\rm E}$ spectra for each epoch (E) of the CNM component exhibiting significant temporal variations. Each spectrum is offset by 0.15 for clarity. \textit{Right panel:} 
The difference spectra $\left[1-e^{-\tau}\right]_{\mathrm{E}} - \overline{\left[1-e^{-\tau}\right]}$ for each epoch (E) relative to the noise-weighted mean spectrum $\overline{\left[1-e^{-\tau}\right]}$. For clarity, each spectrum is offset vertically by 0.15. 
The velocity range over which significant variation was detected is marked in grey.} %The epoch pairs showing variability above a SNR of 5.0 are Epochs 1 and 15, and Epochs 1 and 17. }
\vspace{0.2cm}
\label{fig:tsas_19epoch} 
\end{figure*}

As a result of this analysis, we find that while 12 \hi\ absorption features remained stable over 1.2 years, significant variations (SNR exceeding 5.0) were observed for the absorption feature centered at $-58.6$ \kms\, with the maximum SNR of 5.8 achieved for a velocity range of [$-59.3$, $-57.6$] \kms. Figure~\ref{fig:tsas_19epoch} shows a closeup of this absorption feature over all 19 epochs. 

\begin{figure*}
\centering\includegraphics[width=0.49\linewidth]{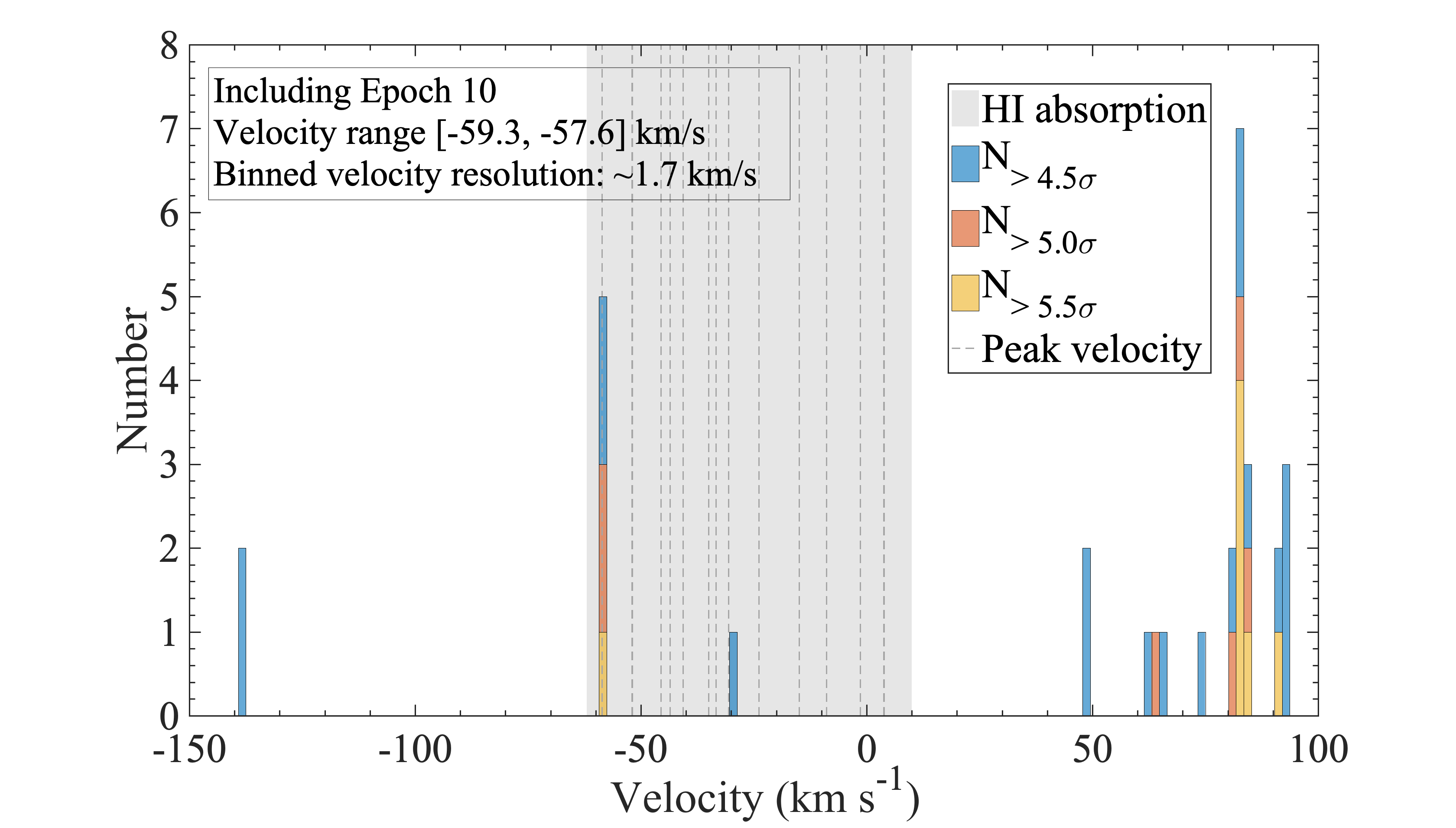}
\includegraphics[width=0.49\linewidth]{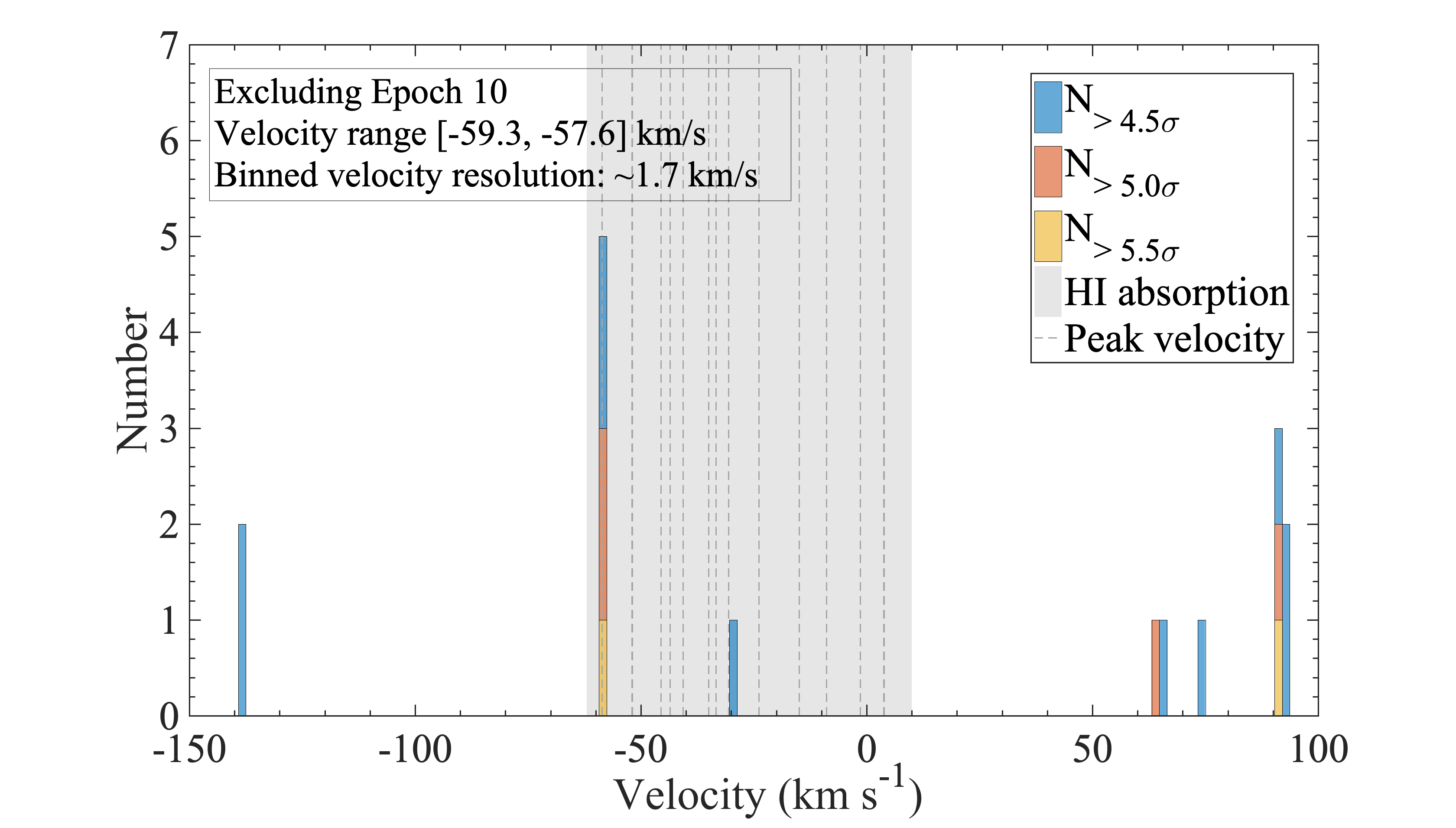}
\includegraphics[width=0.49\linewidth]{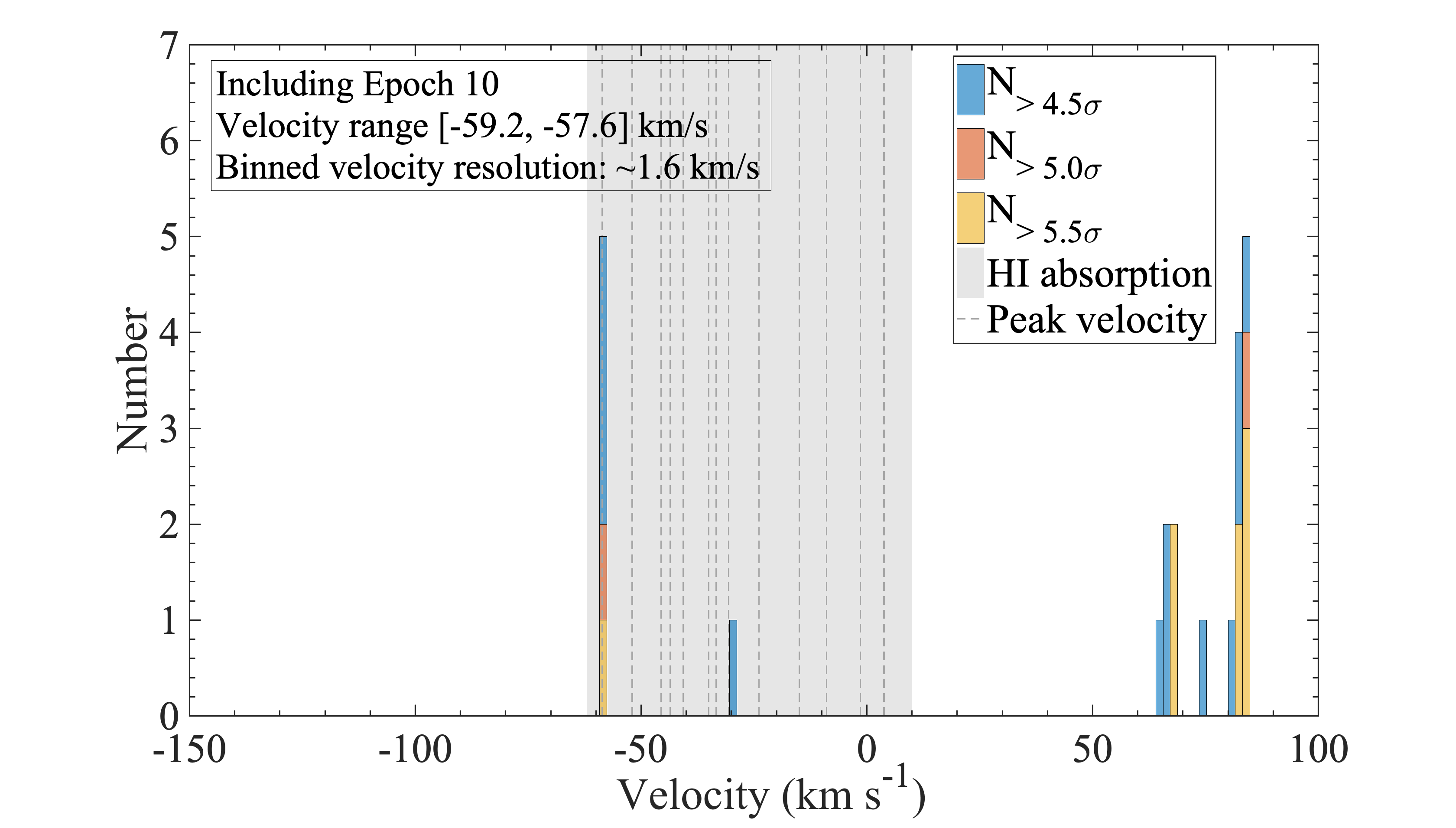}
\includegraphics[width=0.49\linewidth]{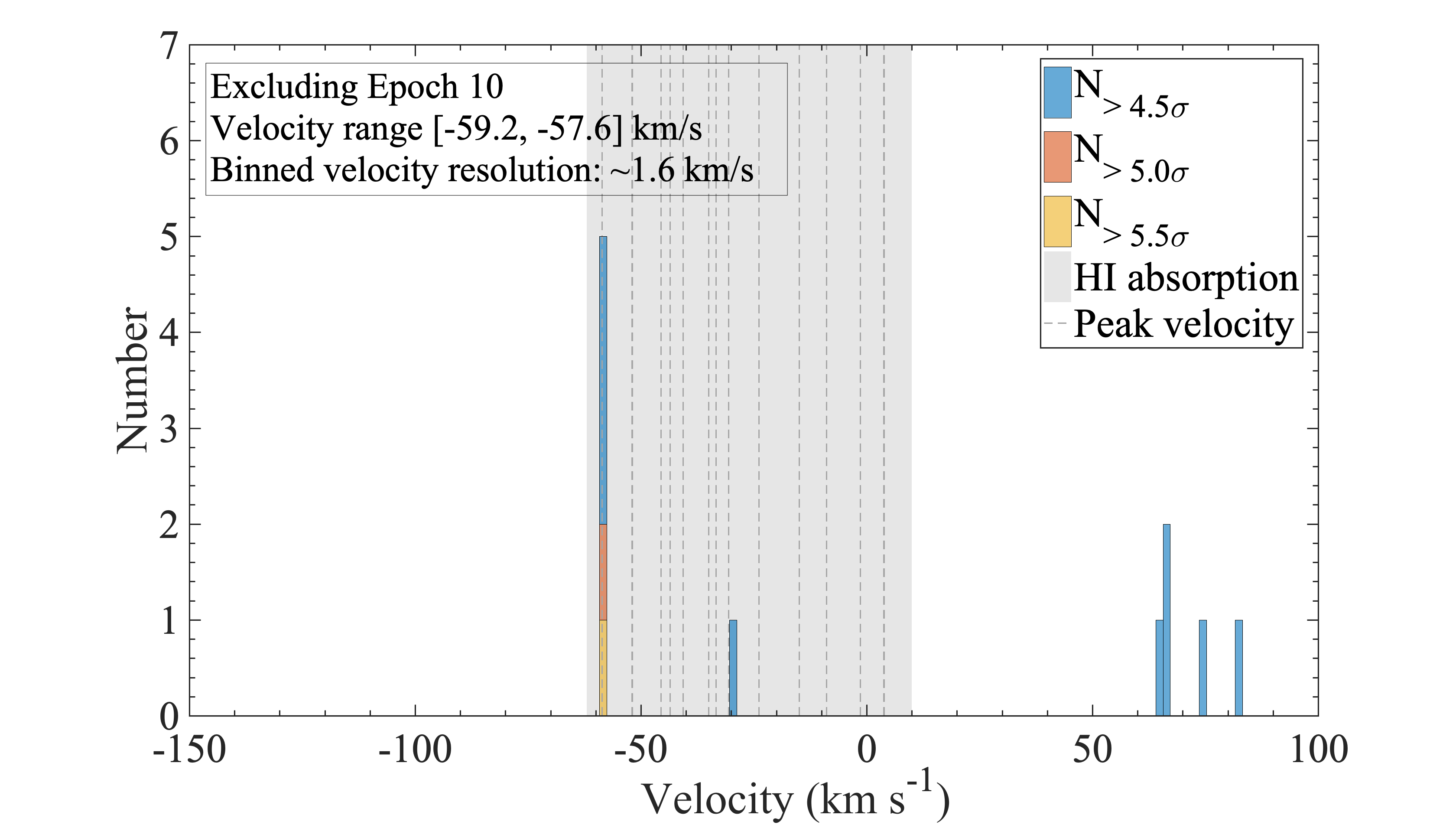}
\caption{\textit{Top left}: Number of detections of \hi\ temporal variability with signal-to-noise ratios of 4.5, 5.0, and 5.5 across all pairs of epoch-to-epoch difference spectra. The bin width is 1.7 \kms, with the bin placement set such that the velocity range of the persistently detected variation at 58.3 \kms\ is 
$-59.3$ to $-57.6$ \kms. \textit{Top right}: As top left panel, but excluding Epoch 10. 
\textit{Bottom left}: Number of detections of \hi\ temporal variability with signal-to-noise ratios of 4.5, 5.0, and 5.5 across all pairs of epoch-to-epoch difference spectra. The bin width is 1.6 \kms, with the bin placement set such that the velocity range of the persistently detected variation at 58.3 \kms\ is $-59.2$ to $-57.6$ \kms. \textit{Bottom right}: As bottom left panel, but excluding Epoch 10. 
The gray area represents the velocity range of \hi\ absorption, while the dashed lines mark the fitted central velocities for the 13 CNM components.
} 
\vspace{0.2cm}
\label{fig:N_old} 
\end{figure*}

To make a rough assessment of the potential for false positive detections, we binned all difference spectra to a velocity resolution of 1.7 \kms (to match the highest SNR bin width of the $-58.6$ \kms\ feature), and evaluated the resulting detection statistics. 
Figure~\ref{fig:N_old} shows histograms of the total number of detections with SNR thresholds of 4.5, 5.0, and 5.5. 

We found that while real \hi\ absorption is only present within the velocity range of [$-70$, 10] \kms,  
a significant number of apparent spectral variations are found outside this range, particularly around 83 \kms. The location of these variations outside of the true velocity range of the \hi\ absorption indicates that they are spurious. This clearly warrants further investigation, since the existence of such prominent false detections could potentially cast doubt on the validity of the $-58.6$ \kms\ detection.  
Upon inspection, we found that the majority of detections at 83 \kms\ were due to comparisons with Epoch 10. 
We ruled out the possibility of contamination from spectral lines with rest frequencies near that of atomic hydrogen, such as the \hi\ recombination line $\textrm{H}_{(209)\beta}$ and formic acid: $t-$H$^{13}$COOH, as detailed in Appendix~\ref{sec:spec_fake}. 
In the absence of an astrophysical explanation, we suspect an instrumental cause and therefore opted to exclude Epoch 10 from further analysis. 
When Epoch 10 is excluded, the number of spurious detections  outside of the absorption velocity range is greatly decreased, as seen in the top right panel of Figure~\ref{fig:N_old}, while the strongest and most persistent detections remain centered on $-58.6$ \kms. 

Next, we recognise that `tuning' the velocity binning of our data to favor the strongest detection in the $-58.6$ \kms\ limits the statistical robustness of our false positive test. We therefore experiment with varying the bin width and placement. The  bottom two panels of Figure~\ref{fig:N_old} show example results for a bin width of 1.6 \kms\ such that the $-58.6$ \kms\ feature range is [$-59.2$,$-57.6$]. Reassuringly, the spurious detections are sensitive to the chosen velocity binning, whereas the $-58.6$ \kms\ detection is not. In the example under discussion, 
%For instance, when changing the bin width such that the $-58.6$ \kms feature range is [$-59.2$,$-57.6$] \kms\ and excluding Epoch 10, 
there are no detections above SNR 5.0 outside of the absorption velocity range, as shown in the bottom two panels of Figure~\ref{fig:N_old}, whereas the $-58.6$ \kms\ detection remains strong. %However, the detection at $-58.6$ \kms\ persists across all configurations, suggesting its likely validity. 

While this rough analysis does not definitively rule out the possibility that the $-58.6$ \kms\ variation is a false detection, there are several other factors that support its validity. Firstly, it occurs exactly at an absorption component peak -- a false detection due to instrumental effects should show no such preference for location. Secondly, as will be discussed in Section \ref{sec:sf}, the host CNM component of the $-58.6$ \kms\ variation is the only \hi\ absorption component that shows evidence of a turbulent power law slope in $\Delta\tau_{\rm int}$, once the optical depth differences of all epoch pairs are considered. An instrumental false positive (e.g. due to a spurious narrow-band correlator feature in this part of the band) should generate no such physical relationship.  

\begin{figure*}
\centering
\includegraphics[width=0.45\linewidth]{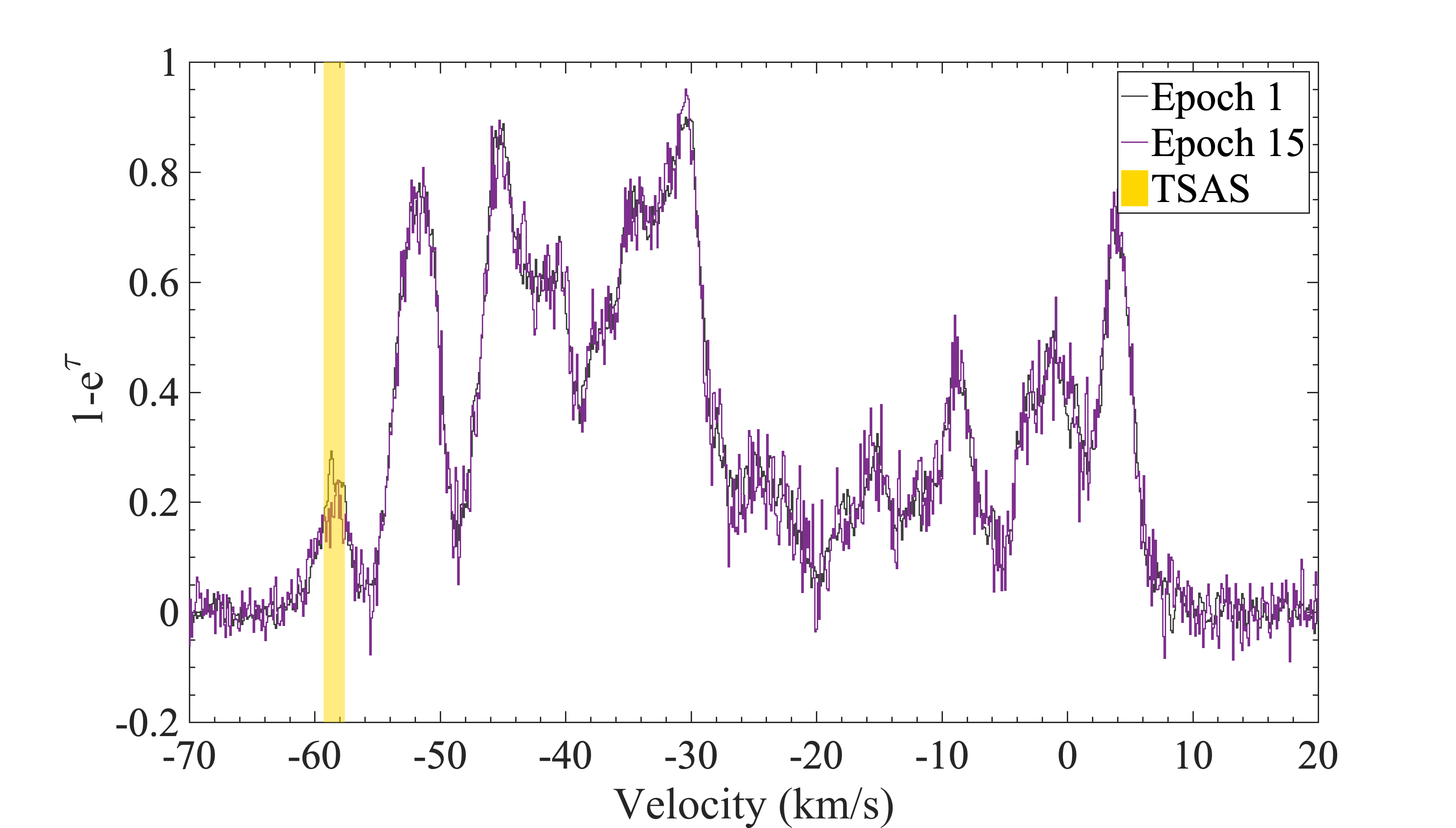}
\includegraphics[width=0.45\linewidth]{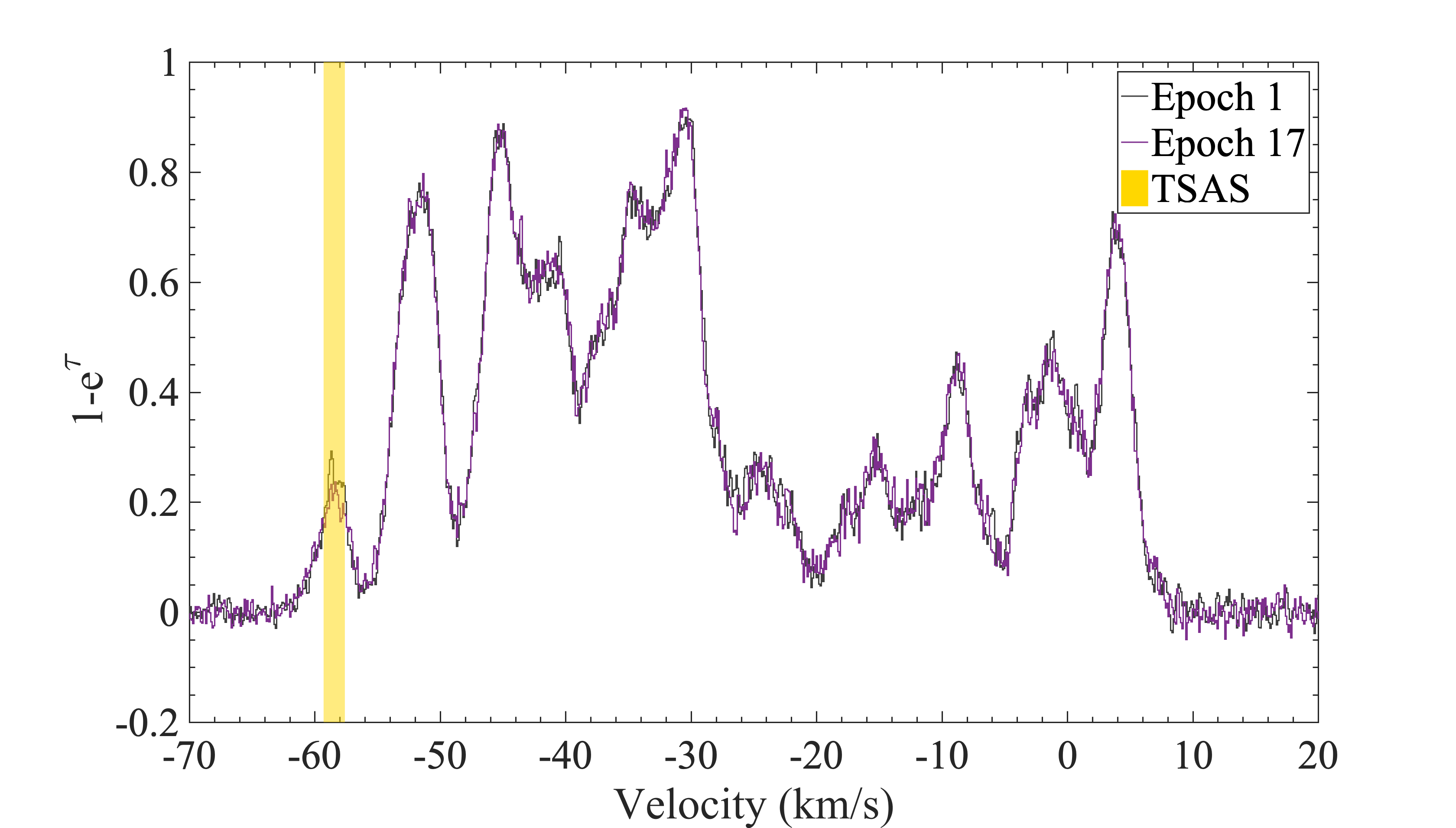}
\includegraphics[width=0.45\linewidth]{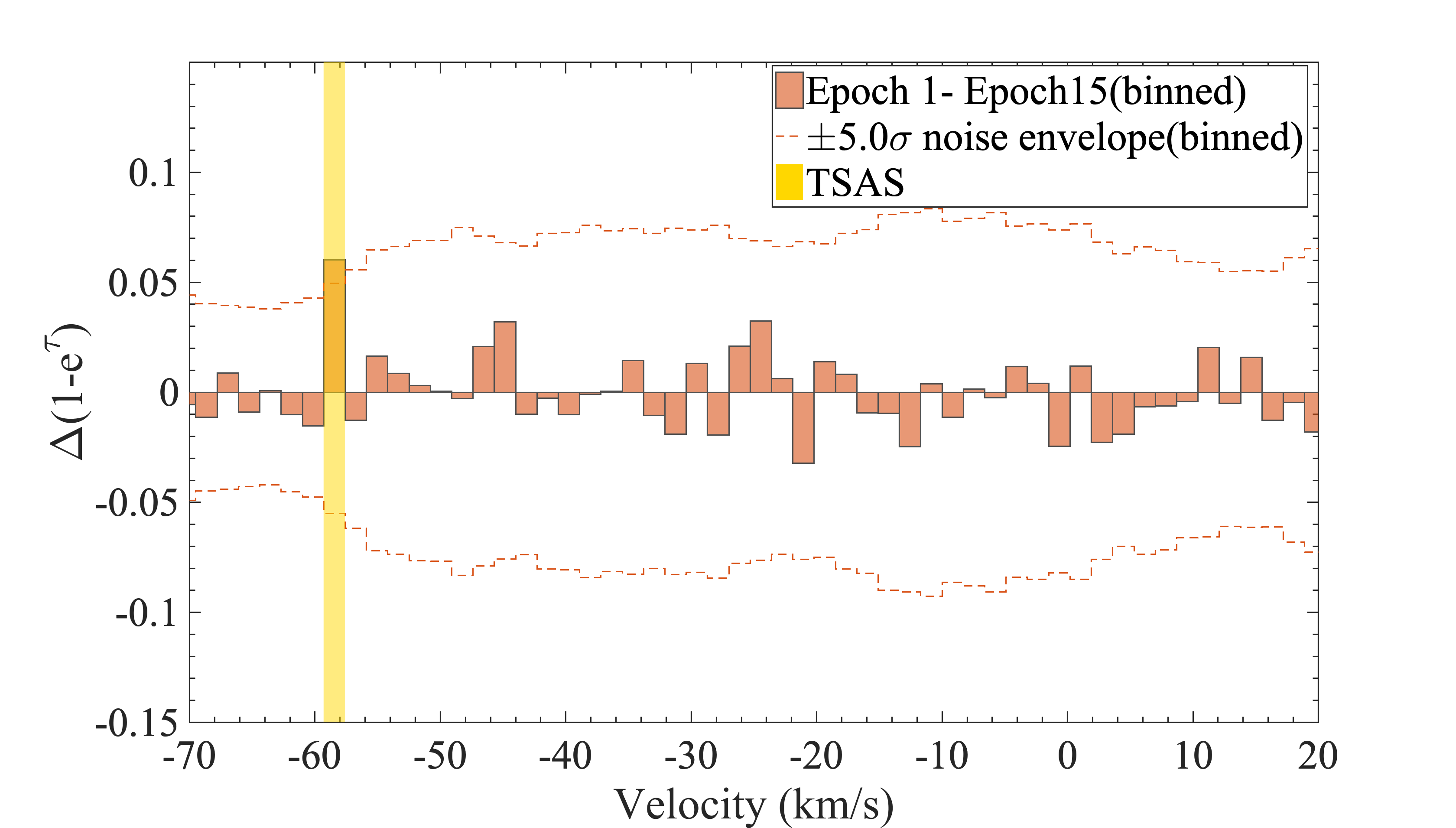}
\includegraphics[width=0.45\linewidth]{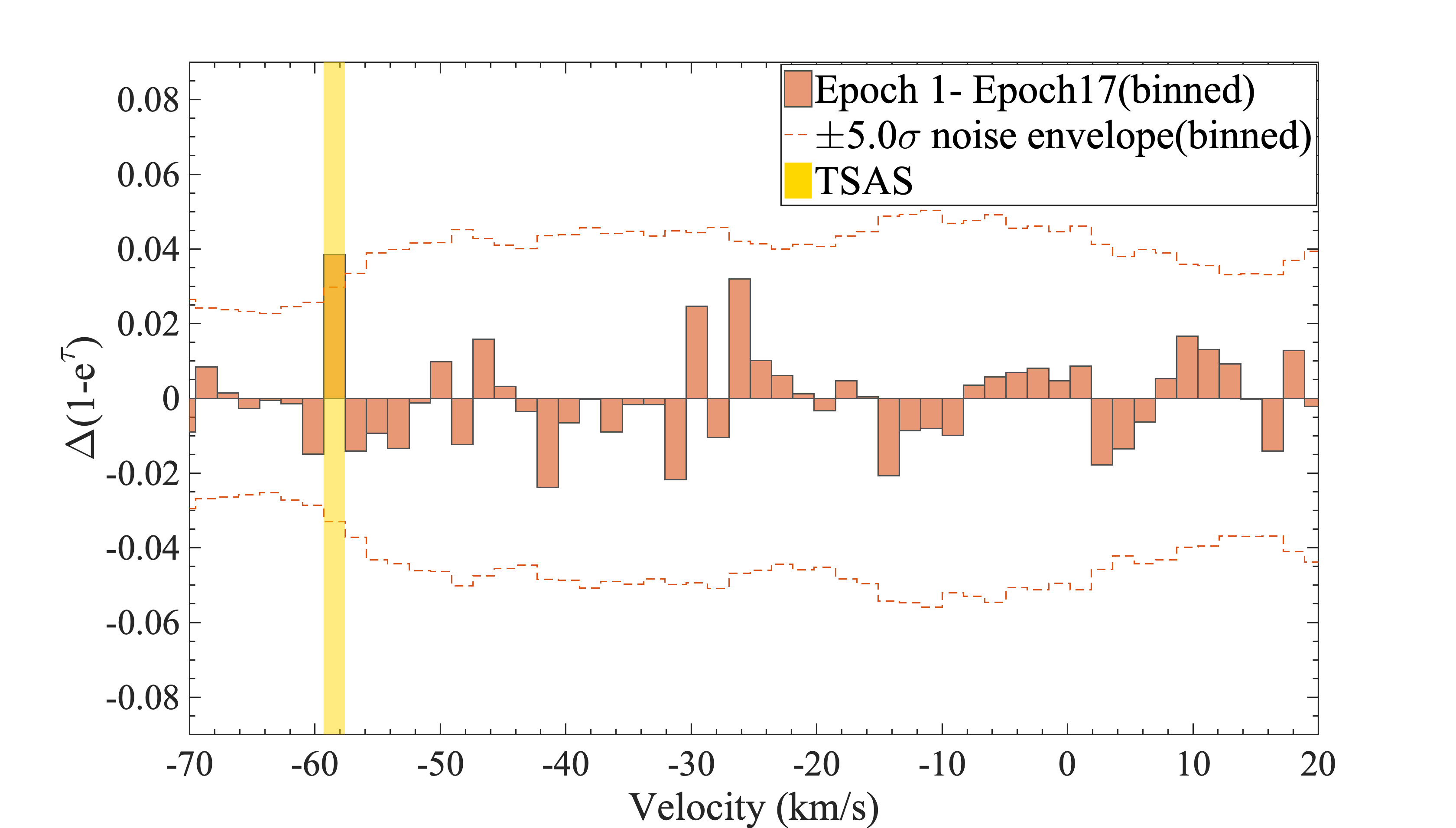}
\caption{ 
{\textit{Left top panel:}} The 0.1 \kms\ resolution, $(1 - e^{-\tau})$ spectra obtained on 2022 Feb 18, Epoch 1 (solid black line) and 2023 Mar 10, Epoch 15 (solid purple line). The velocity range showing a significant difference is highlighted in yellow; otherwise the two are almost indistinguishable.   {\textit{Left bottom panel:}} The epoch-difference, binned 1.7 \kms\ resolution, $\Delta(1 - e^{-\tau}),$ spectrum between the above two epochs is shown in red, with red dashed lines indicating its $\pm$5.0-$\sigma$ noise envelope. 
{\textit{Right top panel:}} The 0.1 \kms\ resolution, $(1 - e^{-\tau})$ spectra obtained on 2022 Feb 18, Epoch 1 (solid black line) and 2023 Mar 13, Epoch 17 (solid purple line). The velocity range showing significant variation is highlighted in yellow; otherwise the two are almost indistinguishable.
{\textit{Right bottom panel:}} As left bottom panel. 
} 
\vspace{0.2cm}
\label{fig:Deltatau_spec_all_1420MHz} 
\end{figure*}

For the purposes of defining a `detection', in line with common practice in the TSAS literature, we adopt a conservative SNR of 5.0. This results in the detection of two significant temporal variations in \hi\ absorption, both centered on $-58.6$ \kms, arising from differences in the absorption spectra of Epochs 1 and 15, and Epochs 1 and 17. 
Their $\Delta\int(1 - e^{-\tau})\,\text{d}v$ values are $\sim$0.099$\pm$0.018 and 0.063$\pm$0.011 \kms, and they have SNR of 5.5 and 5.8, respectively, over a velocity range of [$-59.3$,$-57.6$] \kms. 
These variations correspond to characteristic spatial scales, with a lower limit $\Delta L^{\rm Low}$= 10.5$\pm$0.8 au and 10.6$\pm$0.8 au, and an upper limit $\Delta L^{\rm Up}$= 163.8$\pm$12.3 au and 165.3$\pm$12.7 au, respectively.  
A comparison of the $(1 - e^{-\tau})$ spectra of these two detections and their corresponding difference spectra can be found in Figure~\ref{fig:Deltatau_spec_all_1420MHz}. 
We note that our TSAS detection occurs in a region of the spectrum where the \hi\ emission is weak, and hence the spectral noise is lower (see the dashed line in the lower panels of Fig.~\ref{fig:Deltatau_spec_all_1420MHz}). 
It is possible that the increased noise in the other CNM components is partly responsible for the non-detection of TSAS at their velocities.

\section{Discussion} \label{sec:disc}

We will now discuss these results in the context of two interpretations. The first is a classic ``cloudlet'' picture of TSAS that considers only the detections of high statistical significance and asks what properties a discrete ISM structure would need to possess in order to generate the observed fluctuations. 
The second considers a turbulent origin, and incorporates measurements between all epoch-pairs. 

\subsection{Properties of a classic TSAS cloudlet}\label{sec:TSAS_prop}
If we interpret the \hi\ opacity variations as arising from tiny \hi\ cloudlets along the line of sight, their properties are summarized in Table~\ref{tab:tsas}.  
The characteristic spatial scales inferred for the variations are similar to previous TSAS measurements from pulsar and interferometer observations \citep{2018ARA&A..56..489S}. 
The spin temperature of the CNM component hosting the measured opacity variations is derived from the Gaussian decomposition, as described in Section \ref{sec:fitting}, and ranges from $126.7\pm7.0$ to $161.0\pm6.0$ K. For reference, the Gaussian fits to this component over the 19 epochs are illustrated in Figures~\ref{fig:cnm_TSAS_fit0} through \ref{fig:cnm_TSAS_fit3}. 
Since the spin temperatures derived from the Gaussian fitting process for the host absorption component may be affected by the presence or absence of the TSAS in the spectral profile for any given epoch, here we adopt a spin temperature of $T_\mathrm{s}^{\mathrm{TSAS}} = 141.6 \pm 2.3$ K for both detections, which is the mean of the parent CNM's spin temperatures measured across the 19 epochs. 
The corresponding \hi\ column densities $N{(\hi)}_{\rm TSAS}$ for the two TSAS are 0.3 and 0.2 $\times10^{20}$ $\rm cm^{-3}$, respectively.
These values were computed using 
$N{(\hi)}_{\rm TSAS}= C_0 \cdot \int \Delta\tau\mathrm{d}v \cdot T_s^{\mathrm{TSAS}}$, where $C_0 = 1.823\times10^{18}$ cm$^{-2}/($km s$^{-1}$ K), $\Delta\tau$ is the channel-by-channel sum of the difference in optical depth across the defined TSAS velocity range (see Table~\ref{tab:tsas}), and $\mathrm{d}v$ is the velocity resolution of 0.1 \kms.

Assuming a line of sight dimension comparable to the length-scale on the plane of the sky, the inferred lower and upper limits of \hi\ volume densities and thermal pressures for our hypothetical two TSAS cloudlets are [1.3, 20.2] and [0.8, 13.02]$\times 10^{4}$ cm$^{-3}$ (cloudlet 1), and [0.2, 2.9] and [0.1, 1.8]$\times 10^{7}$ cm$^{-3}$ K (cloudlet 2),  respectively. 
Similar to previous TSAS measurements, these values are three to four orders of magnitude higher than the values expected in the neutral medium. 
To bring down these values to the typical values observed in the ISM, the spatial scale of the line of sight would need to be $\sim$0.3 pc corresponding to an elongation factor of $\sim$1000--5000, which is significantly larger than the maximum aspect ratio observed in \hi\ filaments ($\sim$100) \citep{Yuen_HI_filament_2024}. 
Therefore, the TSAS probed in our study are unlikely to be \hi\ filaments. 
Unfortunately,
the one-dimensional nature of this experiment precludes direct measurements of the elongation.

We estimated a turbulent velocity dispersion for the two TSAS cloudlets by adopting the properties of the parent Gaussian component, and assuming the nonthermal velocity dispersion is dominated by turbulent motions such that $\sigma_{\mathrm{turb}}^2$= $(\Delta v_0/2.355)^2-$ $k_{\rm B}T_{\rm s}^{\rm TSAS}/m_{\rm H_0}$
where $\Delta v_0= 2.8$ \kms\ is the FWHM of the parent CNM component, $k_{\rm B}$ is the Boltzman constant  
and $m_{\rm H_0}$ is the hydrogen mass. 
The resultant turbulent velocity dispersion is 0.5 \kms, whereas the thermal velocity dispersion, $\sigma_{\rm therm}=\sqrt{k_{\rm B}T_{\rm s}^{\rm TSAS}/m_{\rm H_0}}$, is 1.1 \kms.  
Based on the analysis of \citet{2018ARA&A..56..489S}, turbulence can generate significant pressure variations, which are often associated with large velocity fluctuations.
To confine a cloudlet that is over pressured by a factor of 100, the turbulent velocity dispersion would need to be at least 10 times the thermal velocity dispersion \citep{1992ApJ...399..551M,2018ARA&A..56..489S}. 
Clearly this is not achieved for our TSAS detections. Taken at face value, this analysis suggests that if our measured TSAS did indeed arise from tiny overpressurised cloudlets, then this cloudlet would be 
short-lived and eventually expand.

\begin{rotatetable*}
\begin{deluxetable*}{cccccccccccccccccc}
\setlength{\tabcolsep}{0.05in} 
\tabletypesize{\scriptsize}
\tablewidth{1.5pt}
\tablecaption{Properties of TSAS under the assumption of tiny cloud structures \label{tab:tsas} }
\tablehead{\colhead{Epochs}      &
\colhead{$\Delta\int (1 - e^{-\tau})$d$v$}      &
\colhead{Center velocity}      &\colhead{ $L_{\perp}$ }& \colhead{Dist}& \colhead{$\Delta v_0$} &   \colhead{$\int\Delta\tau$d$v$} &\colhead{$T_s^{\mathrm{TSAS}}$} &
\colhead{$N(\rm HI)_{\rm TSAS}$}    & \colhead{$n(\rm HI)_{\rm TSAS}$ } & $P/k$ &\colhead{$\sigma_{\rm turb}$}& \colhead{$\sigma_{\rm therm}$}\\
\colhead{}&\colhead{(\kms)}&\colhead{(\kms)}   &\colhead{(au)}& \colhead{(kpc)}   &   \colhead{(\kms)} &\colhead{ (\kms)}&
\colhead{($\rm\,K$) } &  \colhead{($10^{20}\rm\,cm^{-2}$) }&    \colhead{($10^{4}\rm\,cm^{-3}$) } &  \colhead{($10^{7}\rm\,cm^{-3}\,K$) } &       \colhead{(\kms) }&
\colhead{(\kms) }
}
\startdata
1 and 15 &0.099$\pm$0.018 &$-58.6\pm 0.1$  & [10.5$\pm$0.8, 163.8$\pm$12.3] &3.8 & 1.7 &   0.121$\pm$0.026 & 141.6 $\pm$ 2.3 &  0.3  &  [1.3, 20.2] &  [0.2, 2.9]  &   0.5 &  1.1\\  
1 and 17& 0.063$\pm$0.011 &$-58.6\pm 0.1$  & [10.6$\pm$0.8, 165.3$\pm$12.7] &3.8 & 1.7 &   0.078$\pm$0.016 & 141.6 $\pm$ 2.3 &  0.2  & [0.8, 13.0]  &  [0.1, 1.8]  &  0.5 & 1.1 \\  
\enddata
\tablecomments{ Col. (1): 
Observing Epochs during which TSAS were detected. Col. (2): Integrated $\int(1-e^{-\tau})$ variations. Col. (3): TSAS central velocity. Col. (4): The lower and upper limits of characteristic spatial scale of TSAS, based on its and the pulsar’s relative transverse motion, assuming that both are at rest with respect to their own LSRs, as discussed in Section \ref{sec:obs}. 
Col. (5): TSAS distance. Col. (6): Velocity width of TSAS. Col. (7): Integrated optical depth variations. Col. (8): Spin temperature. 
Col. (9): TSAS \hi\ column density. It is calculated by
$N{(\rm HI)}_{\rm TSAS} = C_0 \cdot \int\Delta\tau$ d$v\cdot T_s$. 
Col. (10): The lower and upper limits of TSAS \hi\ volume density. Col. (11): The lower and upper limits of thermal pressure. Col. (12): The one dimensional turbulent velocity.  Col. (13): The one dimensional thermal velocity.
} 
\end{deluxetable*}
\end{rotatetable*}

\subsection{Optical Depth Structure Function and the Steep Power-law Index $\alpha$} \label{sec:sf}

\citet{2000MNRAS.317..199D} proposed that fluctuations in the optical depth of \hi\ measured on au scales 
may be explained as the result of contributions from structure on all scales, arising from a single turbulent power-law spectrum. 
If the opacity power spectrum is a power law with a slope of $\alpha$ where $2< \alpha <4$, then the opacity structure function will have a power law slope of $\alpha-2$, and the correlation between optical depth fluctuations and spatial scales can be expressed as \citep{1975ApJ...196..695L,2000MNRAS.317..199D,2012ApJ...749..144R} 
\begin{equation}
\Delta\tau=\Delta\tau_{0}\Delta L^{(\alpha-2)/2}.
\label{eq:powerlaw}
\end{equation} 

Several studies have measured the value of $\alpha$ down to au scales, but none had previously done so while simultaneously detecting the significant optical depth fluctuations typically defined as TSAS. By measuring the \hi\ opacity power spectrum from high-resolution interferometric images of \hi\ absorption towards Cygnus A and Cas A, \citet{2000ApJ...543..227D} reported values of $\alpha=2.75\pm 0.25$ for the Perseus arm and Outer arm, and $\alpha=2.5$ in the Local arm, over spatial scales ranging from $\sim$2000 au to $\sim$3 pc. 
\citet{2010MNRAS.404L..45R} employed a robust method to extract fluctuations in the \hi\ opacity power spectrum from interferometric data towards Cas A. 
They found $\alpha = 2.86\pm 0.1$ for both the Perseus arm and the Local arm, over scales of 0.07 to 2.3 pc and 400 au to 0.07 pc, respectively. 
\citet{2012ApJ...749..144R} found a slightly shallower value of $\alpha=2.33\pm 0.52$ over spatial scales of $\sim$5--100 au by measuring the opacity structure function from high angular resolution observations of \hi\ absorption toward 3C 138. 
With an unbiased estimator, \citet{2014MNRAS.442..647D} derived $\alpha=2.81\pm 0.59$ on spatial scales ranging from 5 to 40 au, from similar observations toward the same source.

We derived integrated optical depth variation, $\Delta\tau_{\rm int}=\Delta\tau$d$v$ (where d$v$ is the velocity resolution of $\sim0.1$ \kms), for each of our 13 Gaussian components, for all 171 pairs of 19 epochs. 
For the $-58.6$ \kms\ CNM component associated with our TSAS detections, we used the same velocity range as before, over which the SNR of the measured variations was highest (see Section \ref{sec:var_HI_abs}).  
For the other 12 features, we adopted the average fitted FWHM velocity range over the 19 epochs. 
Kendall's Tau Correlation Test was used to assess the association between $\Delta\tau_{\rm int}$ and $\Delta L$, the inferred length scale \citep{2010ApJ...720..415S}. The CNM component associated with TSAS is the only component showing a clear correlation, with a corresponding probability of (1--$0.5\times10^{-11}$). 
This represents the first time that a correlation between optical depth variation and length scale has been reported simultaneously with the detection of TSAS, and we consider this strong evidence that turbulence is implicated in TSAS production.

\begin{figure*}
\centering
\includegraphics[width=1.0\linewidth]{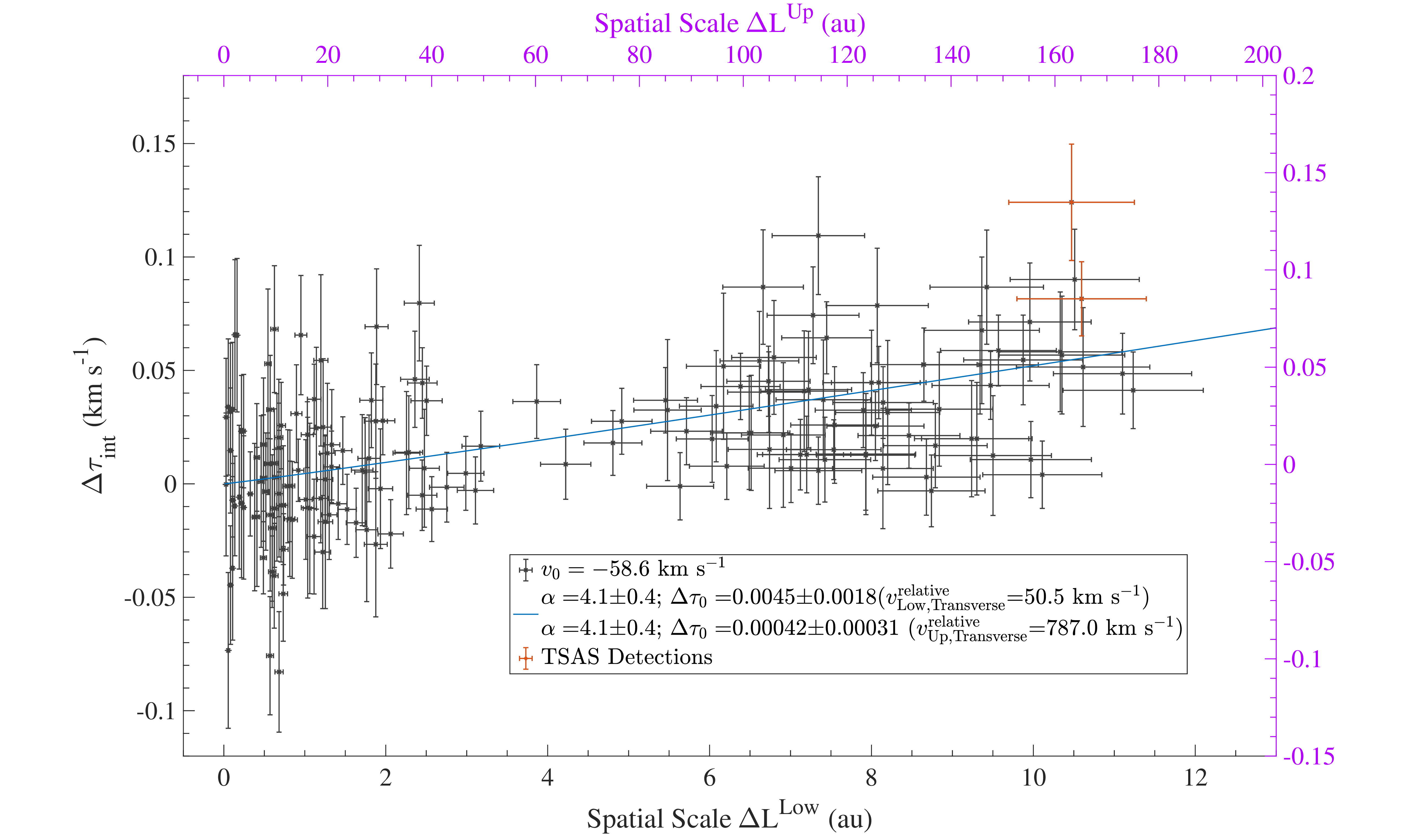}
\caption{Relationship between the integrated optical depth variations, $\Delta\tau_{\rm int}$, and spatial scale, $\Delta L$, for the CNM associated with TSAS centering at $-58.6$ \kms\ over 19 epochs. The channels used to estimate $\Delta\tau_{\rm int}$ share the same velocity range as TSAS detections, ranging from $-59.3$ to $-57.3$ \kms. Error bars represent $\pm1\sigma$ uncertainties. The blue line illustrates the fitting results for the power-law relation of $\Delta\tau_{\rm int}$ and $\Delta L$. 
Two red errorbars represent the TSAS detections. 
The black and purple x- and y-axes represent the results with $v^{\rm relative}_{ \rm Low, Transverse}$ and $v^{\rm relative}_{ \rm Up, Transverse}$, respectively. }
\vspace{0.2cm}
\label{fig:tur_fit} 
\end{figure*}

For this component, we fitted $\alpha$ and $\Delta\tau_0$ for Equation \ref{eq:powerlaw}, the power-law relation describing the channel-by-channel sum of the difference in optical depth as a function of spatial scales. 
In order to assess the best fit and associated uncertainties, we conducted 1000 Monte Carlo simulations of $\Delta \tau$ and $\Delta L$, adding random noise to these parameters based on a Gaussian probability distribution function, with a standard deviation corresponding to our measurement uncertainties. 
The estimates of $\alpha$ and $\Delta\tau_0$, including their uncertainties, were derived by computing the mean and standard deviations across all simulations. 
The resulting power-law index, with $\alpha \approx 4.1\pm0.4$ for spatial scales, from 0.03 to 11.2 au (lower limit) and from 0.4 to 174.7 au (upper limit) is presented in Figure \ref{fig:tur_fit}. 

Based on numerical simulations of magnetohydrodynamic turbulence, \citet{Lazarian_2000}, \citet{Lazarian_2001}, and \citet{2004ApJ...616..943L} demonstrated that both the turbulence-induced 3D density variations and 3D velocity fluctuations contribute to the observed column density power spectrum \citep{Yuen_2021}.
When the column density power spectrum index is estimated from 2D position-position-velocity datacubes integrated over a velocity range larger than the turbulence velocity dispersion (in thick slicing), it is identical to that of the underlying 3D turbulence density power spectrum. 
In the case of our detected TSAS sample and its hosting CNM component, the velocity integration for detecting the TSAS is 1.7 \kms, which is larger than the turbulence velocity dispersion of the host CNM (0.5 \kms, as listed in Table~\ref{tab:tsas}). 
Consequently, our measurements of the power spectrum index $\alpha=4.1\pm0.4$ from optical variations (or column density variations, in our case, since the spin temperature of the \hi\  component hosting the TSAS remains stable over 19 epochs, and opacity variation can be treated as column density variation) over different spatial scales is expected to reflect the 3D turbulence density spectrum. 
The estimated $\alpha$ value is significantly steeper than those reported in previous studies with $\alpha=2.3$-- 2.9, but close to the power-law index of the electron column density power spectrum with $\alpha$=11/3 ranging from 10$^6$ to 10$^{18}$ m, known as the ``giant power-law in the sky'' \citep{1995ApJ...443..209A,2010ApJ...710..853C,2019NatAs...3..154L}.  

\subsection{The Steep Power-law Index $\alpha$ and the Sonic Mach Number $M_{\rm s}$} \label{sec:alpha_ms}

Previous magnetohydrodynamic (MHD) turbulence simulations have shown that the power-law slope of column density power spectra is highly dependent on the sonic Mach number $M_{\rm s}$ \citep{2005ApJ...630L..45K,2005ApJ...624L..93B,2007ApJ...666L..69K,2010ApJ...708.1204B}. 
(Note that although our estimation of the power-law index is based on opacity fluctuations, under the reasonable assumption we have made of a constant spin temperature, the observed opacity variations correspond directly to changes in \hi\ column density). 
In supersonic turbulence, the column density power spectrum is dominated by turbulence-induced density fluctuation; thus it is not expected to follow a Kolmogorov spectrum. 
The observed power-law index is shallower than the Kolmogorov value of 11/3 \citep{2017ApJ...835....2X,2012A&ARv..20...55H}. 
As the sonic Mach number $M_{\rm s}$ decreases, the power-law slope $\alpha$ becomes steeper. 
In transonic ($M_{\rm s}\sim 1$) and subsonic ($M_{\rm s}<1$) regimes, turbulence-induced density fluctuations behave as a passive scalar and are expected to follow the same Kolmogorov spectrum as turbulent velocity \citep{2017ApJ...835....2X}. Thus, 
the column density fluctuations are expected to follow the same Kolmogorov scaling law as the velocity fluctuations, characterized by $\alpha \approx 11/3 \sim 3.7$ 
\citep{2003MNRAS.345..325C,2017ApJ...835....2X}.

To assess whether the detection of a steep power-law index in our study might be due to lower values of $M_{\rm s}$ than those of previous \hi\ opacity structure function studies, 
we applied the methods outlined in \citet{2003ApJ...586.1067H} to estimate $M_{\rm s}$ for the CNM component associated with significant variability in our observations, as well as for previous studies that derived the \hi\ opacity structure function. 
By adopting the same mean atomic weight of $1.4 m_{\rm H}$ and a corresponding fractional helium abundance of 0.15 as in \citet{2003ApJ...586.1067H}, the sonic Mach number, $M_{\mathrm{s}}$, can be expressed as:
\begin{equation}
M_t^2 = {V_{t, \mathrm{3D}}^2 \over C_s^2} = 4.2  \left( {T_{k,\mathrm{max}} \over T_{s}} -1 \right).
\end{equation}

\noindent \citet{2003ApJ...586.1067H} derived the one-dimensional mean square turbulent velocity in terms of the ratio of the maximum kinetic temperature (an upper limit assuming purely thermal line broadening), $T_{k,\mathrm{max}}$, to the \hi\ spin temperature, $T_{\mathrm{s}}$:
\begin{equation}
V_{t, \mathrm{1D}}^2 =  \frac{k T_s}{m_H} \left( {T_{k,\mathrm{max}} \over T_{s}} -1 \right),
\end{equation}

\noindent where $m_H$ is the mass of a hydrogen atom, and $k$ is the Boltzmann constant. 
$T_{k,\mathrm{max}}$, is computed from the FWHM of the \hi\ absorption component as $T_{k,\mathrm{max}} = \left(\frac{\mathrm{FWHM}}{0.215}\right)^2$. 
The mean square three-dimensional turbulent velocity is then given by $V_{t, \mathrm{3D}}^2 = 3V_{t, \mathrm{1D}}^2$.

For our TSAS-hosting CNM component, 
we find a maximum kinetic temperature of $T_{k,\mathrm{max}} = 174$ K (from its mean FWHM of 2.8 \kms), while the mean spin temperature over 19 epochs is $T_{\mathrm{s}} = 142$ K. 
The corresponding sonic Mach number is $M_{\mathrm{s}} = 1.0$, indicating a transonic state. 
This is consistent with the estimated $M_{\mathrm{s}}$ in the warm ionized gas, which is also of the order of unity \citep{1999ApJ...523..223H,2008ApJ...686..363H,2011Natur.478..214G,2012ApJ...749..145B}. 
For previous \hi\ opacity structure function studies, we obtained the FWHM of the \hi\ absorption components from published spectra \citep{2000ApJ...543..227D,2010MNRAS.404L..45R,2012ApJ...749..144R,2014MNRAS.442..647D}, finding a range of 5 to 10 \kms. 
Assuming a typical CNM spin temperature of 100 K, the sonic Mach numbers for these studies range from 4.2 to 9.3, suggesting supersonic conditions. 
This comparison suggests that the much steeper power-law observed in our study may indeed be due to a relatively low sonic Mach number.

\begin{figure}
\centering
\includegraphics[width=1.0\linewidth]{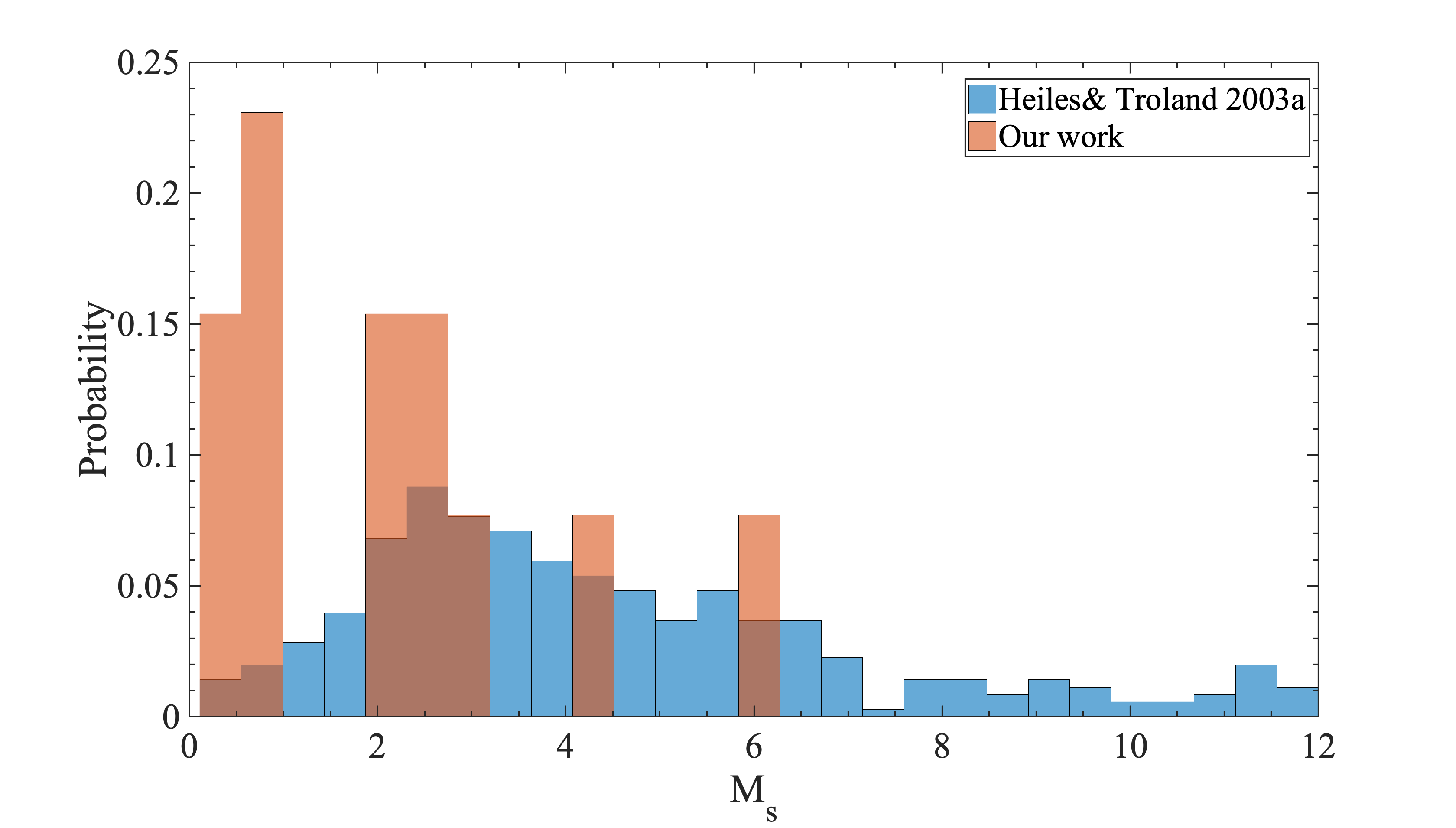}
\caption{Sonic Mach number ($M_{\mathrm{s}}$) probability density distributions for CNM from \citet{2003ApJS..145..329H} (blue) and for the 13 CNM components of our work (red).}
\vspace{0.2cm}
\label{fig:dist_ms} 
\end{figure}

To explore the potential reasons behind the detection of low Mach number values in the CNM associated with TSAS in our study, we compared the $M_{\mathrm{s}}$ distribution for the 13 CNM components in our \hi\ absorption spectra with the results from the Arecibo Millennium Survey \citep{2003ApJS..145..329H}, which traced CNM in absorption against 79 continuum radio sources.
For each of the 13 CNM components, we calculated the mean $M_{\mathrm{s}}$ value across our 19 epochs. 
According to Table 2 of \citet{2003ApJS..145..329H}, 341 out of 353 CNM components exhibit $M_{\mathrm{s}} > 1$, with a peak around 2.5, indicating predominantly supersonic conditions in the CNM. 
In contrast, our distribution is skewed toward sub/transonic values, with most $M_{\mathrm{s}}$ measurements falling below 1. The comparison of these distributions is shown in Figure~\ref{fig:dist_ms}.

It is possible that the overabundance of sub/transonic detections in our study results from the extremely small angular size of our background source. 
This small angular size may allow us to resolve individual \hi\ absorption components down to the
size of 0.03" (see our derivation in section~\ref{sec:obs}), which were previously blended together in the Millennium Survey (and similar studies)\citep{2003ApJS..145..329H}, where absorption was measured against larger background sources (generally not resolved by the telescope beam). However, this does not explain the higher Mach numbers observed in the past VLBI TSAS studies cited above, where the effective spatial resolution of the absorption measurements was similar to ours.

\subsection{Stability of Power-Law Index $\alpha$ and Potential Turbulence Dissipation Scale} \label{sec:alpha_ms}

\begin{figure*}
\centering
\includegraphics[width=1.0\linewidth]{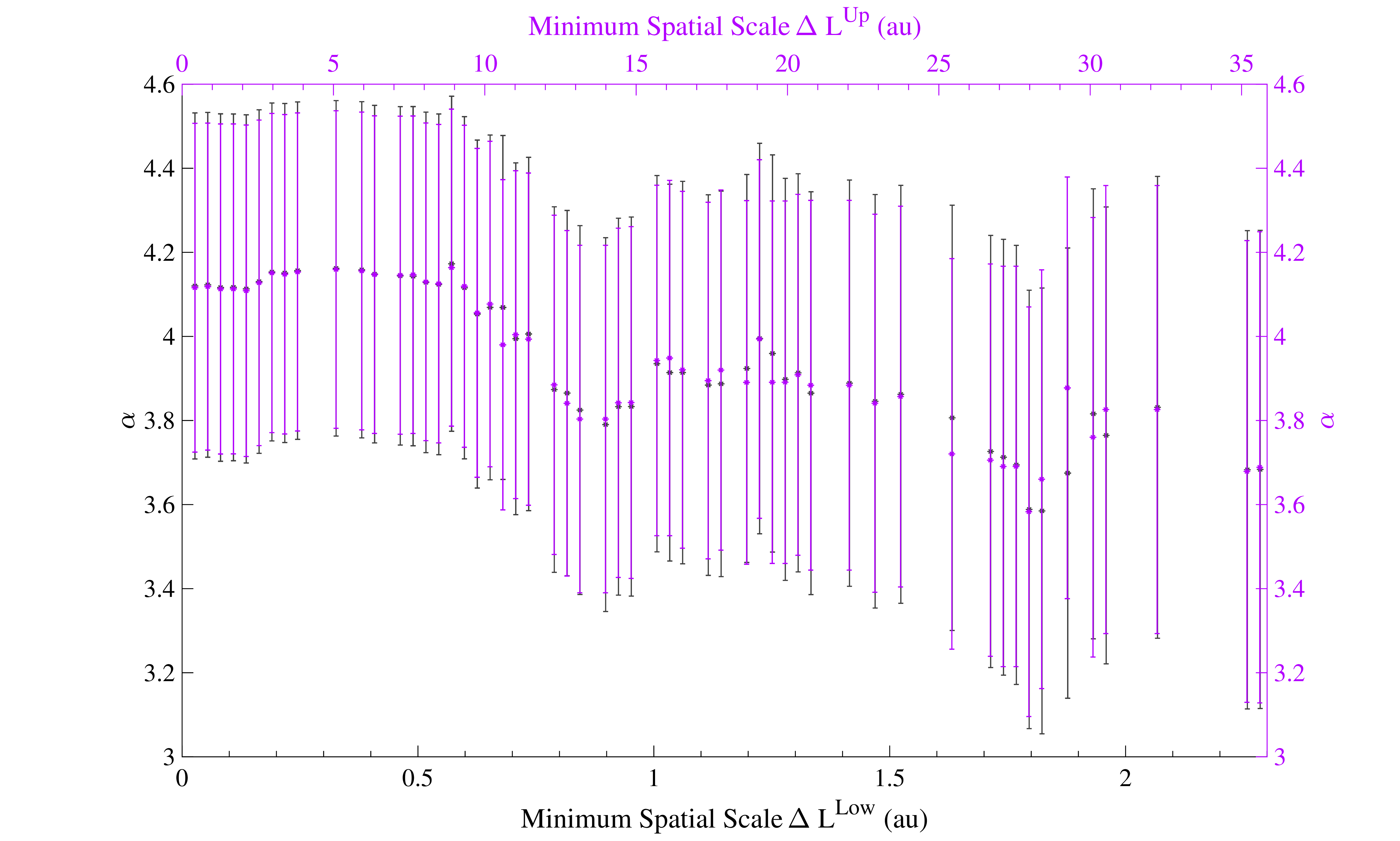}
\caption{Fitted parameter \(\alpha\) as a function of the minimum spatial scale cutoffs. The lower spatial scale limit ranges from 0.03 to 2.3 au (black), and the upper limit ranges from 0.4 to 36 au (purple). Error bars indicate the \(\pm 1\sigma\) uncertainties associated with each measurement. The black and purple x- and y-axes correspond to results for $v^{\rm relative}_{ \rm Low, Transverse}$ and $v^{\rm relative}_{ \rm Up, Transverse}$, respectively. 
}
\vspace{0.2cm}
\label{fig:alpha_fit} 
\end{figure*}

As spatial scales approach the turbulent dissipation scale, we expect to observe a steeper power-law index $\alpha$ \citep{2024MNRAS.527.3945H}. 
Due to incomplete and irregular time sampling, our observations can only probe a limited number of spatial scales, and there is a notable gap in our coverage that occurs between $\sim$ 2.3-- 5.0 au for the lower limit pulsar relative transverse velocity, and around 36--78 au for the upper limit. 
To investigate potential variations in $\alpha$, we applied different cut-off points for the minimum spatial scales (by excluding all datapoints below a given scale), ranging from 0.03 to 2.3 au for the lower limit and 0.5 to 36 au for the upper limit, and repeated our process of fitting for $\alpha$. The results are illustrated in Figure \ref{fig:alpha_fit}. 
We find that $\alpha$ remains relatively stable across these minimum spatial scale ranges. 
While there is a gradual steepening from $\alpha\sim3.7\pm0.7$ (consistent with the expected value for turbulence with $M_{\mathrm{s}}=1$) to $\alpha\sim4.1\pm0.4$ as the cut-off spatial scale decreases from 2.3 to 0.03 au (lower limit) and 36 to 0.4 au (upper limit), the differences remain within the 1$\sigma$ uncertainties. 
This detection of a relatively stable $\alpha$ suggests that turbulence in the CNM may cascade down to a smaller length scale, with a dissipation scale less than 0.03 au (lower limit) to 0.4 au (upper limit).

\section{Conclusion} \label{sec:concl}
We conducted 19-epoch observations of \hi\ absorption over 1.2 years towards PSR J1644-4559 to search for temporal variations in \hi\ absorption that might indicate Tiny-Scale Atomic Structure, constituting the highest temporal sampling rate ever utilized in such studies. 
The main conclusions are

\begin{enumerate}
    \item We report two significant detections of \hi\ temporal variability, associated with a single Cold Neutral Medium component and with three observing epochs, each corresponding to characteristic spatial scales with a lower limit of 11 au and an upper limit of 165 au. The integrated optical depth fluctuations are $0.121\pm0.026$ and $0.078\pm0.016$ \kms, integrated over a velocity range of 1.7 \kms, and both are detected with SNRs exceeding 5.0. 
    If the observed TSAS are \hi\ cloudlets, their corresponding volume density is on the order of $\sim10^5$ cm$^{-3}$ (assuming the line-of-sight length scale is similar to that in the plane of the sky), with the thermal pressure on the order of $10^7$ cm$^{-3}$ K. These cloudlets would therefore be both overdense and overpressurised.

    \item We fit the structure function of integrated optical depth fluctuations, $\Delta\tau_{\rm int}$, across length scale, $\Delta L$, for the CNM component hosting the detected TSAS, by considering $\Delta\tau_{\rm int}$ across all epoch-pairs. The fit is  
    characterized by the relationship $\Delta\tau_{\rm int} = \Delta\tau_0 (\Delta L)^{(\alpha-2)/2}$, with $\Delta\tau_0 = 0.0045 \pm 0.0018$ and $\alpha = 4.1 \pm 0.4$(for the lower limit), and  $\Delta\tau_0 = 0.00042 \pm 0.00031$ and $\alpha = 4.1 \pm 0.4$(for the upper limit). This is the first time that a correlation between optical depth variation and length scale has been reported simultaneously with the detection of TSAS, and it provides strong observational evidence that turbulence is implicated in TSAS production.

    \item The power-law index, $\alpha = 4.1 \pm 0.4$, is significantly steeper than previously reported values for the CNM, but comparable to that in ionized gas. 
    We suggest that this steep index likely results from the trans-sonic Mach number of $M_{\mathrm{s}} = 1.0$ of the CNM component associated with our TSAS detections.

    \item There is no significant variation in $\alpha$ across the spatial scales probed in this study, with lower limits ranging from 0.03 to 10.9 au and upper limits from 0.4 to 174.6 au. This suggests that if the \hi\ opacity fluctuations are driven by turbulence, the cascade may extend to smaller scales, potentially down to dissipation lengths with a lower limit of 0.03 and an upper limit of 0.4 au.

\end{enumerate}

%% Also note that the akcnowlodgment environment does not support long amounts of text. If you have a lot of people and institutions to acknowledge, do not use this command. Instead, create a new \section{Acknowledgments}.
\begin{acknowledgments}
This work is supported by the National Natural Science Foundation of China (NSFC) program No.\ 11988101, 12203044, 12041303, 12473023, and 11903003, by the Leading Innovation and Entrepreneurship Team of Zhejiang Province of China grant No.\ 2023R01008, by Key R\&D Program of Zhejiang grant No.\ 2024SSYS0012, by the CAS Youth Interdisciplinary Team, the Youth Innovation Promotion Association CAS (id. 2021055), by the Cultivation Project for FAST Scientific Payoff and Research Achievement of CAMS-CAS, and by the University Annual Scientific Research Plan of Anhui Province (No.\ 2023AH030052, No.\ 2022AH010013), and by Cultivation Project for FAST Scientific Payoff and Research Achievement of CAMS-CAS. 
DL is a new Cornerstone investigator. 
Murriyang, CSIRO's Parkes radio telescope, is part of the Australia Telescope National Facility (https://ror.org/05qajvd42) which is funded by the Australian Government for operation as a National Facility managed by CSIRO. We acknowledge the Wiradjuri people as the Traditional Owners of the Observatory site.
We express our thanks to Andrew Jameson, Willem van Straten, John  Sarkissian, and Naomi McClure-Griffiths for help with proposal submission, observations, and data processing. 
We are grateful to Christoph Federrath for his useful discussion on turbulence. 

\end{acknowledgments}


%\vspace{5mm}
%\facilities{HST(STIS), Swift(XRT and UVOT), AAVSO, CTIO:1.3m,


%\software{astropy \citep{2013A&A...558A..33A,2018AJ....156..123A},  
%          Cloudy \citep{2013RMxAA..49..137F}, 
%          Source Extractor \citep{1996A&AS..117..393B}


\appendix
% Redefine figure numbering for appendices
\renewcommand{\thefigure}{\thesection.\arabic{figure}}
\setcounter{figure}{0}
% Redefine table numbering for appendices
\renewcommand{\thetable}{\thesection.\arabic{table}}
\setcounter{table}{0}

\section{Lower Limit on Relative Transverse Velocity of the Pulsar and an \hi\ Cloud }\label{sec:tan_vel}

We use different procedures to estimate the lower and upper limits on the relative transverse velocity of the pulsar and an  \hi\ cloud located along the same line of sight.   Our rationales for each limit are given in Section \ref{sec:obs}, where the actual determination of the upper limit is also found.  
We derive and determine the lower limit here.

For the lower limit on the relative transverse velocity of the pulsar and an absorbing \hi\  cloud along the same line of sight, we must first express both objects' transverse velocities in the same frame of reference before subtracting them.  For simplicity in calculation, we chose this common frame to be the LSR at the Sun's Galactocentric radius $R_0$, which we call LSR($R_0$).

First, we find the transverse (to the line of sight)  lower limit to the velocity of the pulsar in the frame of LSR($R_0$) as follows: The measured velocity lower limit is taken to be zero in the Solar System frame (see Section \ref{sec:obs} for details), which must then be transformed to the LSR($R_0$) frame, yielding $\vec{v}_{\rm PSR, Low} = [U_\sun, V_\sun, W_\sun]$ (where the latter quantity is called the Solar Motion); and then we must find the  part of it transverse to the line of sight,  $\vec{v}_{\rm PSR, Low, Transverse}$, as follows:  First, note that $ \Delta l$, the Galactic longitude difference between $\vec{v}_{\rm PSR, Low}$ and the line of sight to the pulsar at $l_{\rm PSR}$, is given by  $\Delta l = \arctan{\dfrac{V_{\odot}}{U_{\odot}}}-l_{\rm PSR}$, where $l_{\rm PSR} $ is the Galactic longitude of the pulsar. Next, by approximating the pulsar's actual Galactic latitude of  $-0.195^{\circ}$ as zero, we force the plane transverse to the line of sight to be perpendicular to the Galactic plane.  We then have  $\vec{v}_{\rm PSR, Low, Transverse_{[U-V]}}$, the Galactic-planar ( i.e., [U-V] ) component of  $v_{\rm PSR, Low, Transverse}$: 
\begin{equation}
   \vec{v}_{\rm PSR, Low, Transverse_{[U-V]}} = v_{\rm PSR, Low_{[U-V]}} \sin{\Delta l},
\label{equ:v_pl} 
\end{equation}
with $v_{\rm PSR, Low_{[U-V]}} = \sqrt{U_{\odot}^2 + V_{\odot}^2}$; while the component normal to the Galactic plane, $v_{\rm PSR, Low, Transverse_{[W]}}$, is given by
\begin{equation}
    v_{\rm PSR, Low, Transverse_{[W]}}= W_\sun.
\label{equ:WSun}
\end{equation}

\noindent With $l_{\rm PSR} = 339.193^{\circ}$ and $[U_\sun, V_\sun, W_\sun]=[10.1, 12.3, 7.3]$ \kms\ \citep{2019ApJ...870L..10M}, we find that $\Delta l = -288.6^{\circ}, v_{\rm PSR, Low, Transverse_{[U-V]}} = 15.1$ \kms and $v_{\rm PSR, Low, Transverse_{[W]}}=7.3$ \kms.

Second, we calculate in the frame of our own LSR($R_0$), the transverse velocity of the absorbing \hi\ cloud, which is assumed to be moving parallel to the Galactic plane at Galactocentric radius $R$ in conjunction with its own LSR, called LSR($R$). The following relationships are
based on \citet[][pp. 375-376]{2017fuas.book.....K} and \citet[][p. 908]{2017imas.book.....C}.  
In the desired frame of our LSR($R_0$), the component of velocity  transverse to the line of sight to such an absorbing \hi\ cloud at  $l_{\rm cloud}\ (\equiv\ l_{\rm PSR})$ and Galactocentric radius $R$,
\begin{equation}
    \vec{v}_{ \rm cloud, Transverse_{[U-V]} }= |\Theta(R)\sin{\alpha}-\Theta_0\cos{l_{\rm cloud}}|,
\label{equ:dVR} 
\end{equation}
where $\Theta(R)$ is the circular velocity of an LSR($R$) (and hence of the cloud at Galactocentric radius $R$),  $\Theta_0$ is the circular velocity of LSR($R_0$) at $R=R_0$,  and  $\alpha$ is the angle between the cloud's circular Galactic velocity vector and the line of sight.
Equation \ref{equ:dVR} may be explicitly solved with the substitution of the trigonometric relationship $\cos{\alpha}=R_0\sin{l}/R$, and a rotation curve $\Theta(R)$.
We adopted the same linear rotation curve as \citet{2021ApJ...911L..13L},
\begin{equation}\Theta(R) =\Theta_0+ \frac{d \Theta} { d R} \ (R-R_0).
\label{equ:RotCurve}
\end{equation}
with their $\Theta_0= 233.6$ \kms, $d \Theta / d R$ = -1.34 \kms\ kpc$^{-1}$, and $R_0=8.09$ kpc.

\begin{figure}
\centering
\includegraphics[width=0.5\linewidth]{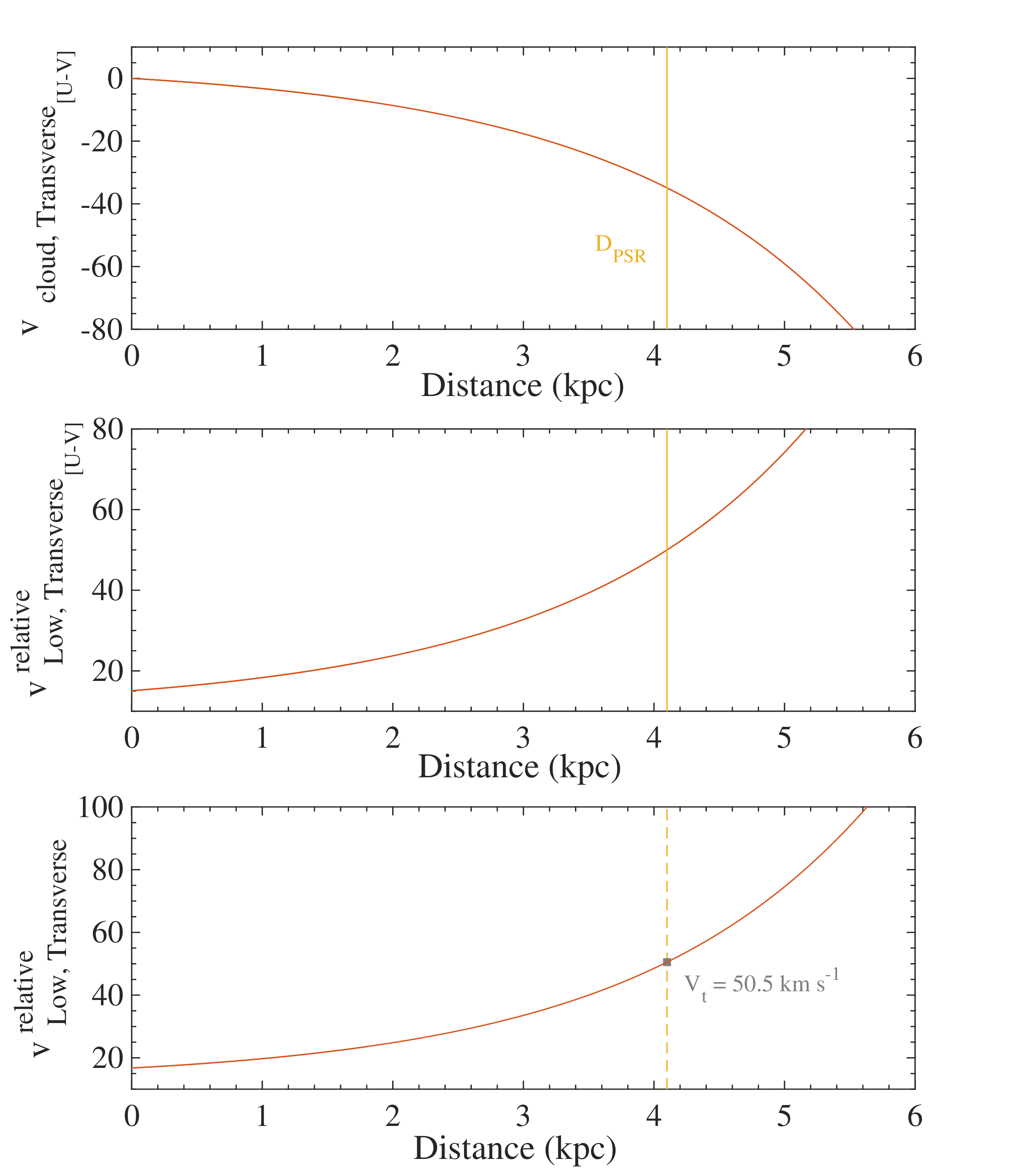}
\caption{\textit{Top}: Planar transverse velocity of the cloud as a function of distance. 
\textit{Middle}: The relative transverse planar velocity of the pulsar and an absorbing cloud along
the same line of sight as a function of distance. 
\textit{Bottom}: The final transverse velocity as a function of distance. The transverse velocity at the pulsar's kinematic distance is $\sim 50.5$ \kms. 
The yellow dashed lines indicate the pulsar's kinematic distance of 4.1 kpc. The frame of reference for all displays is our LSR, LSR($R_0)$.} 
\vspace{0.2cm}
\label{fig:vt_diff} 
\end{figure}

Since $\vec{v}_{\rm PSR, Low, Transverse_{[U-V]}}$ and $\vec{v}_{\rm cloud, Transverse_{[U-V]}}$ are antiparallel, then the desired lower limit on the {\it{relative}} transverse planar velocity of the pulsar and an absorbing cloud along the same line of sight is obtained from Equations \ref{equ:v_pl} and \ref{equ:dVR}, respectively :
\begin{equation}
v^{\rm relative}_{ \rm Low, Transverse_{[U-V]} }=|\vec{v}_{\rm PSR, Low, Transverse_{[U-V]}}| + |\vec{v}_{ \rm cloud, Transverse_{[U-V]} }|.
\end{equation}
as shown in the middle panel of Figure~\ref{fig:vt_diff}. 
Finally, accounting for the lower limit on the perpendicular 
component of the relative  transverse velocity, 
$W_{\odot}$, we have
\begin{equation}
 v^{\rm relative}_{ \rm Low, Transverse}  = \sqrt{  (v^{\rm relative}_{ \rm Low, Transverse_{[U-V]} })^2 + v_{\rm PSR, Low, Transverse_{[W]}}^2   }   
\end{equation}
as illustrated in the bottom panel of Figure~\ref{fig:vt_diff}. 

It is also possible  to determine  a cloud's  ``kinematic'' distance $d$ from Earth,  using the trigonometric relation
\label{equ:RofDistAndLong}
\begin{equation}
R=\sqrt{R_0^2+d^2-2R_0d\cos{l}}. 
\label{equ:RofDistAndLong}
\end{equation}
With a cloud's transverse planar velocity from Equations \ref{equ:dVR} and \ref{equ:RotCurve}, and its distance from Equation \ref{equ:RofDistAndLong}, one can
determine $v_{ \rm cloud, Transverse_{[U-V]}}(d)$, as
shown in the top panel of Figure~\ref{fig:vt_diff}.

\section{Possible Spectral Lines for the Components at $\sim$83 and 105 \kms\ from Epoch 10 }\label{sec:spec_fake}

\begin{figure*}
\centering
\includegraphics[width=0.8\linewidth]{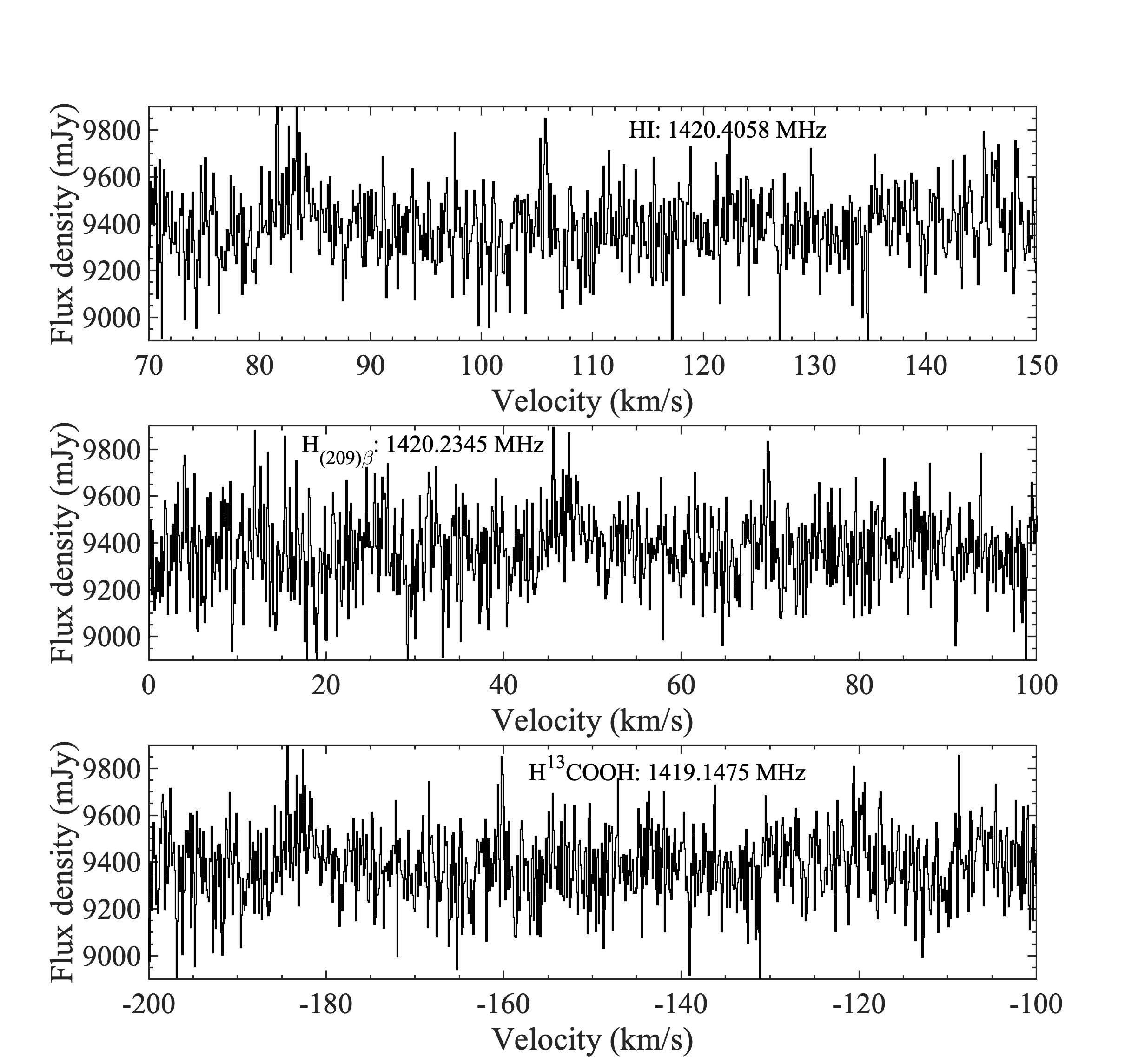}
\caption{Three subsections of a single 1420 MHz pulsar ON-OFF spectrum from epoch 10, where a given horizontal velocity scale is calculated from  the rest frequency of a particular molecule: \hi\ (top panel), $\textrm{H}_{(209)\beta}$ (middle panel), and $t-\mathrm{H^{13}COOH}$ (bottom panel),  respectively.} 
\vspace{0.2cm}
\label{fig:hi_hbeta} 
\end{figure*}

We detected two suspicious components at $\sim$ 83 and 105 \kms\ in the pulsar ON-OFF spectrum of Epoch 10, which contributed to the majority of false positive detections in the difference spectra, as shown in the top panel of Figure~\ref{fig:hi_hbeta}. 
These two velocities are well outside normal HI velocities in this direction, as evidenced by the distance versus radial velocity curve in Figure \ref{fig:HI_dist}.

There are several spectral lines close to the emission frequency of \hi. 
We selected two of the closest spectral lines-formic acid ($t-\mathrm{H^{13}COOH}$) at 1419.1475 MHz and the hydrogen recombination line ($\textrm{H}_{(209)\beta}$) at 1420.2355 MHz-as the central frequencies to test whether these components could originate from these spectral lines. 

The pulsar ON-OFF spectra, after Doppler correction using the rest frequencies of 1419.1475 MHz and 1420.2355 MHz, are shown in the middle and lower panels of Figure~\ref{fig:hi_hbeta}. 
The central velocities are $-185$ and $-160$ \kms\ for the $t$-$\mathrm{H^{13}COOH}$ line, and 45 and 70 \kms\ for the $\textrm{H}_{(209)\beta}$ line, respectively, both of which are more distant than the pulsar. Therefore, the components observed in Epoch 10 at \hi\ velocities of $\sim$83 and 105 \kms\ are unlikely to be associated with these spectral lines.

\section{Baseline Removal for \hi\ absorption spectra}\label{sec:app_blrm}
\begin{figure*}
\centering
\gridline{\includegraphics[width=0.5\linewidth]{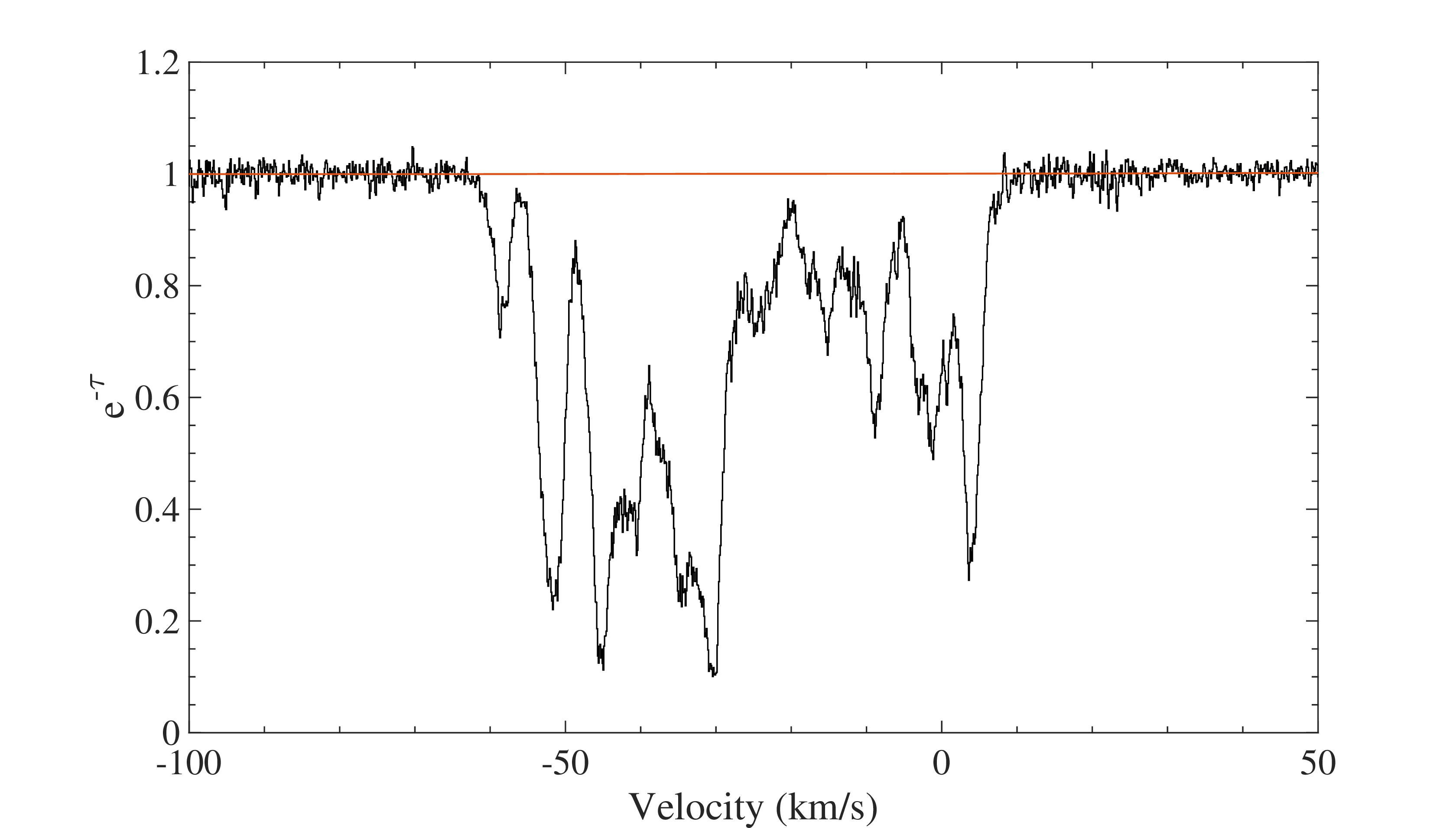}
\includegraphics[width=0.5\linewidth]{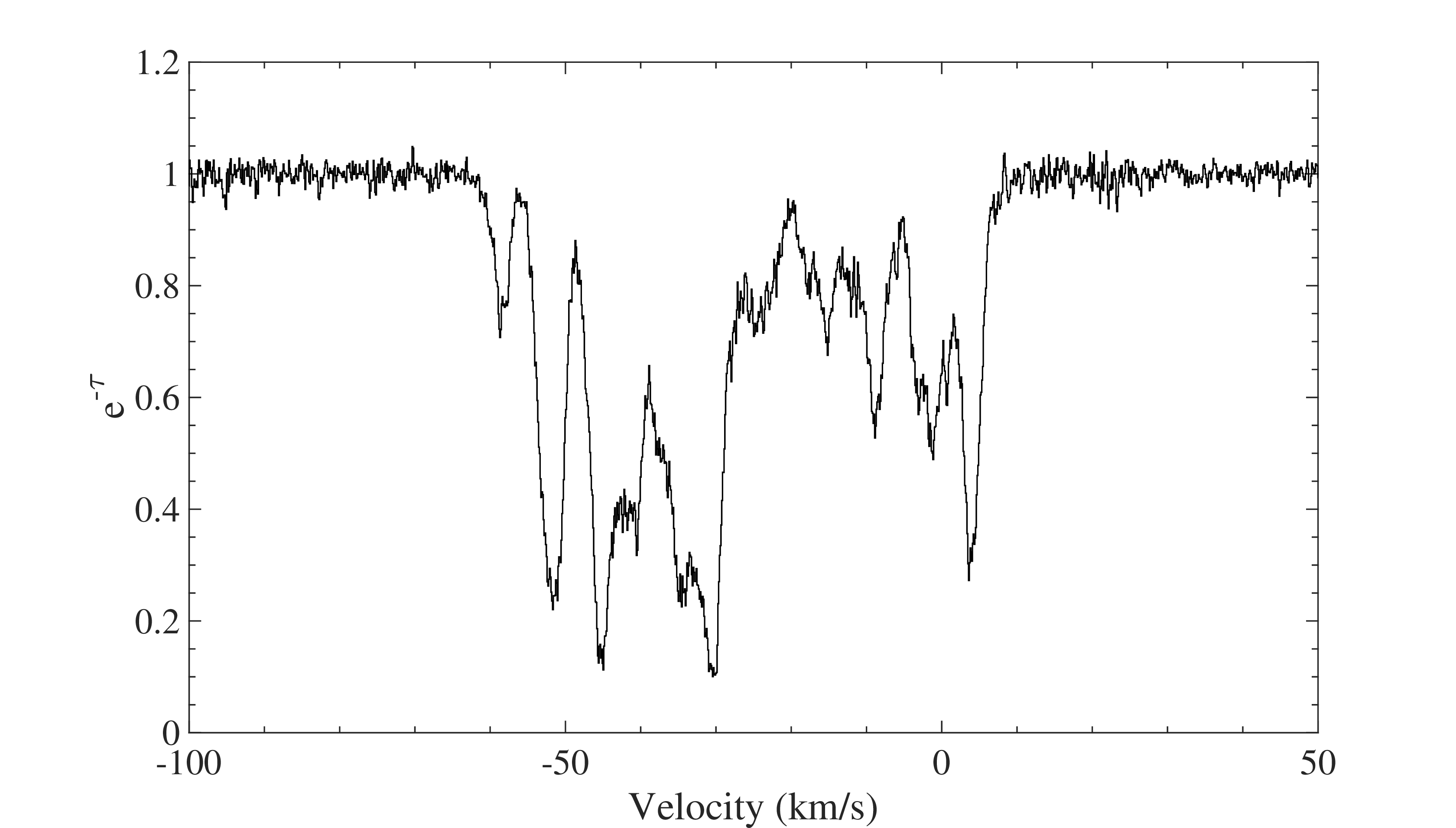}}
\caption{\textbf{Left panel:} the original scaled absorption spectrum from Epoch 1 with the baseline shown in orange.  \textbf{Right panel:} the baseline-removed, scaled absorption spectrum for Epoch 1.
}
\vspace{0.2cm}
\label{fig:blrm_hi_abs} 
\end{figure*}

The baselines of these scaled absorption spectra are not perfectly flat, resulting in an imperfect baseline in the scaled absorption, as shown in the left panel of Figure~\ref{fig:blrm_hi_abs}. 
To flatten the baseline, we used a median filter, which involves the following steps:
1. Mask the online region for the scaled absorption spectrum from -170.2 to 118.6 \kms. 
2. Estimate the mean and standard deviation values of the data collected in the scaled absorption spectrum with velocities of [-170.2,-149.6 \kms] and [118.6, 128.9 \kms]. 
Although the first velocity range is within the masked region, it remains outside the \hi\ absorption feature. 
3. Generate Gaussian random noise with the estimated mean and standard deviation values to replace the online region. 
4. Construct the median filter with a width of 413 \kms. 
5. Subtract the constructed median filter from the scaled absorption spectrum and then add 1.0 to generate the final baseline-removed absorption spectrum, as shown in the right panel of Figure~\ref{fig:blrm_hi_abs}.

To quantify the impact of replacing random noise in the absorption components to generate a median filter on the final differential absorption spectrum, we conducted 1000 Monte Carlo simulations by replacing random noises to generate 1000 median filters. 
The final standard deviation across 1000 trials of the $\Delta(1-e^{-\tau})$ is less than 0.7\% of the mean value.
The pulsar-off spectra were less sensitive to baseline removal because we used only the combined 19-epoch pulsar-off spectrum to estimate the spin temperatures. 
We simply adopted a third-order polynomial to fit and remove baselines.

\section{HI Gaussian Decomposition, Spin Temperature Fitting, and Column Density Estimation}

\begin{figure*}
\centering
\gridline{\includegraphics[width=0.5\linewidth]{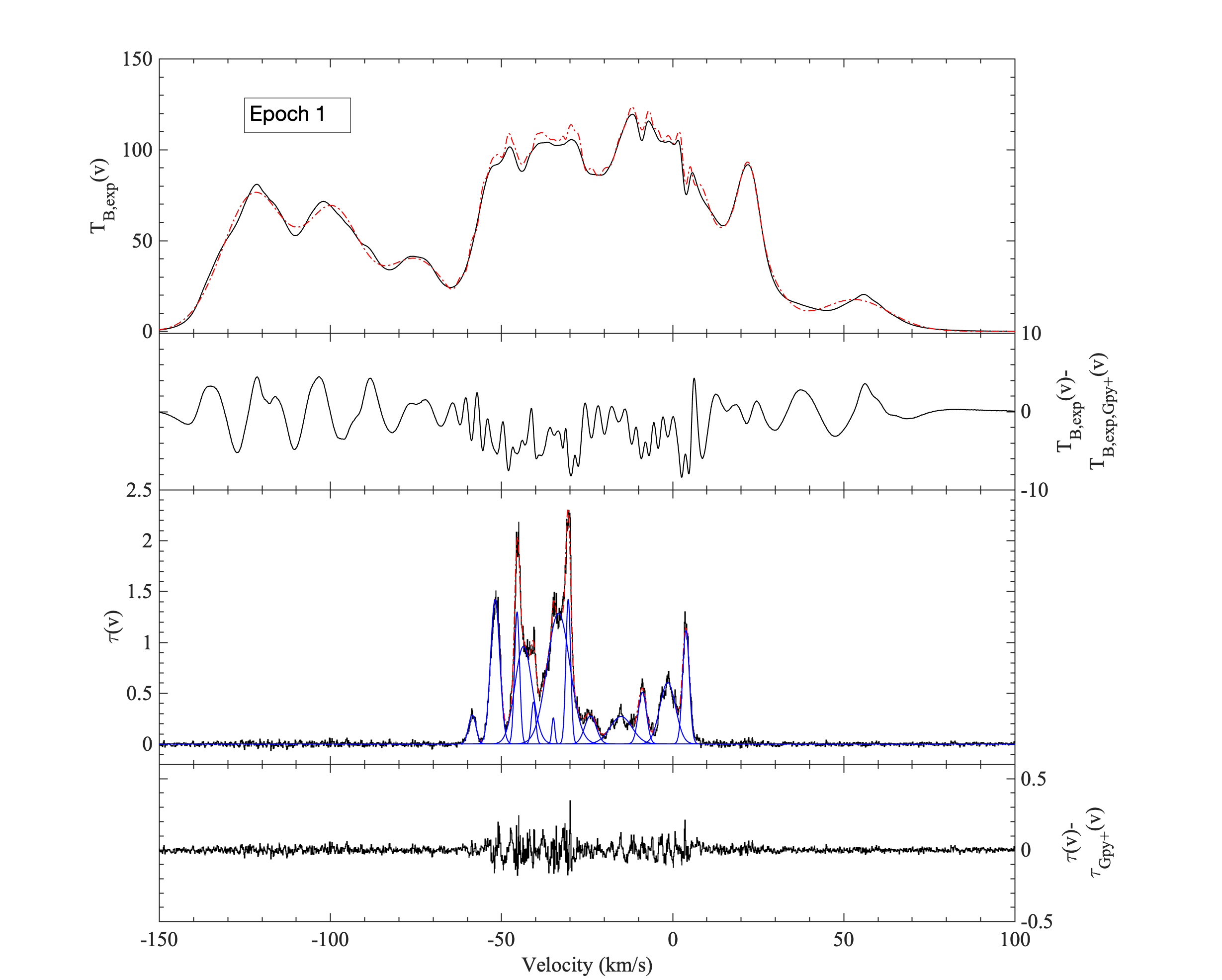}
\includegraphics[width=0.5\linewidth]{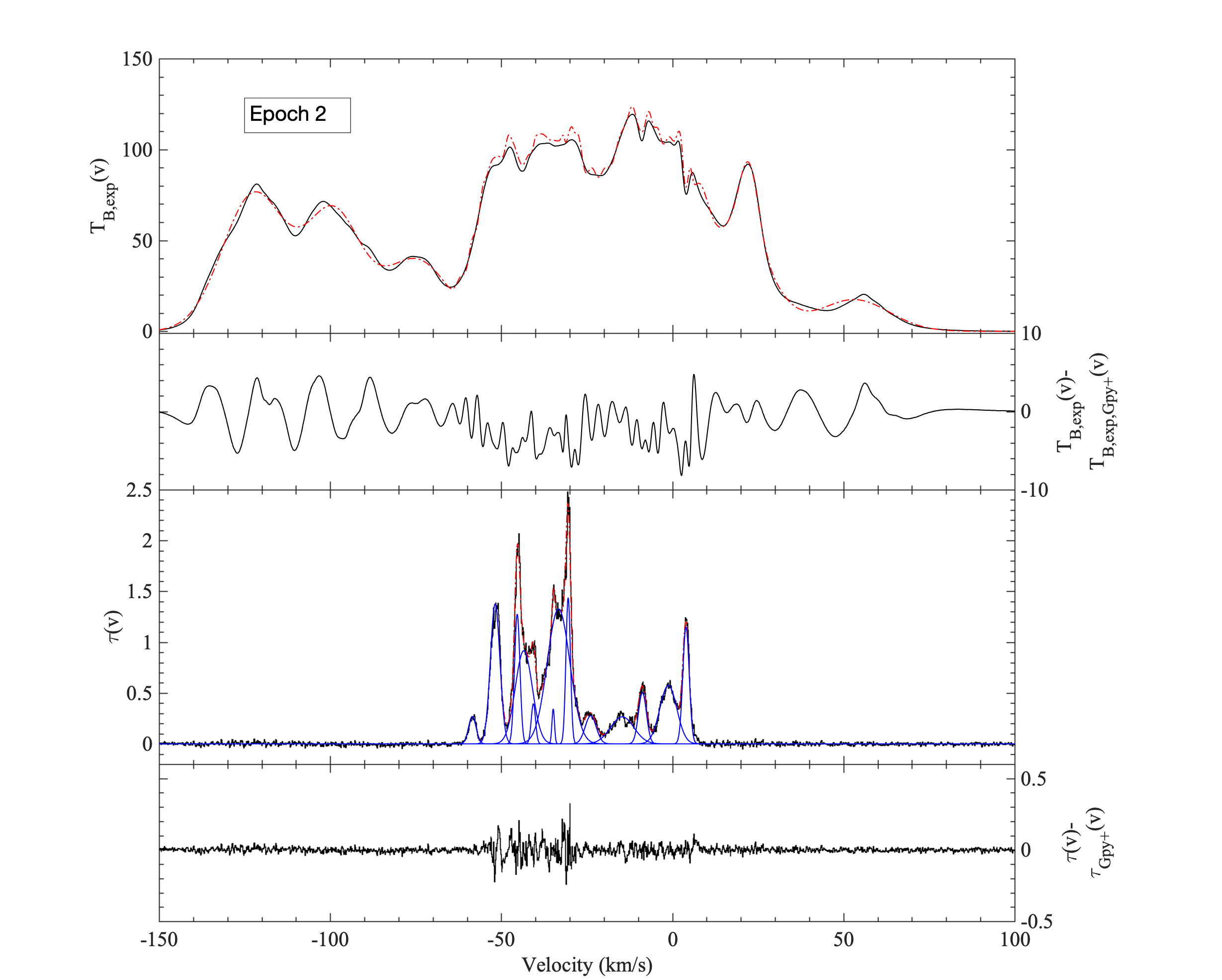}}
\gridline{\includegraphics[width=0.5\linewidth]{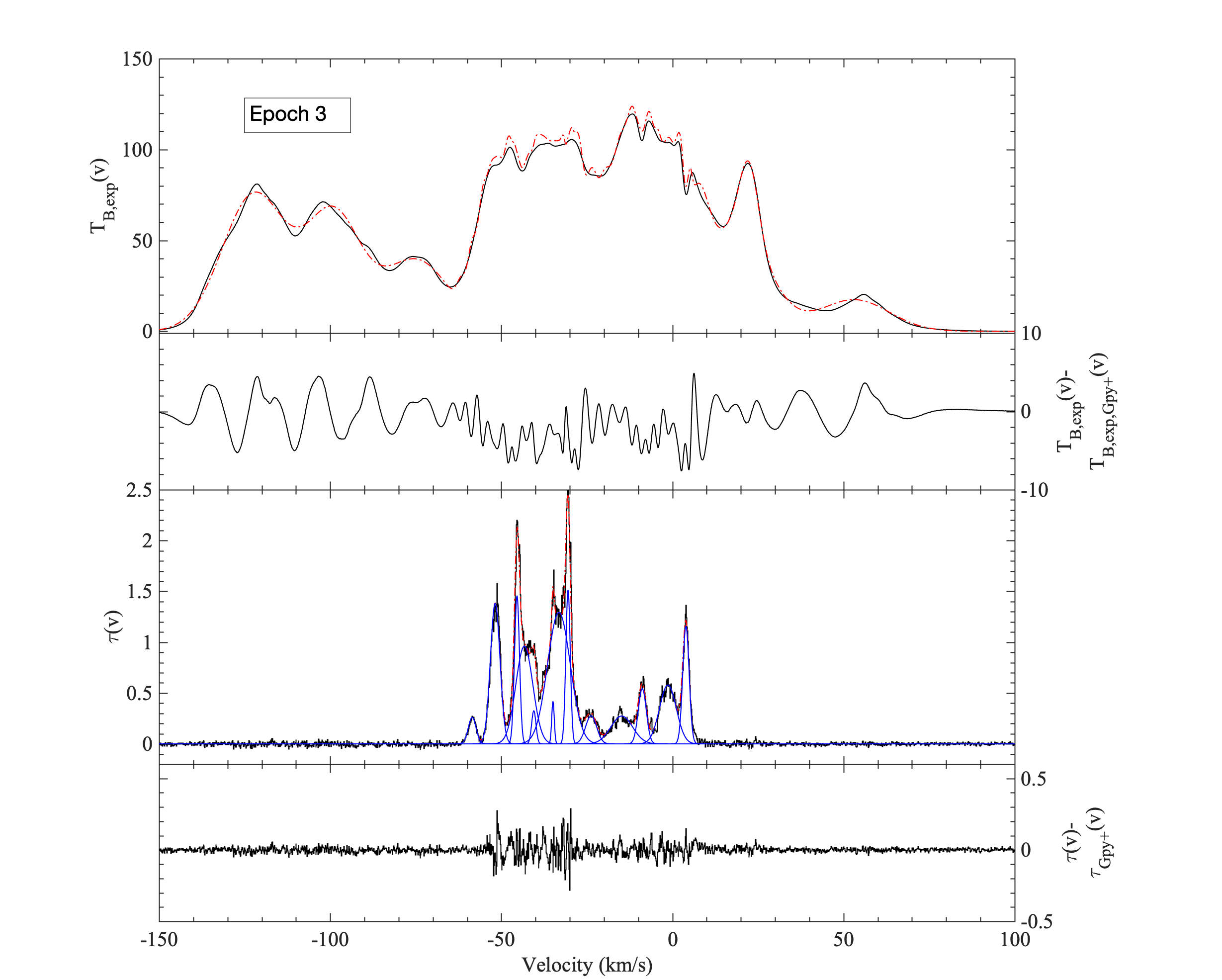}
\includegraphics[width=0.5\linewidth]{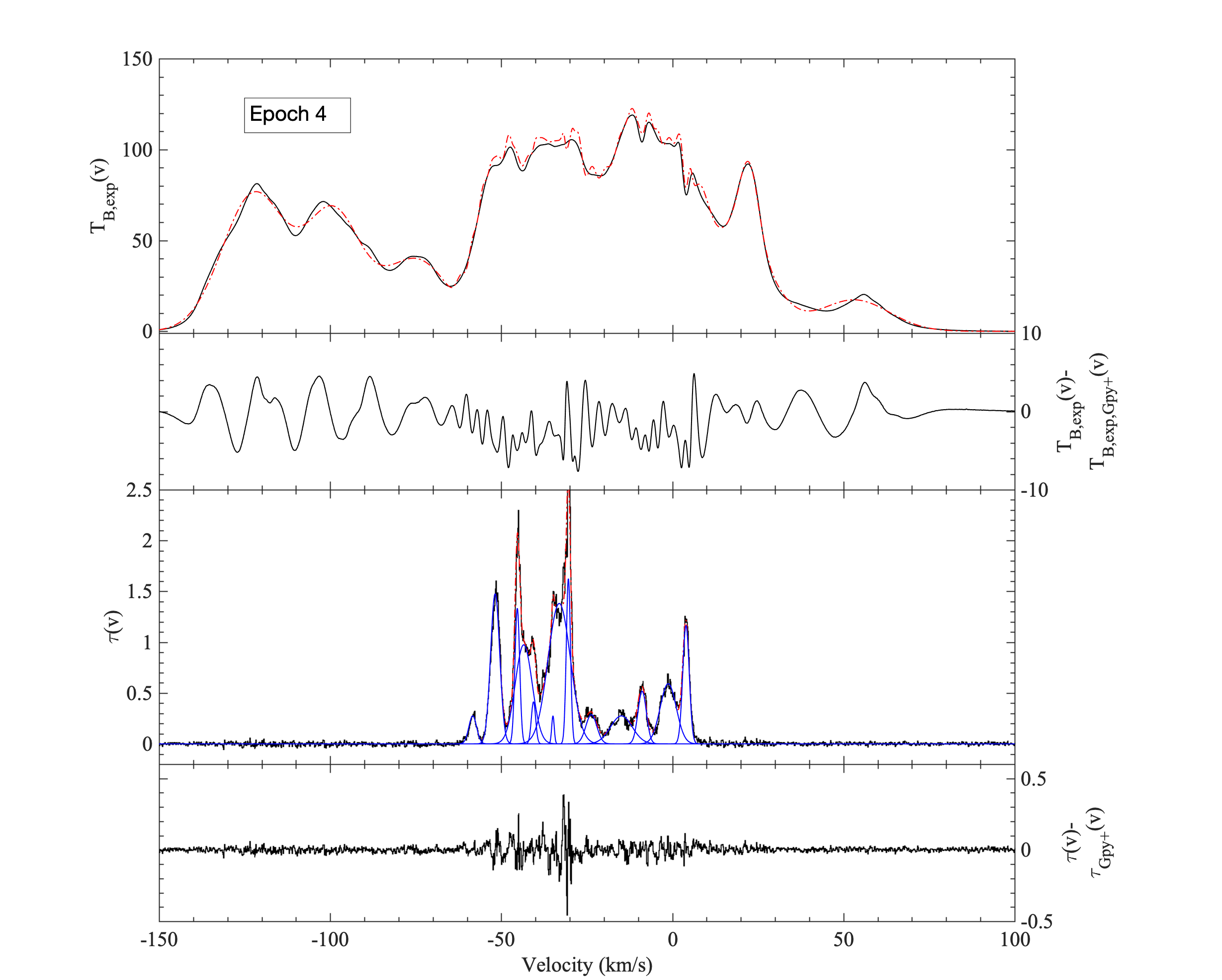}}
\gridline{\includegraphics[width=0.5\linewidth]{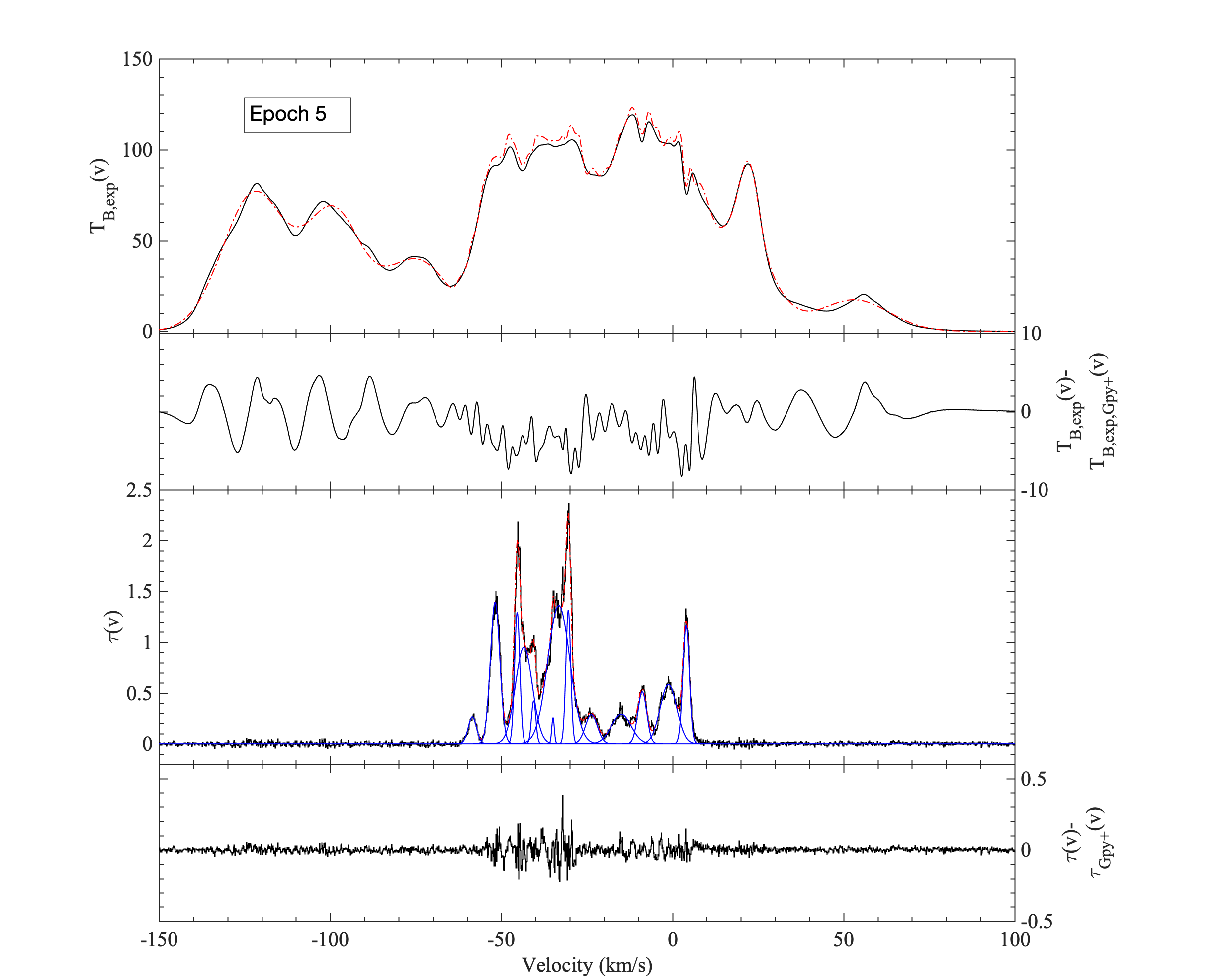}
\includegraphics[width=0.5\linewidth]{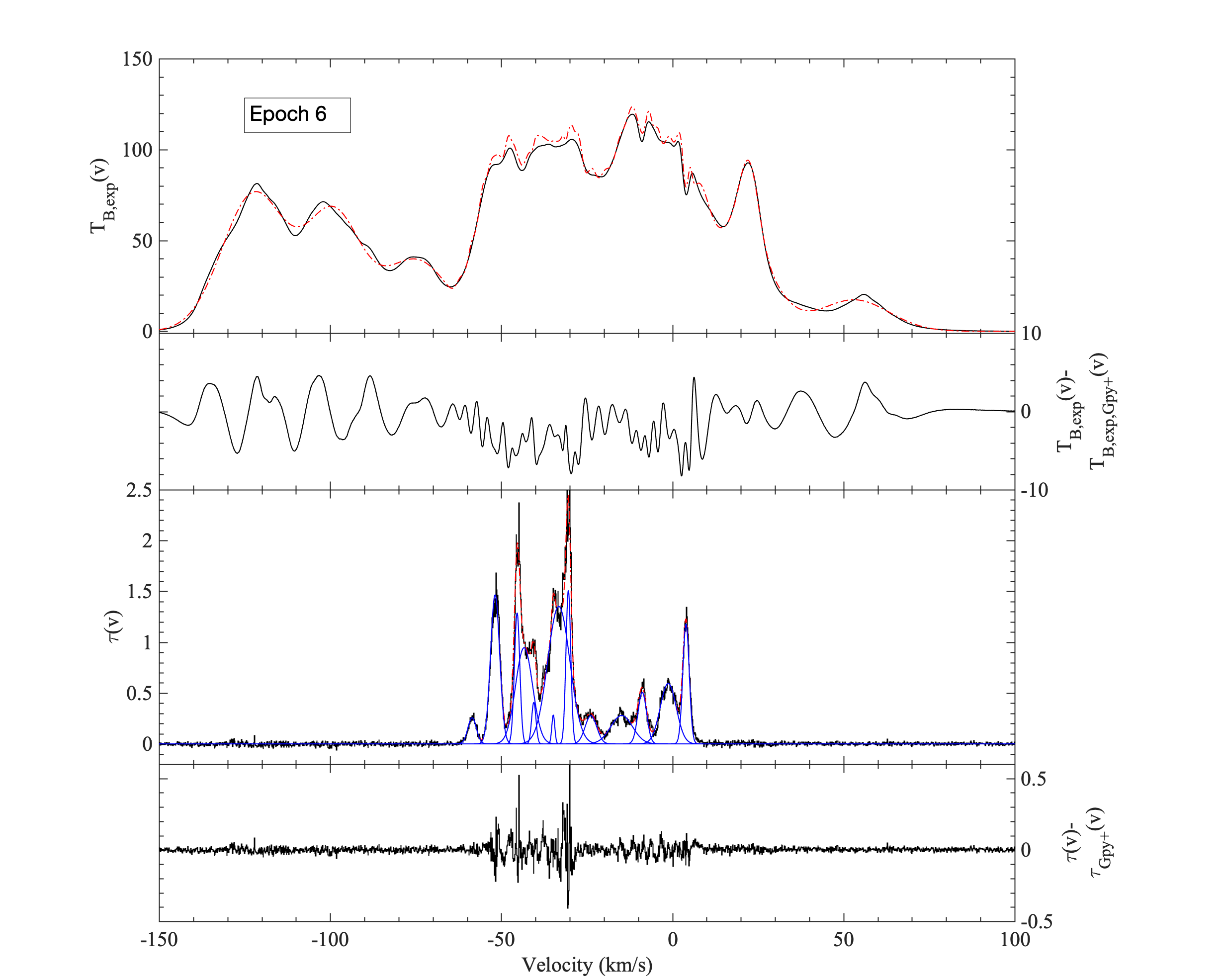}}
\caption{
From epoch 1 to 6: for each observing epoch:
The top panel shows the \hi\ emission brightness temperature $T^{\rm off}(v)$ (black line) and the fitted spectrum (red line).
The second panel presents the corresponding fitting residuals.
The third panel displays the \hi\ optical depth spectrum $\tau(v)$ (black line), the decomposed Gaussian features (blue lines), and the final fitted $\tau(v)$ (red line).
The fourth panel shows the corresponding fitting residuals.
}
\vspace{0.2cm}
\label{fig:compB_gaussian_Tex_1420_fit0} 
\end{figure*}

\begin{figure*}
\centering
\gridline{\includegraphics[width=0.5\linewidth]{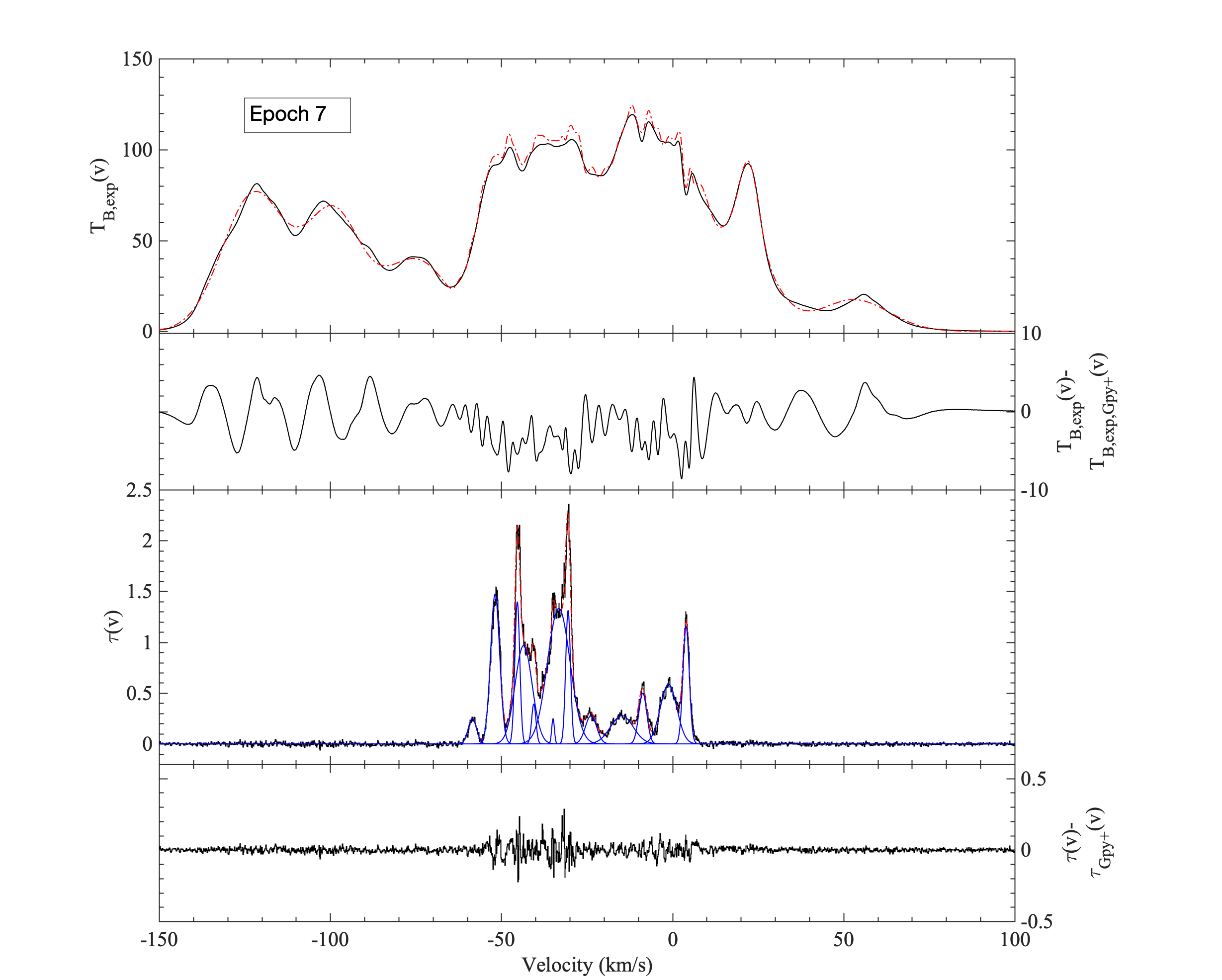}
\includegraphics[width=0.5\linewidth]{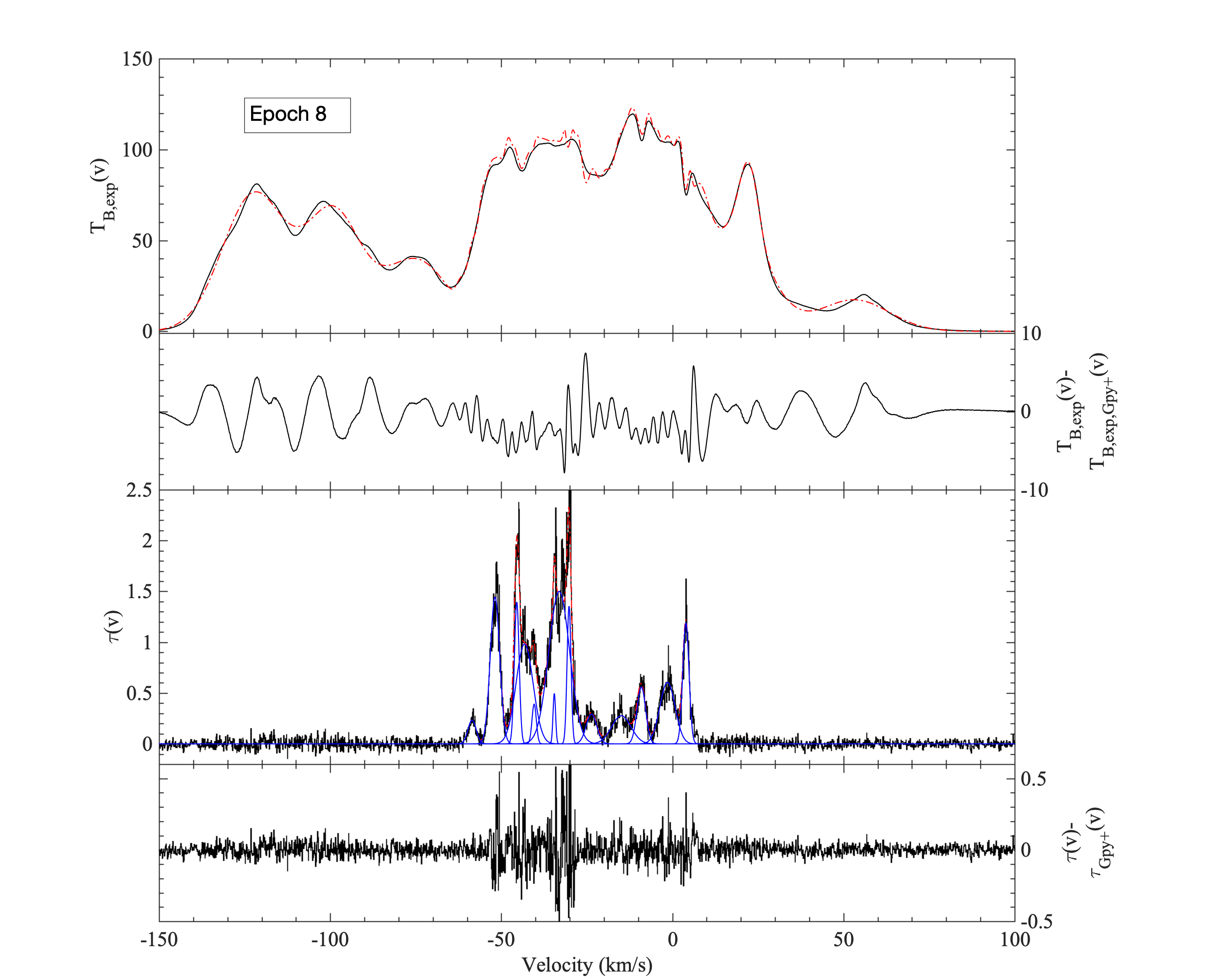}}
\gridline{\includegraphics[width=0.5\linewidth]{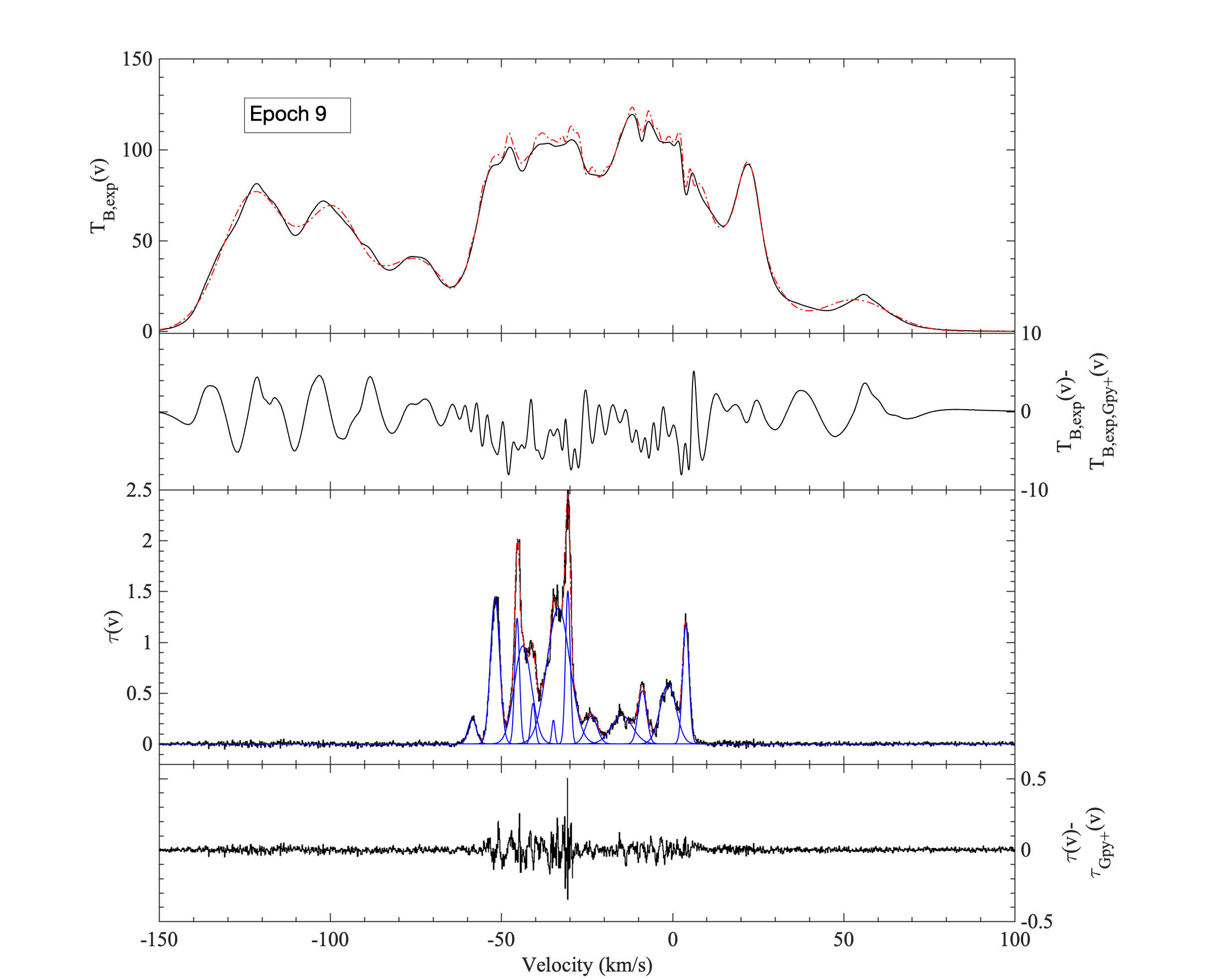}
\includegraphics[width=0.5\linewidth]{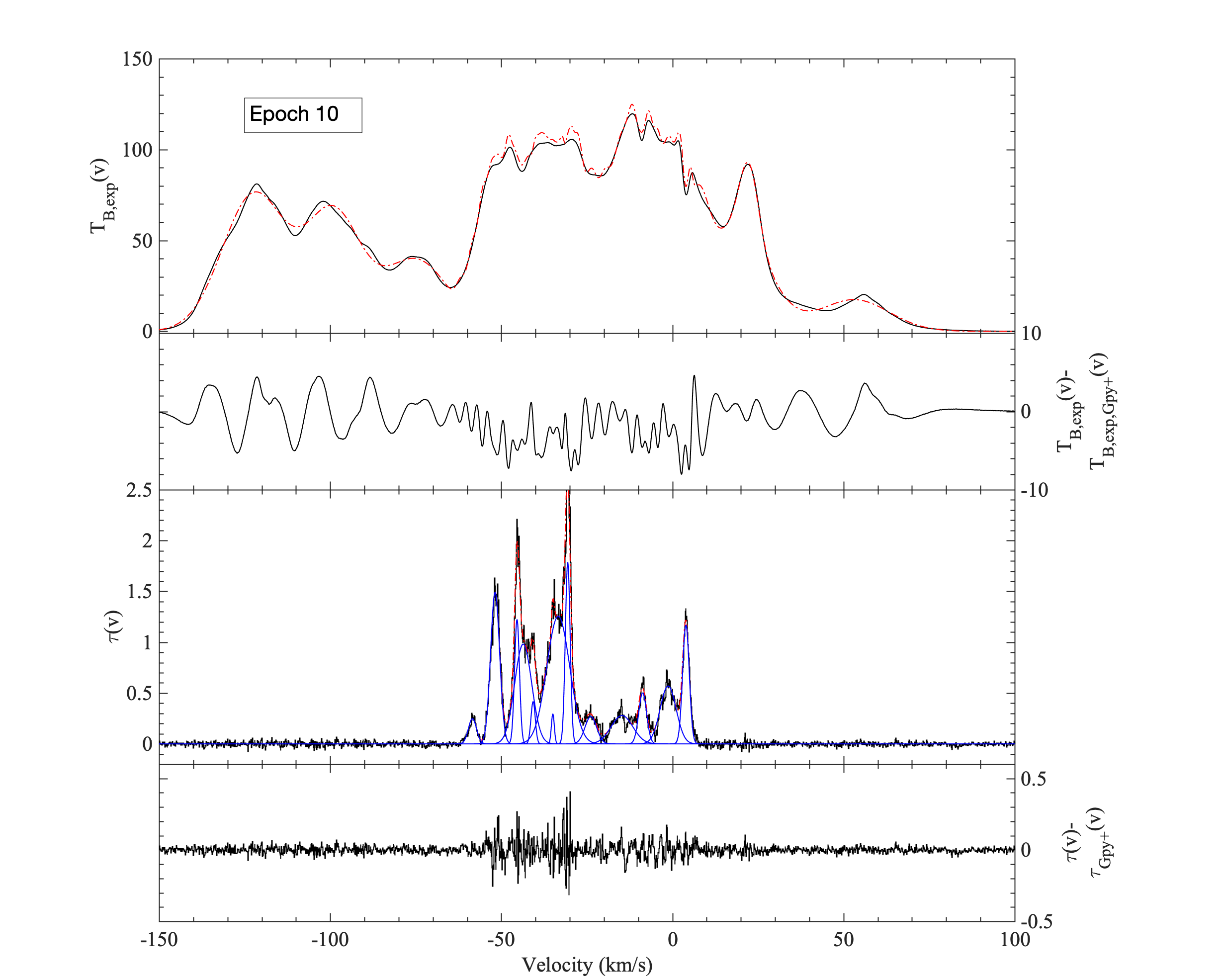}}
\gridline{\includegraphics[width=0.5\linewidth]{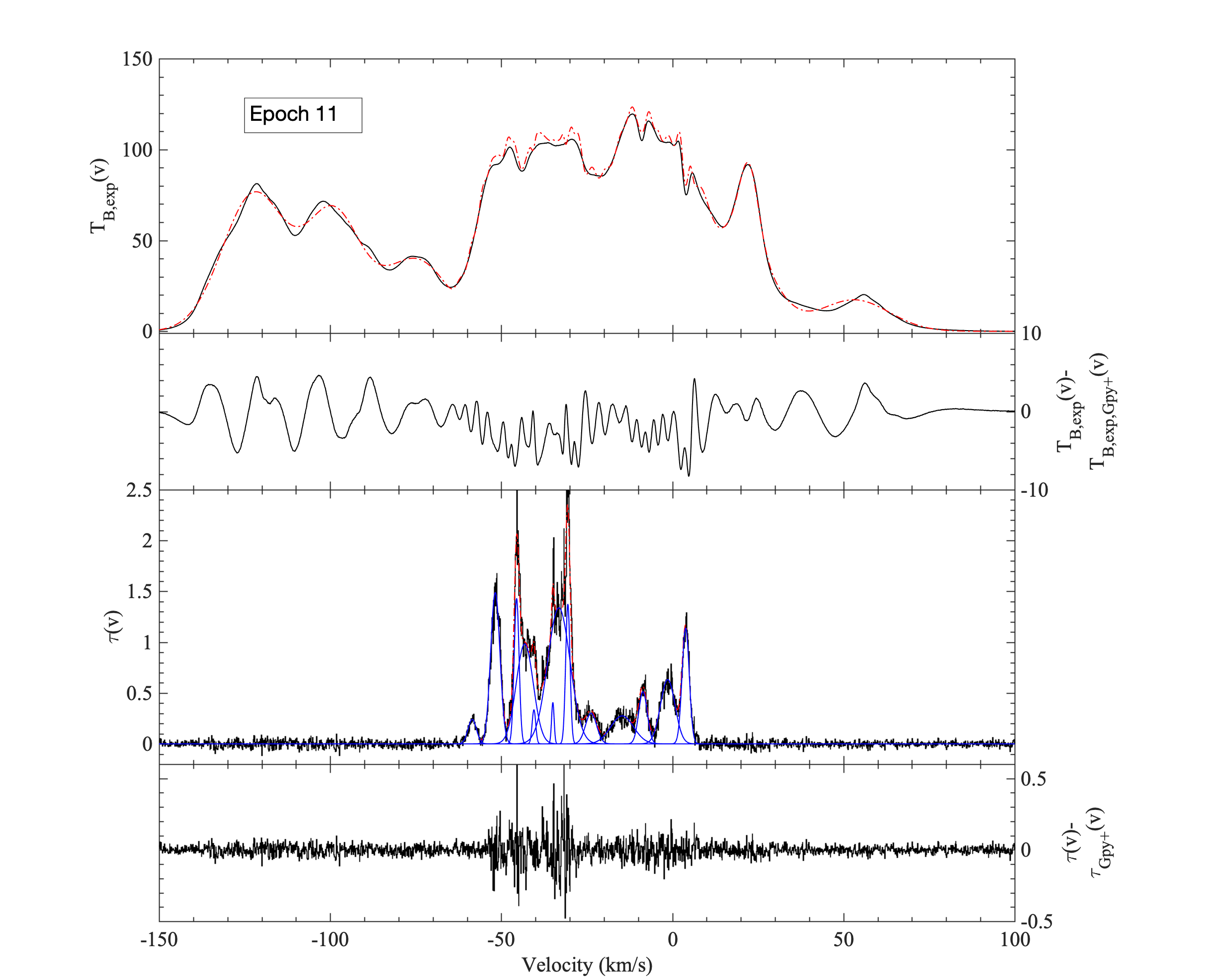}
\includegraphics[width=0.5\linewidth]{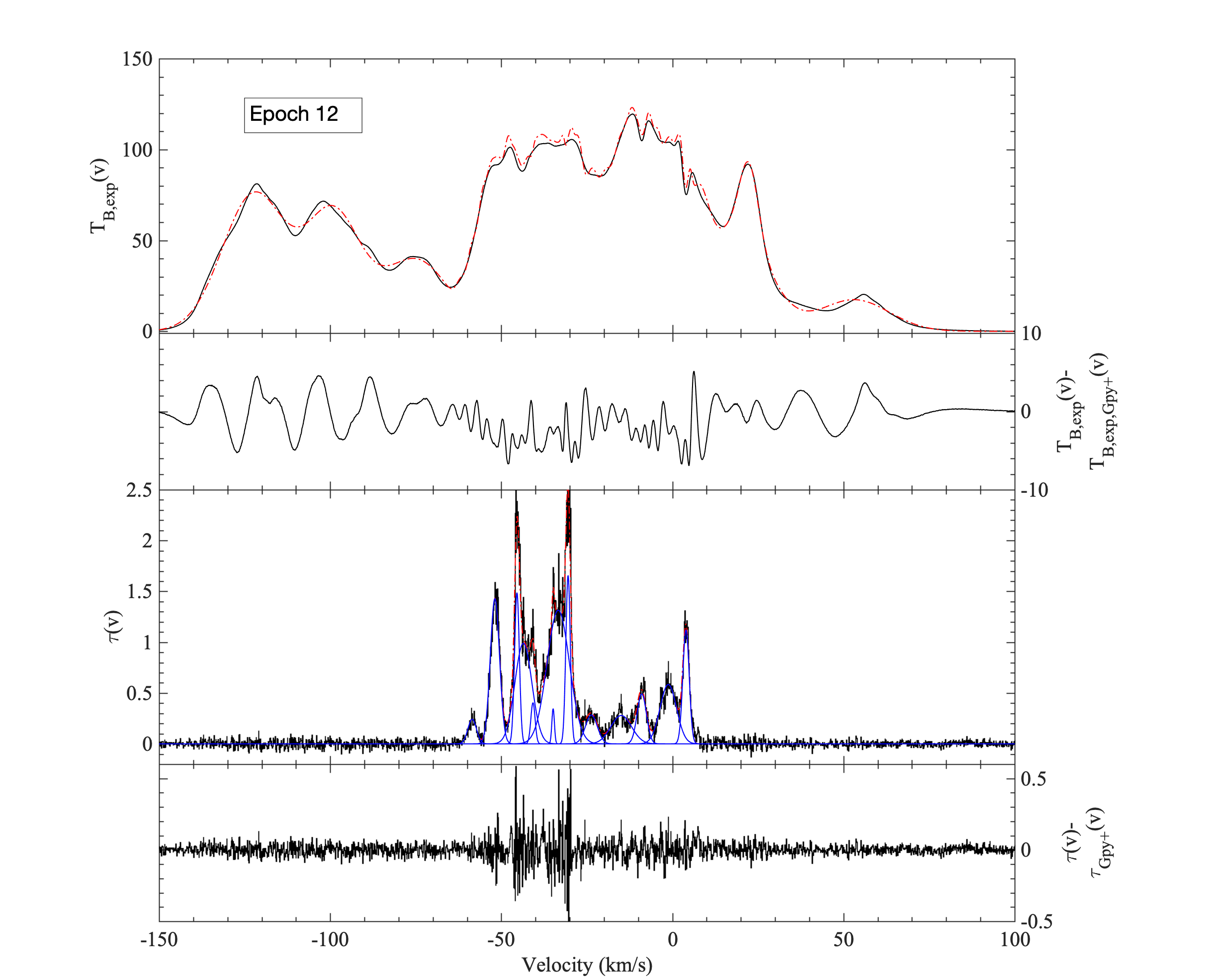}}
\caption{From epoch 7 to 12: similar to Figure~\ref{fig:compB_gaussian_Tex_1420_fit0}.} 
\vspace{0.2cm}
\label{fig:compB_gaussian_Tex_1420_fit1} 
\end{figure*}

\begin{figure*}
\centering
\gridline{\includegraphics[width=0.45\linewidth]{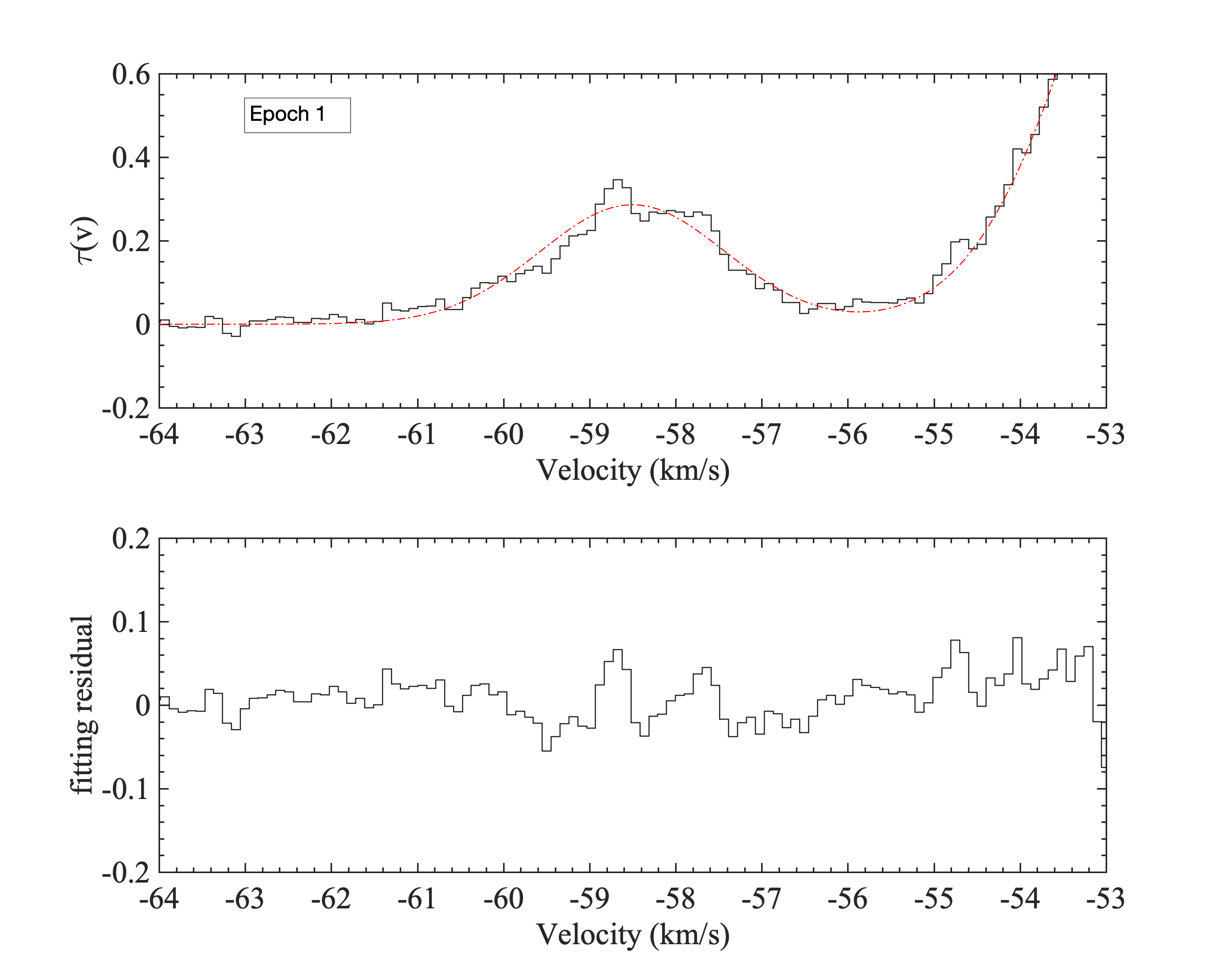}
\includegraphics[width=0.45\linewidth]{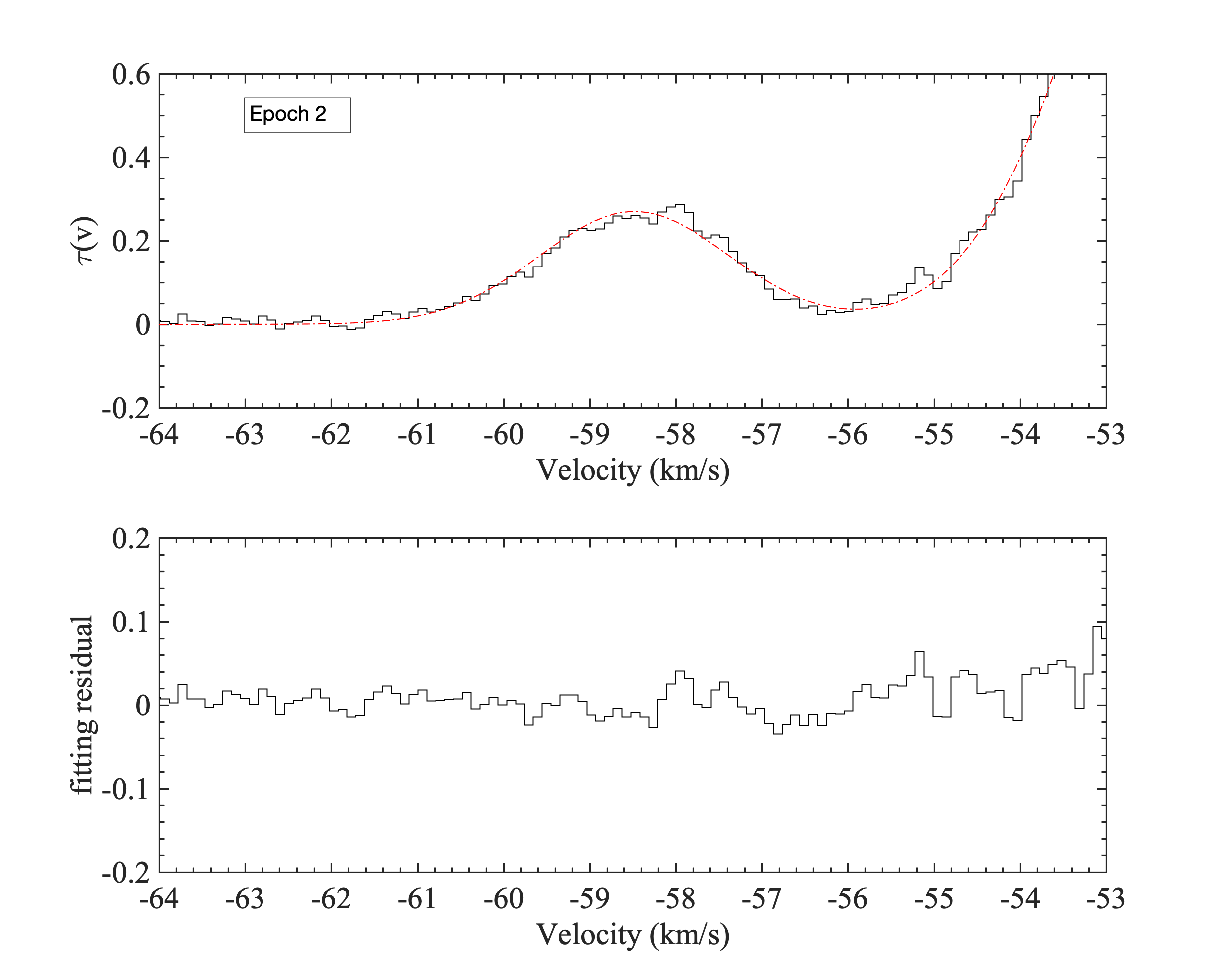}}
\gridline{\includegraphics[width=0.45\linewidth]{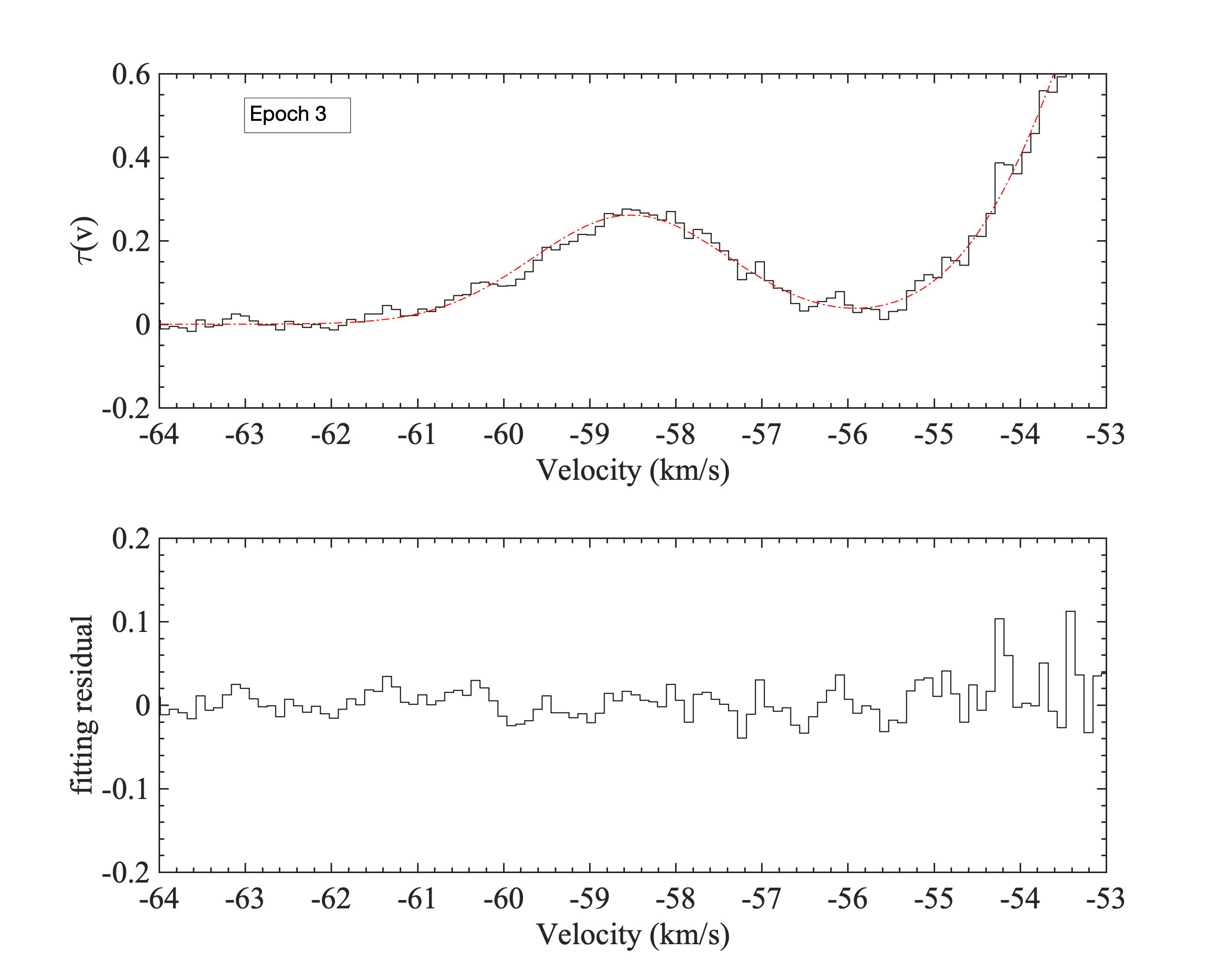}
\includegraphics[width=0.45\linewidth]{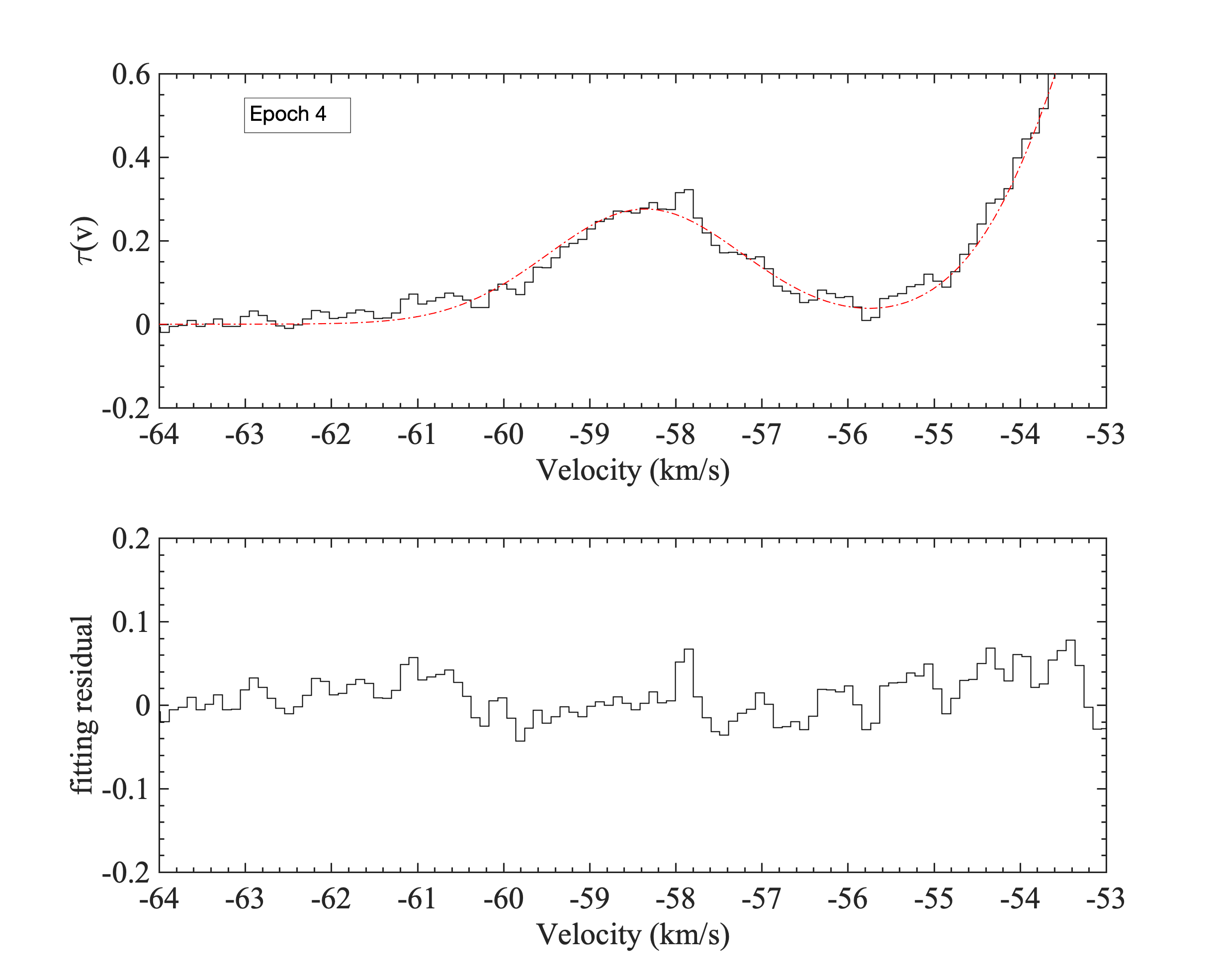}}
\gridline{\includegraphics[width=0.45\linewidth]{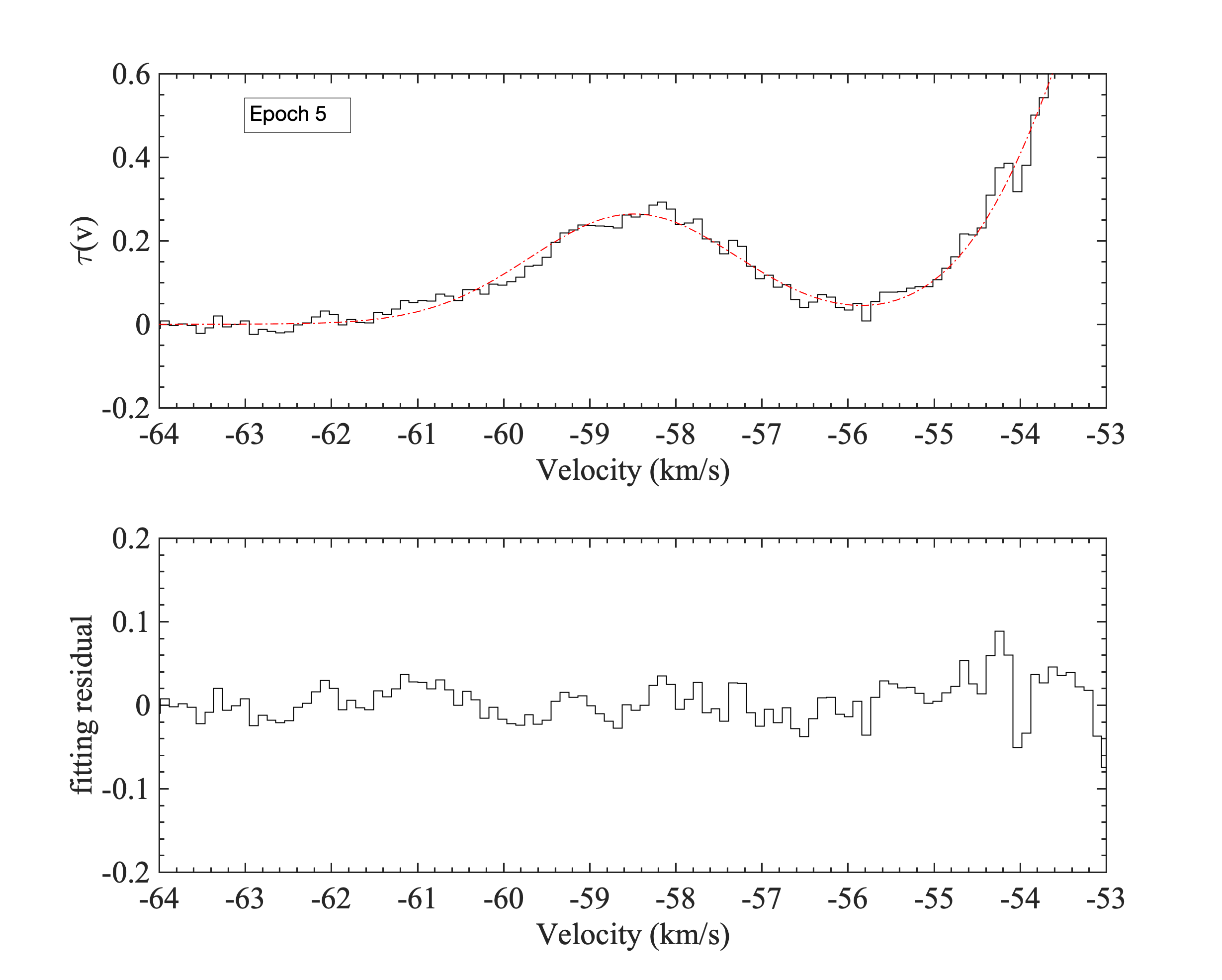}
\includegraphics[width=0.45\linewidth]{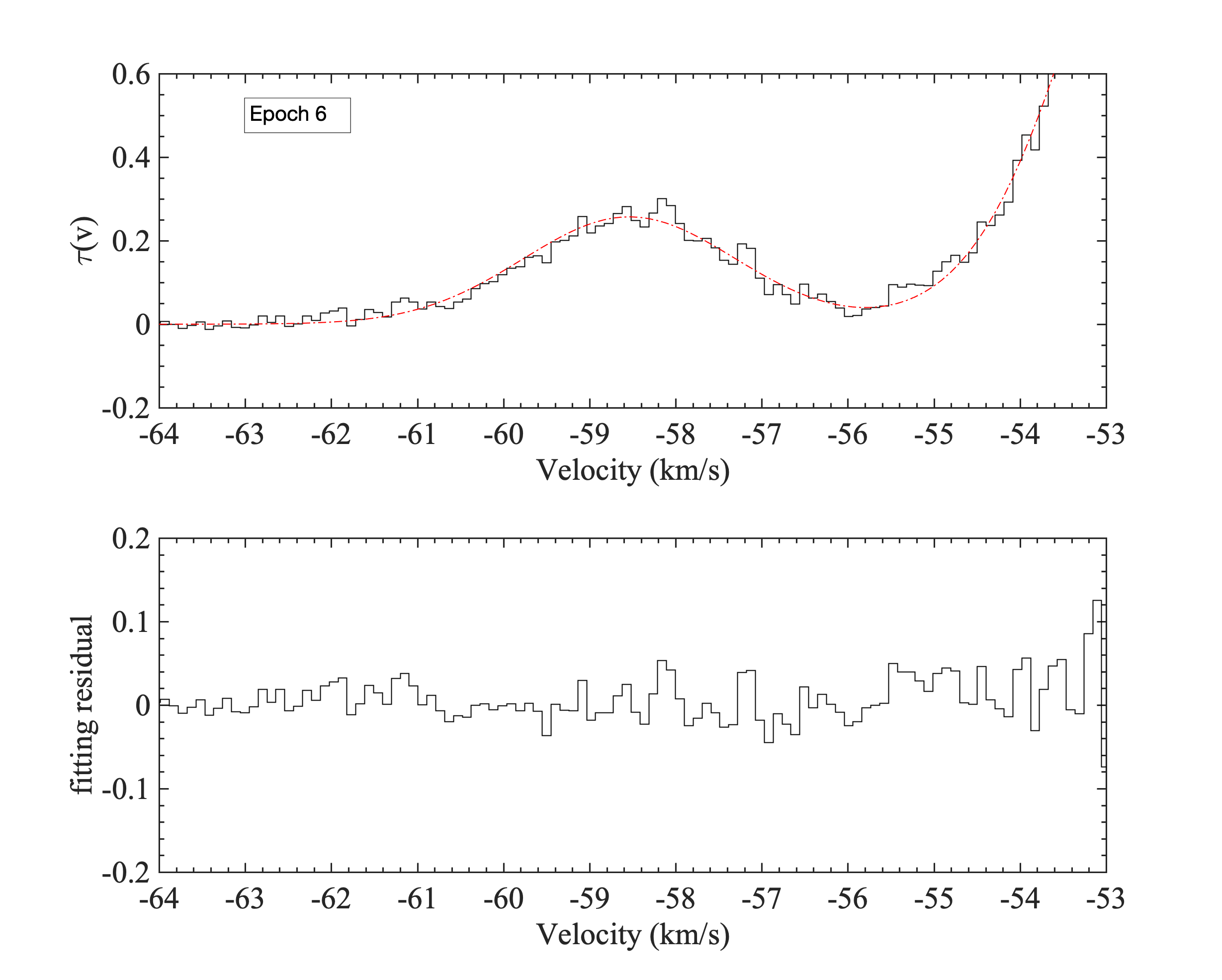}}
\caption{
From epoch 1 to 6: for each observing epoch:
The top panel shows the \hi\ optical depth spectrum $\tau(v)$ for the CNM associated with TSAS detections(black line) and the final fitted $\tau(v)$ (red line).
The bottom panel shows the corresponding fitting residuals. 
}
\vspace{0.2cm}
\label{fig:cnm_TSAS_fit0} 
\end{figure*}

\begin{figure*}
\centering
\gridline{\includegraphics[width=0.45\linewidth]{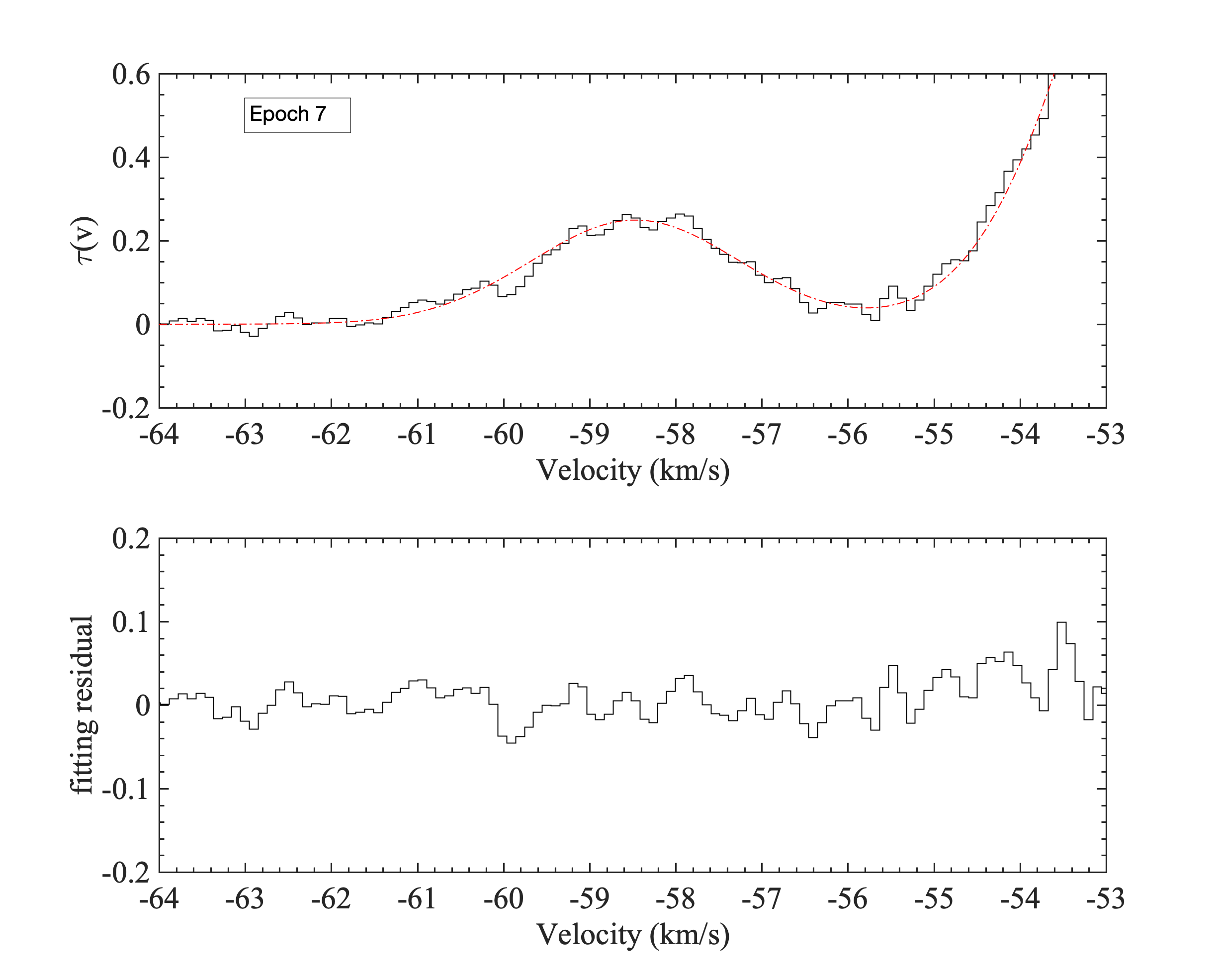}
\includegraphics[width=0.45\linewidth]{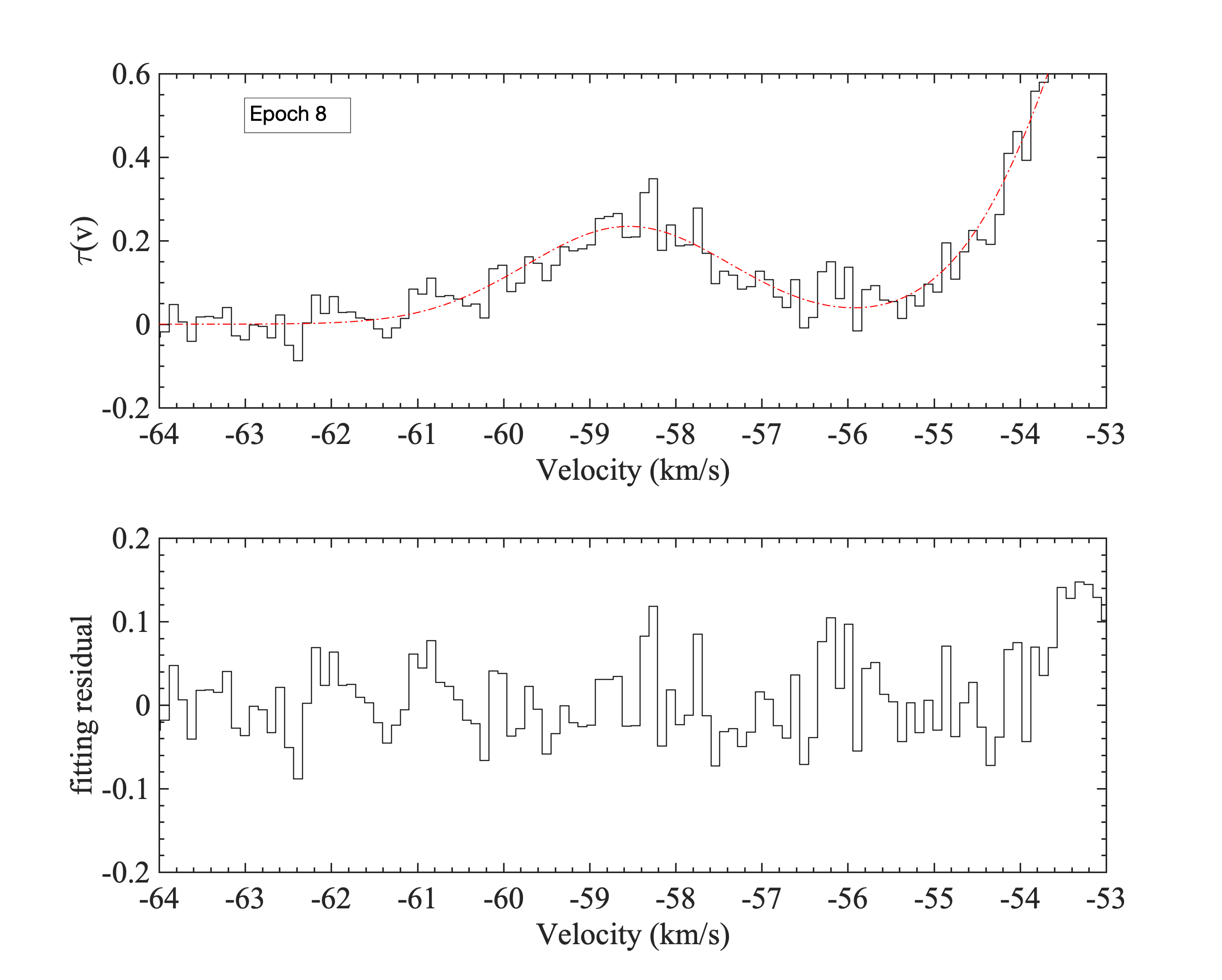}}
\gridline{\includegraphics[width=0.45\linewidth]{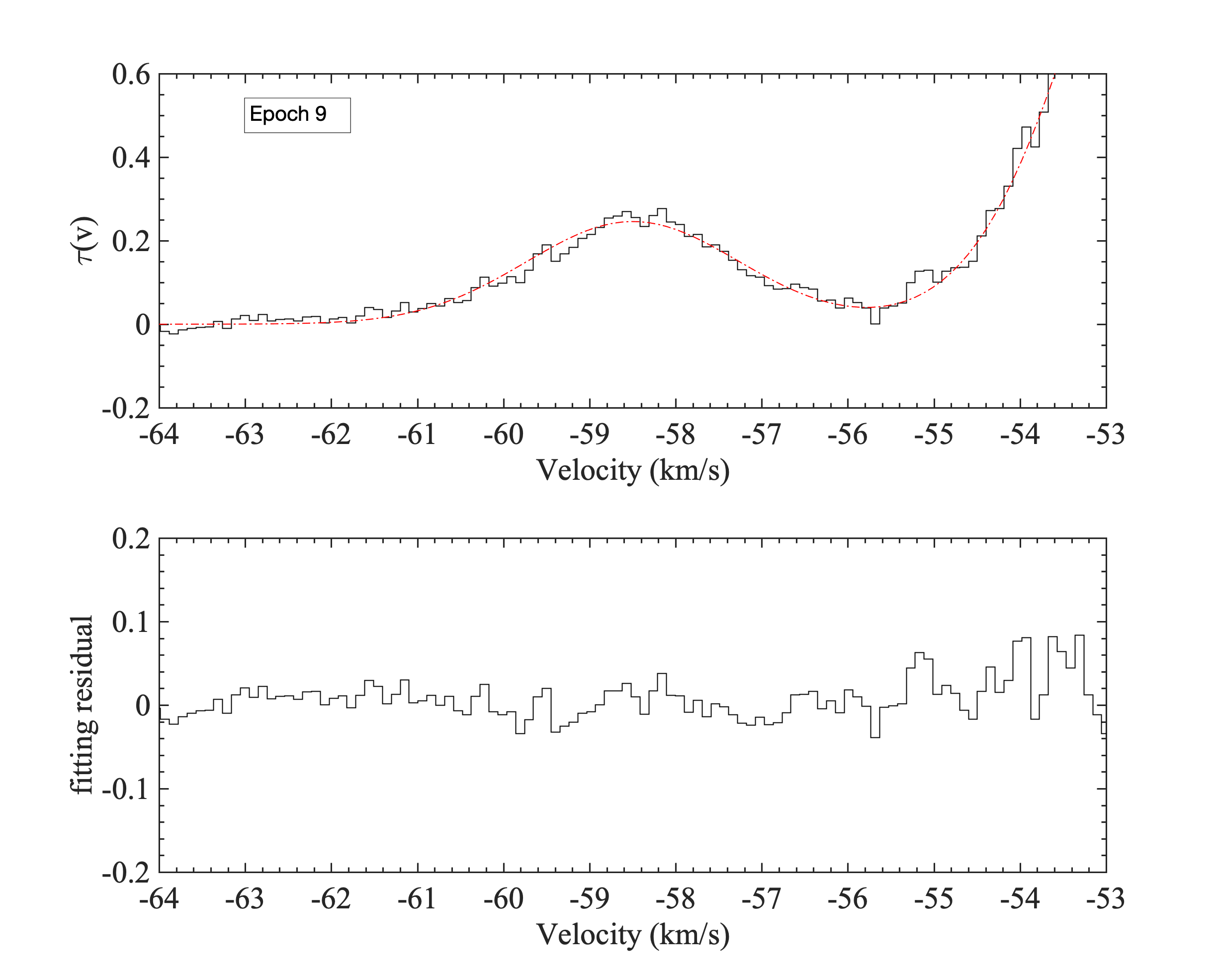}
\includegraphics[width=0.45\linewidth]{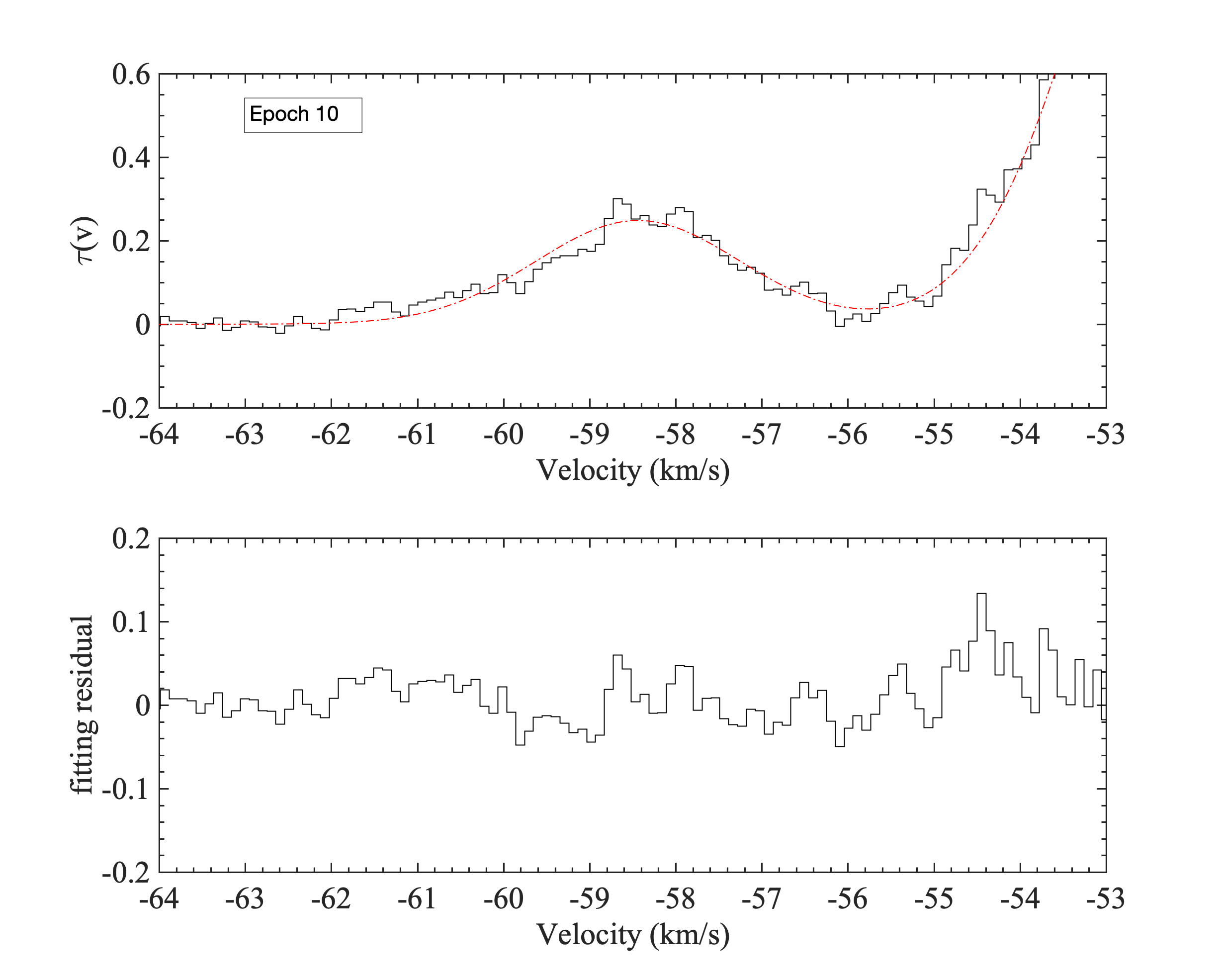}}
\gridline{\includegraphics[width=0.45\linewidth]{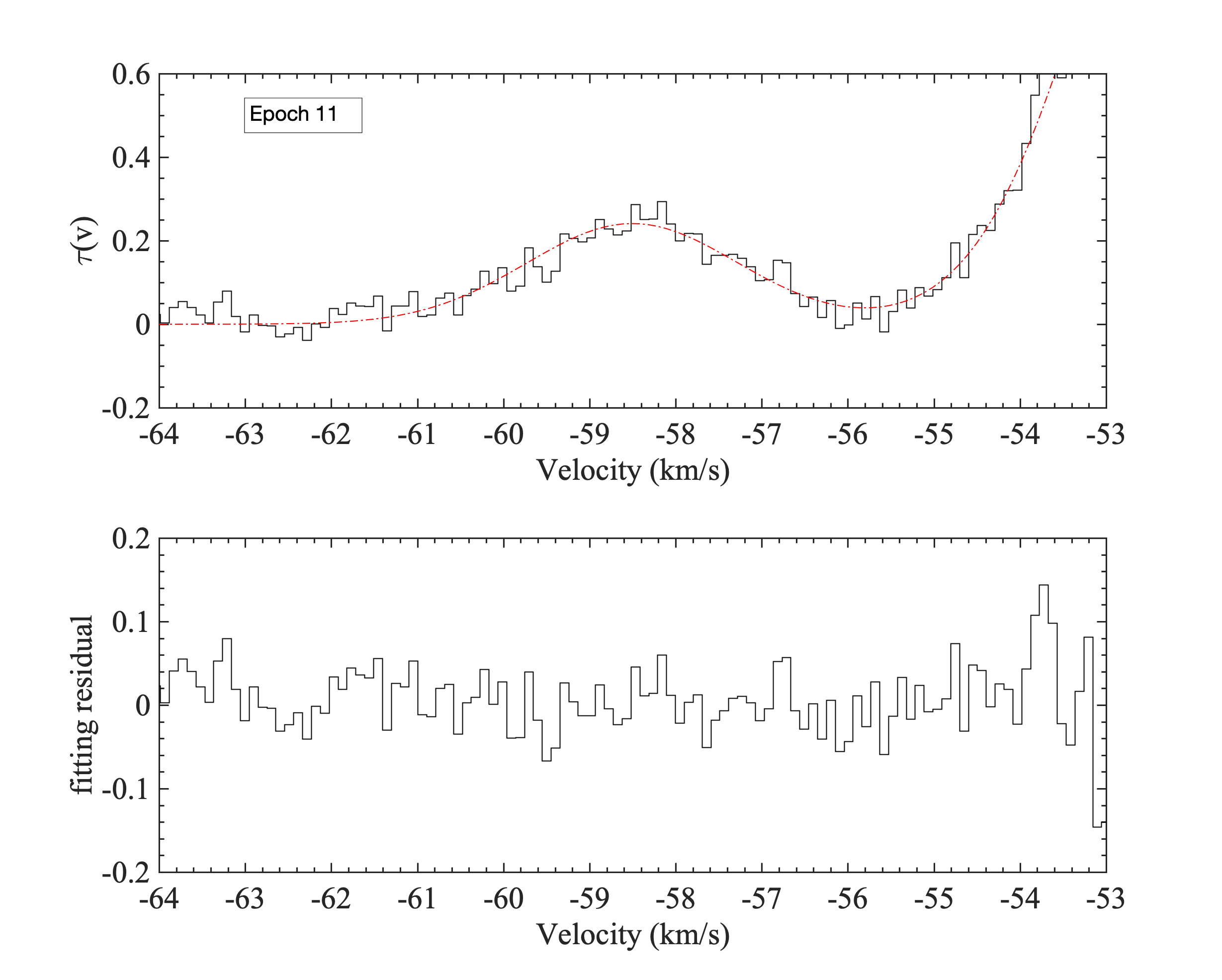}
\includegraphics[width=0.45\linewidth]{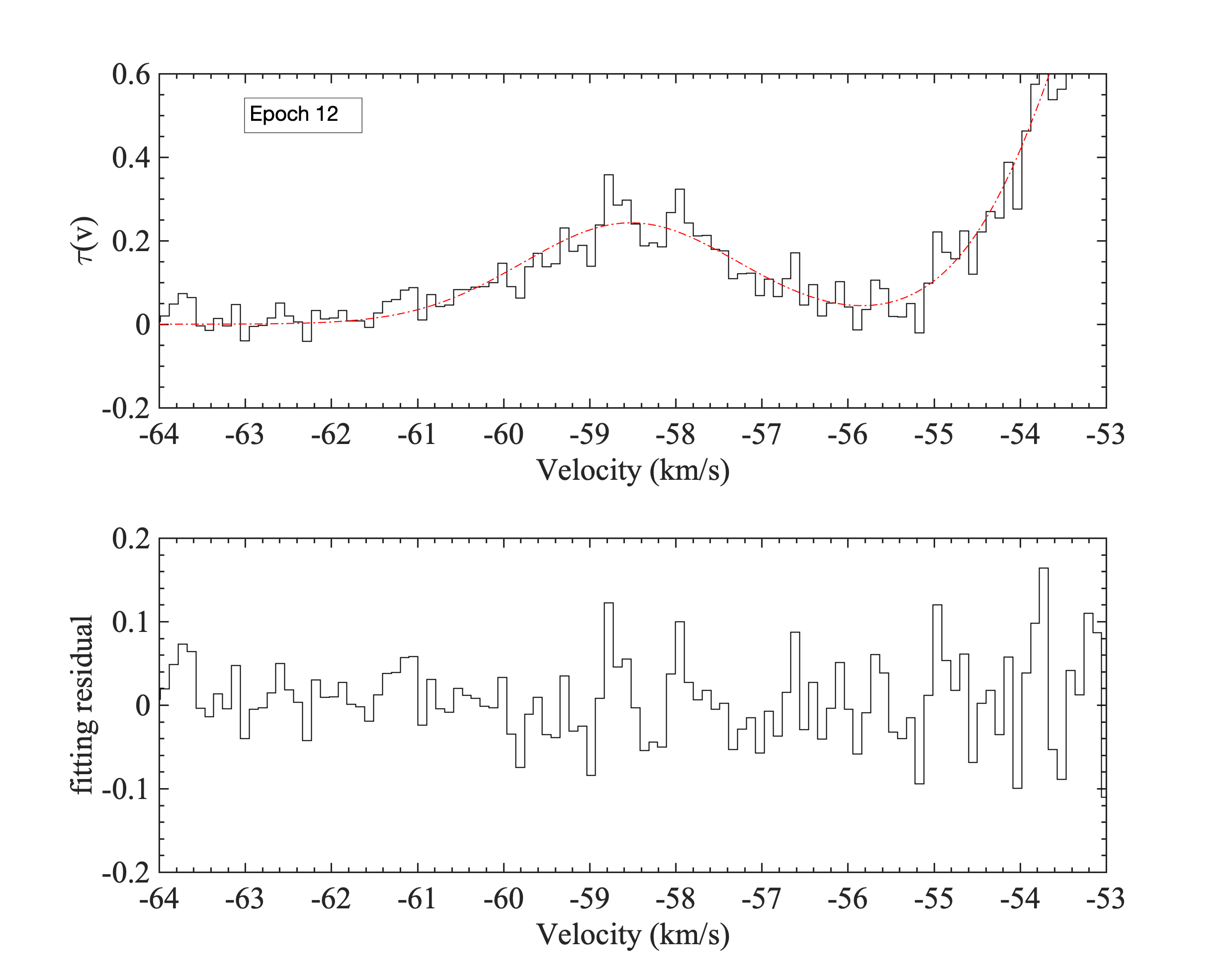}}
\caption{From epoch 7 to 12: similar to Figure~\ref{fig:cnm_TSAS_fit0}.} 
\vspace{0.2cm}
\label{fig:cnm_TSAS_fit1} 
\end{figure*}

\begin{figure*}
\centering
\gridline{\includegraphics[width=0.45\linewidth]{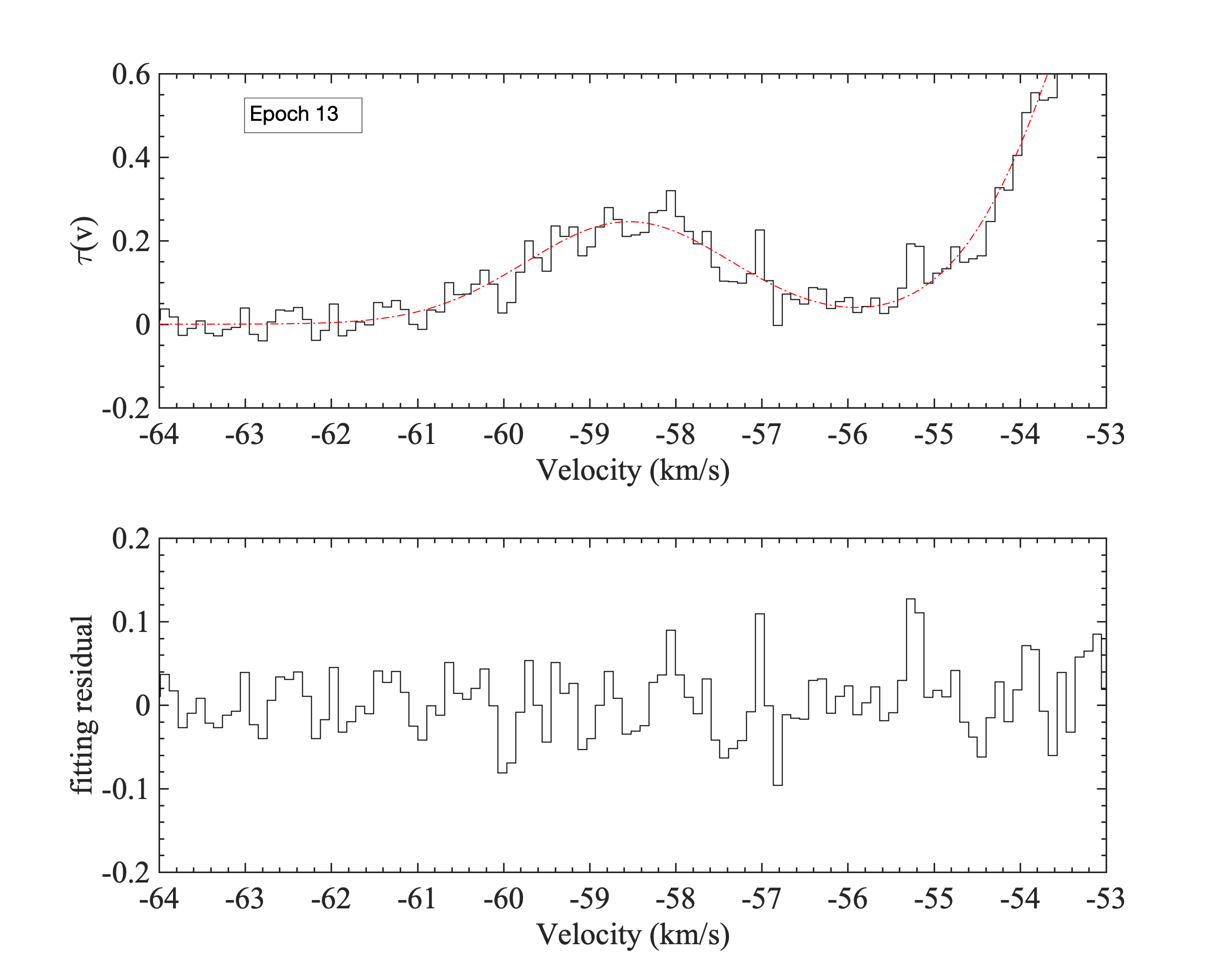}
\includegraphics[width=0.45\linewidth]{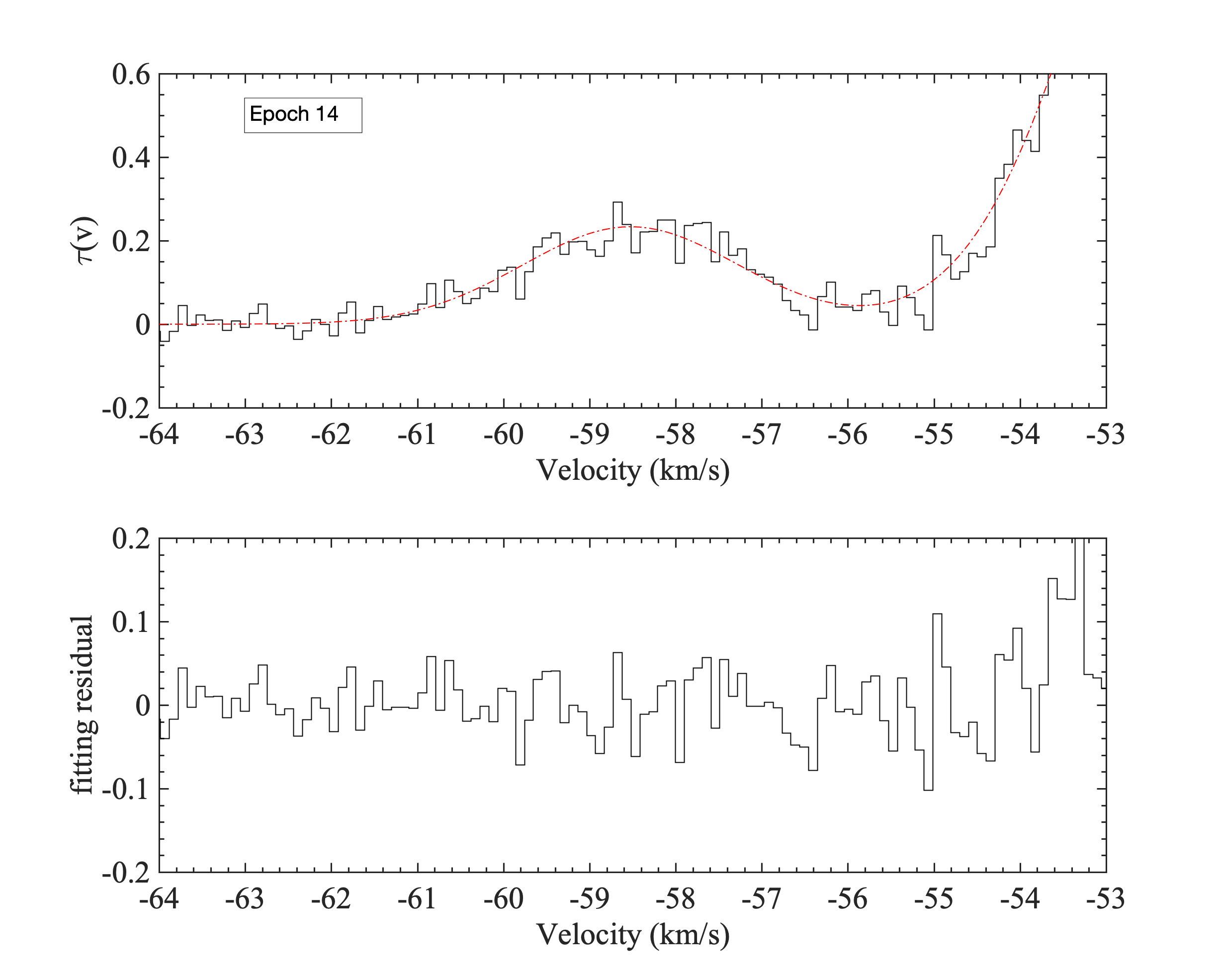}}
\gridline{\includegraphics[width=0.45\linewidth]{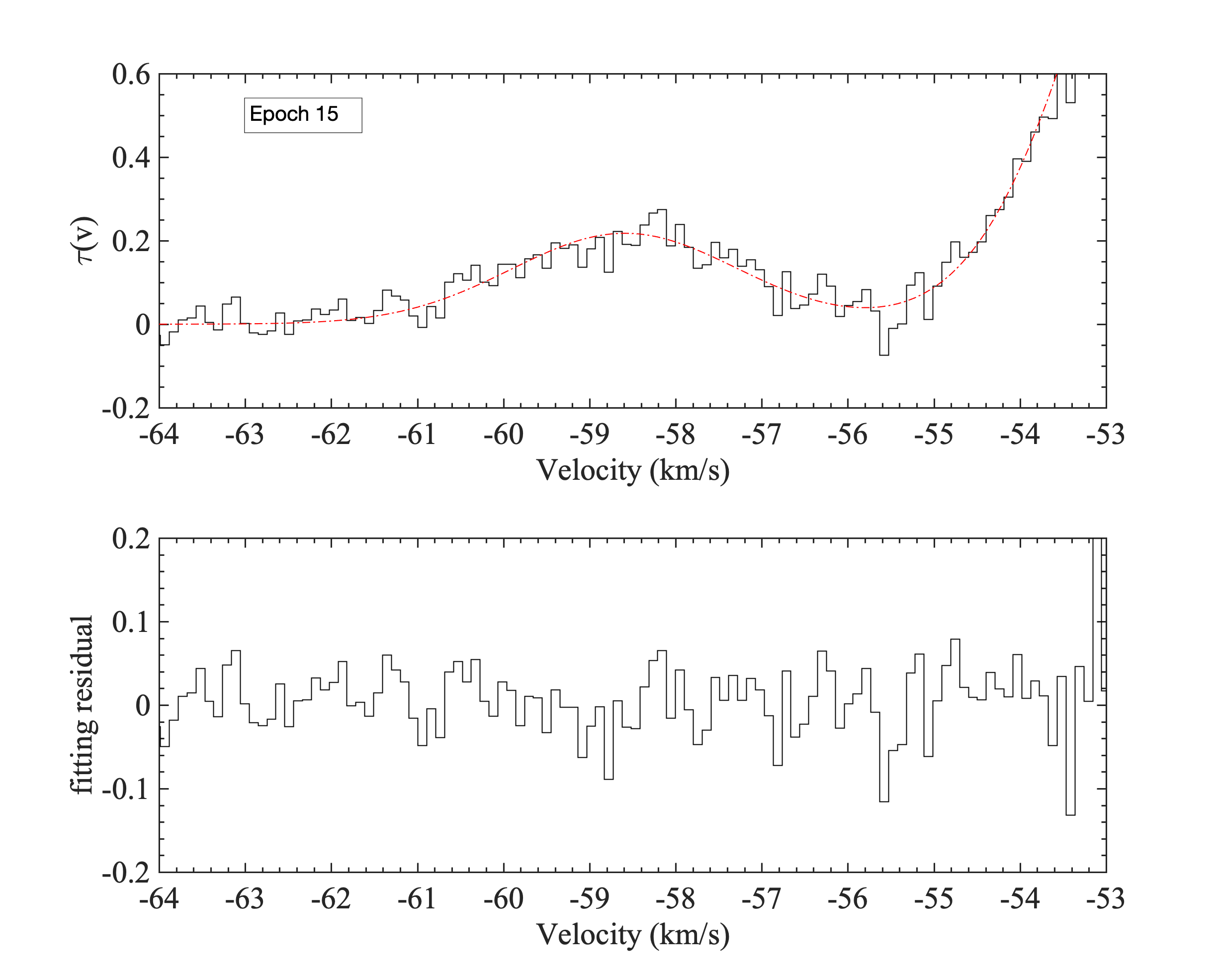}
\includegraphics[width=0.45\linewidth]{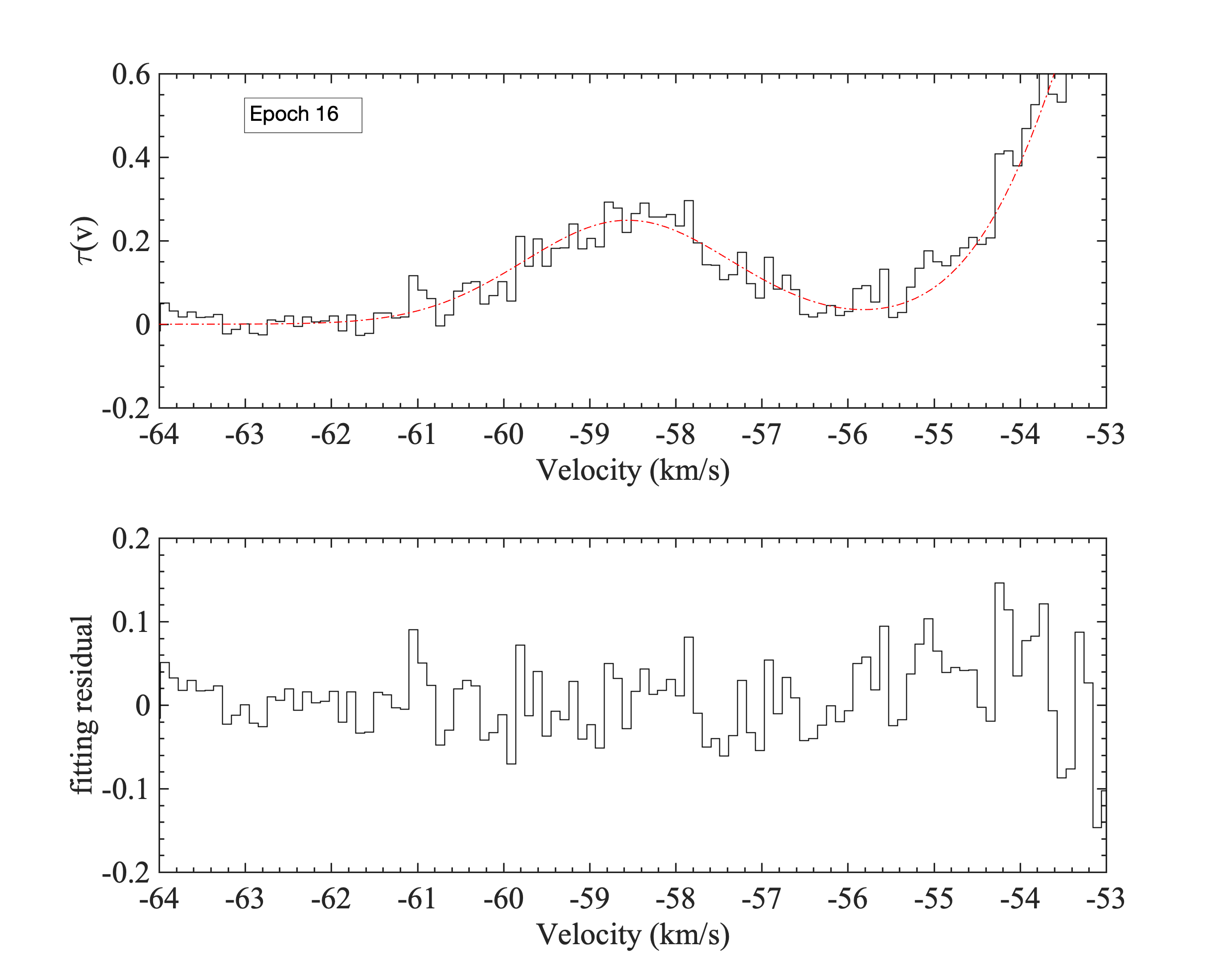}}
\gridline{\includegraphics[width=0.45\linewidth]{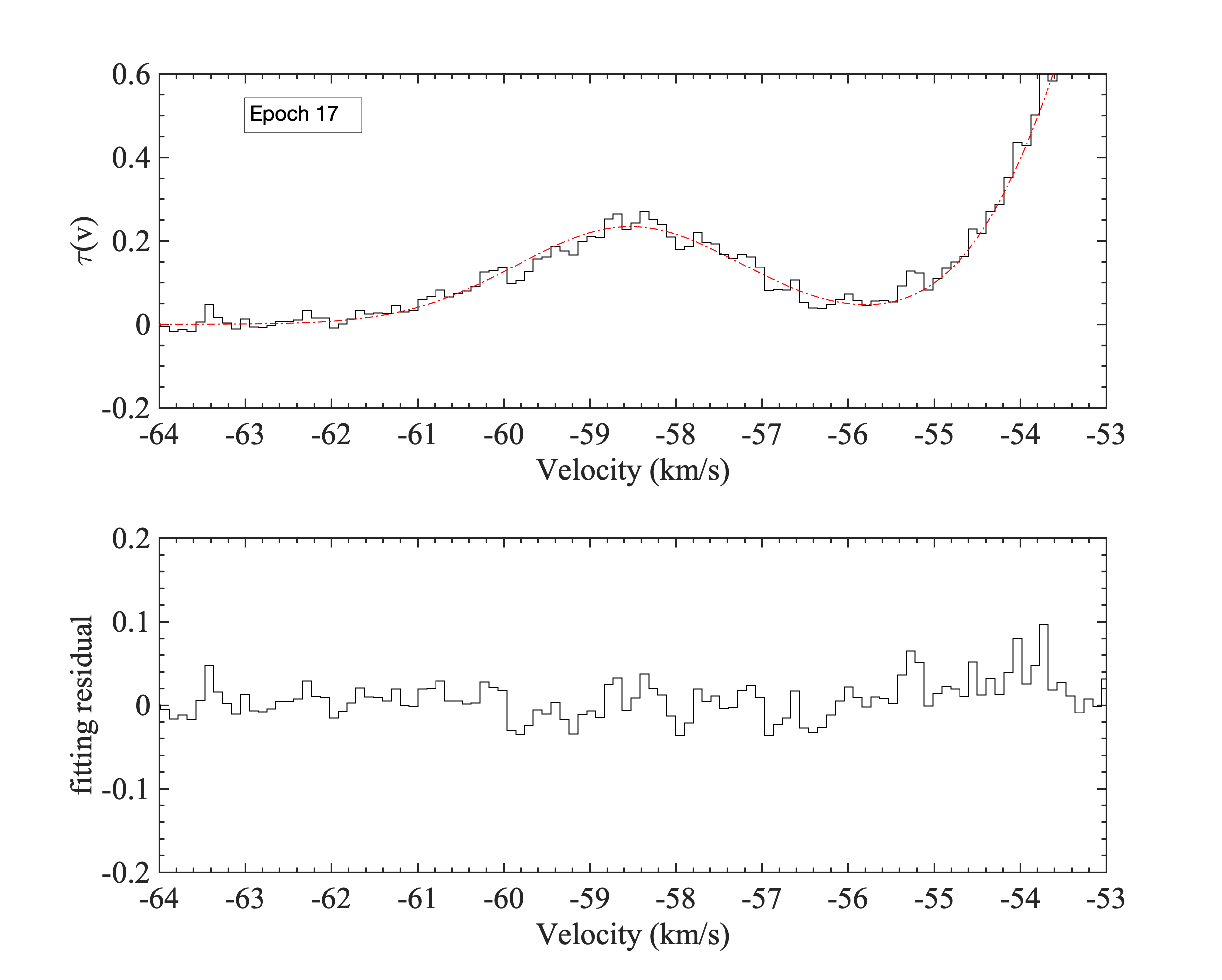}
\includegraphics[width=0.45\linewidth]{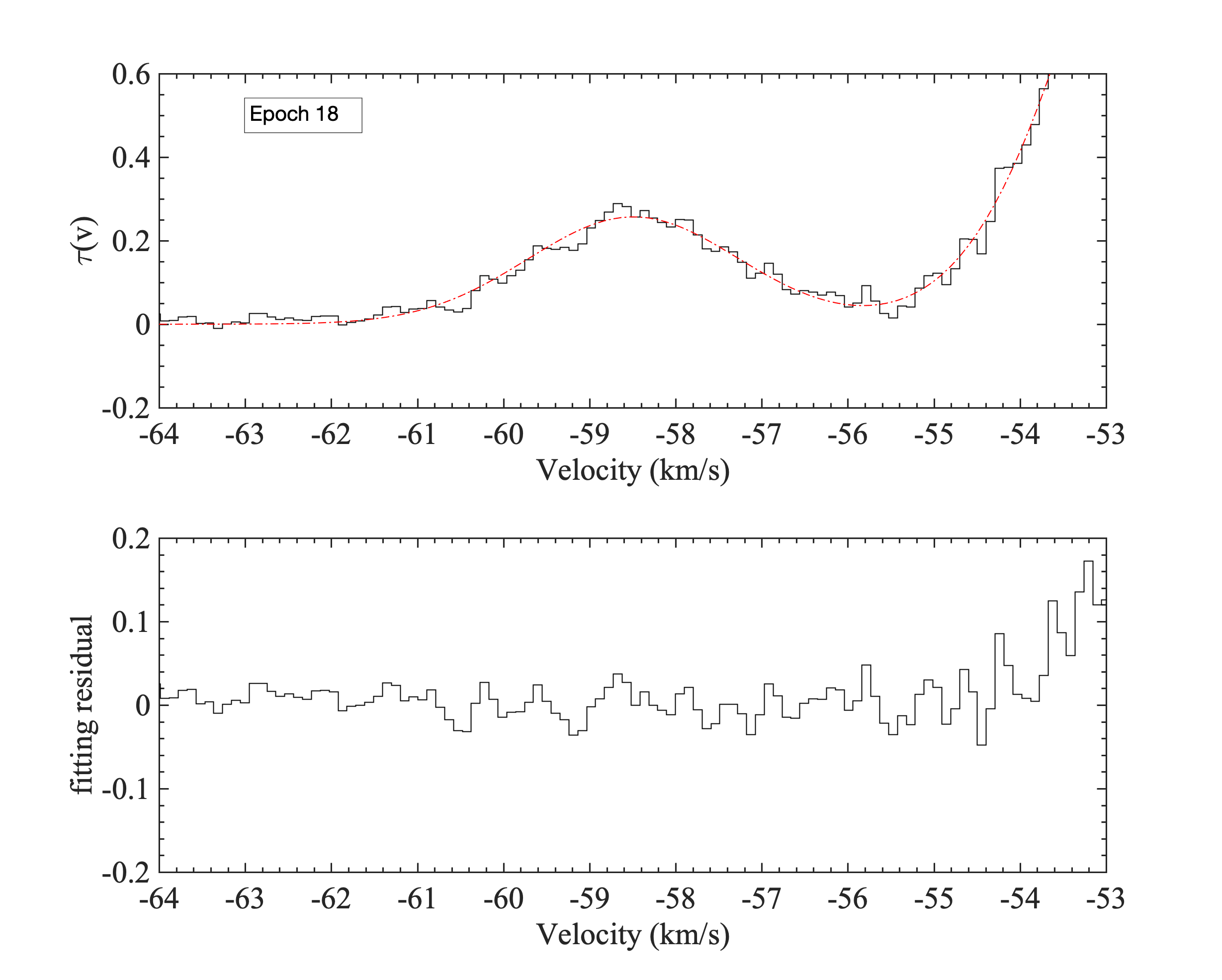}}
\caption{From epoch 13 to 18: similar to Figure~\ref{fig:cnm_TSAS_fit0}.} 
\vspace{0.2cm}
\label{fig:cnm_TSAS_fit2} 
\end{figure*}

\begin{figure*}
\centering
\includegraphics[width=0.45\linewidth]{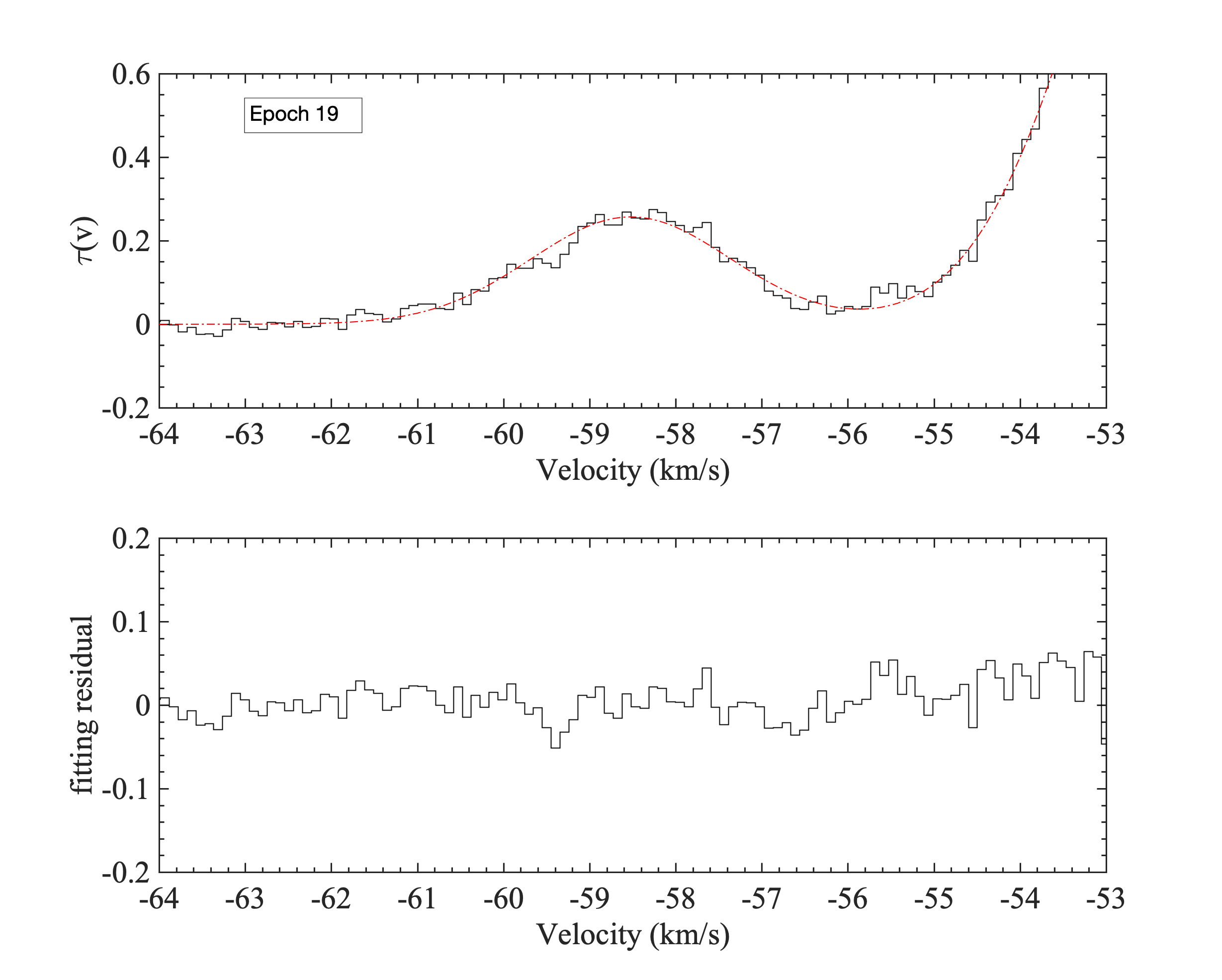}
\caption{For epoch 19: similar to Figure~\ref{fig:cnm_TSAS_fit0}.} 
\vspace{0.2cm}
\label{fig:cnm_TSAS_fit3} 
\end{figure*}

Table~\ref{tab:fit_tau_1420_amp_compB} to  \ref{tab:fit_tau_1420_N_compB} listed the Gaussian decomposition results, the estimated spin temperature, and the column density for all 13 CNM over 19 epochs. 
Figure~\ref{fig:compB_gaussian_Tex_1420_fit0} and \ref{fig:compB_gaussian_Tex_1420_fit1} illustrate the final fitting results over 19 epochs. 
Figure~\ref{fig:cnm_TSAS_fit0} to \ref{fig:cnm_TSAS_fit3} demonstrate the Gaussian fitting results for the CNM associated with the TSAS across 19 epochs.

\begin{figure*}
\centering
\gridline{\includegraphics[width=0.5\linewidth]{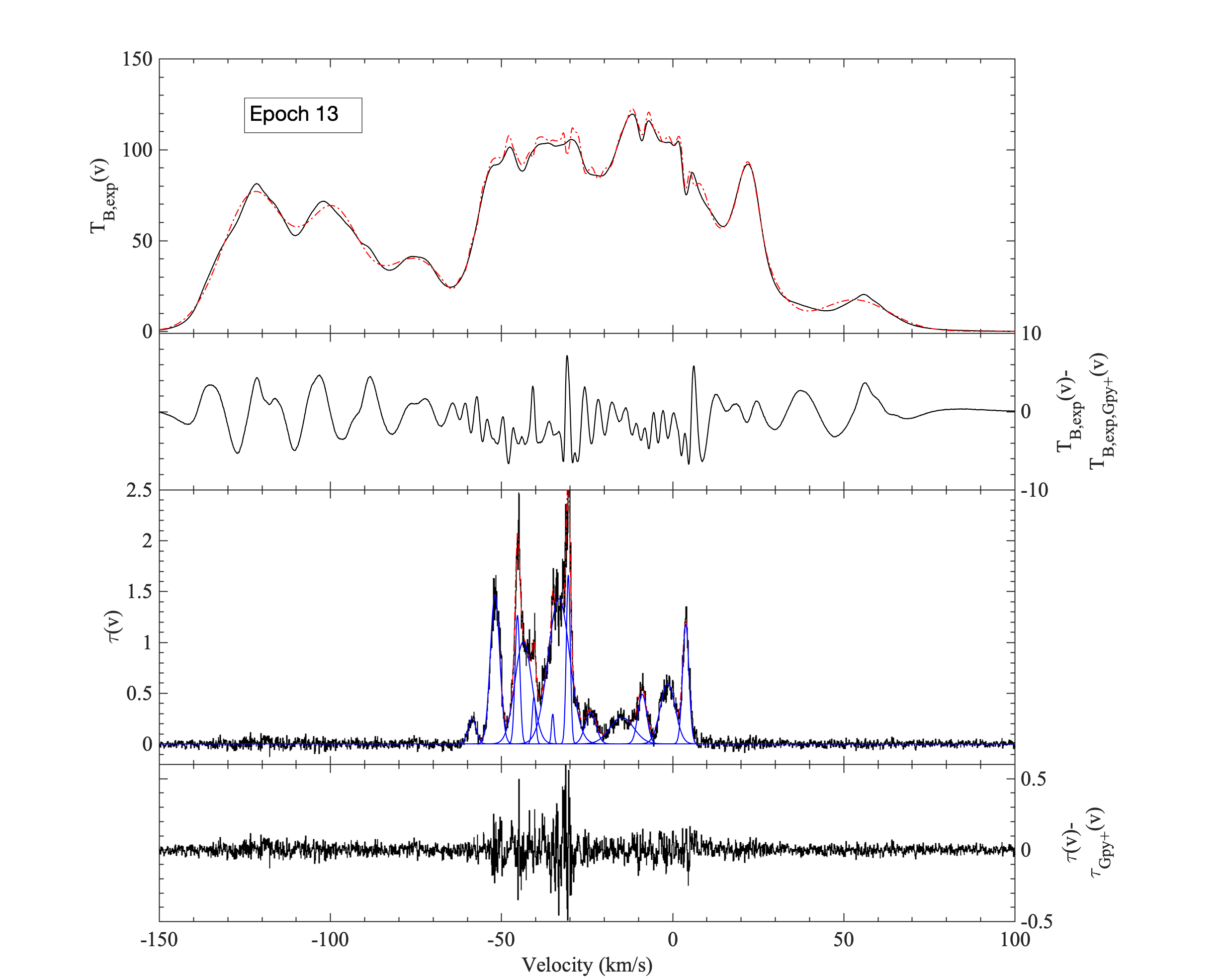}
\includegraphics[width=0.5\linewidth]{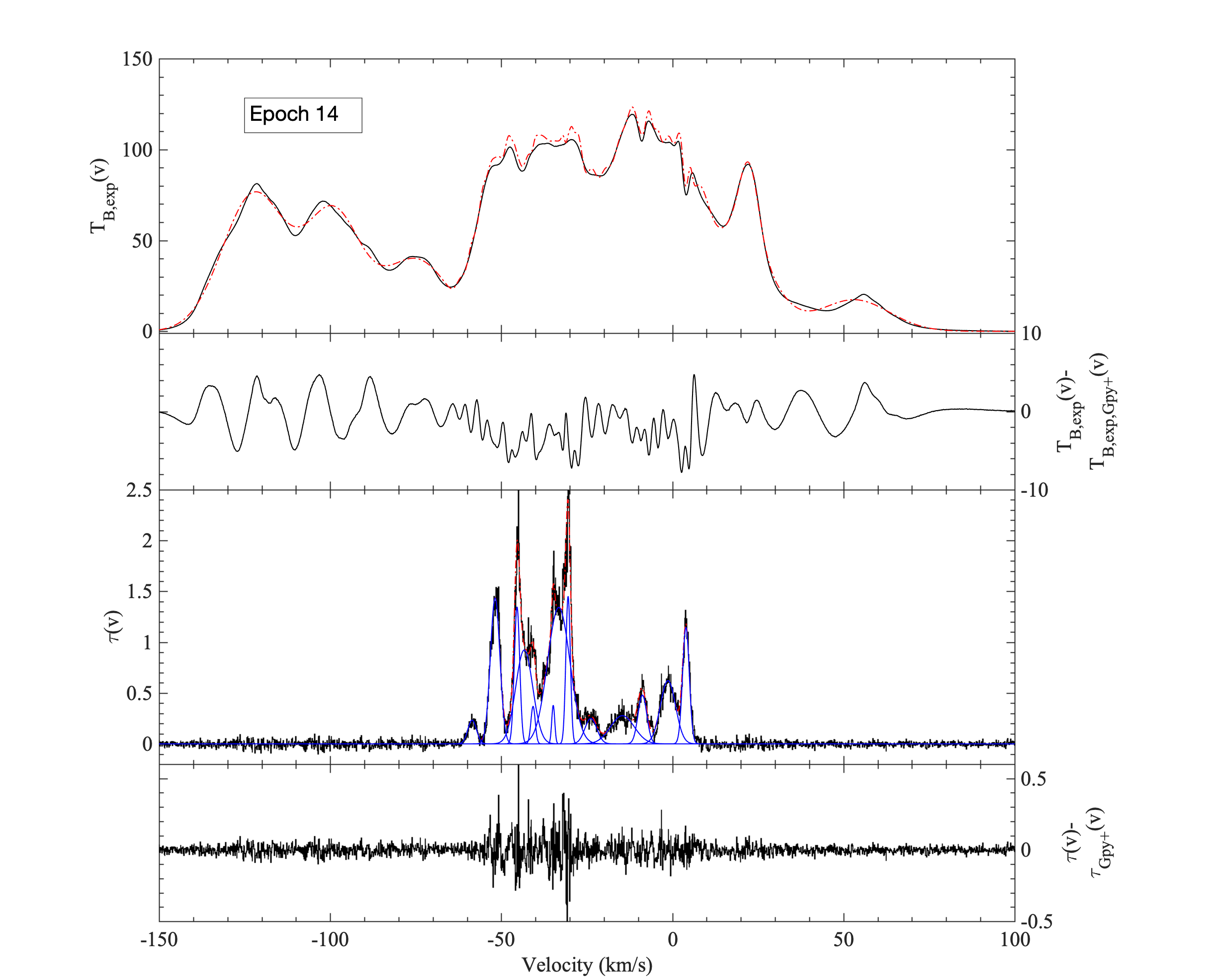}}
\gridline{\includegraphics[width=0.5\linewidth]{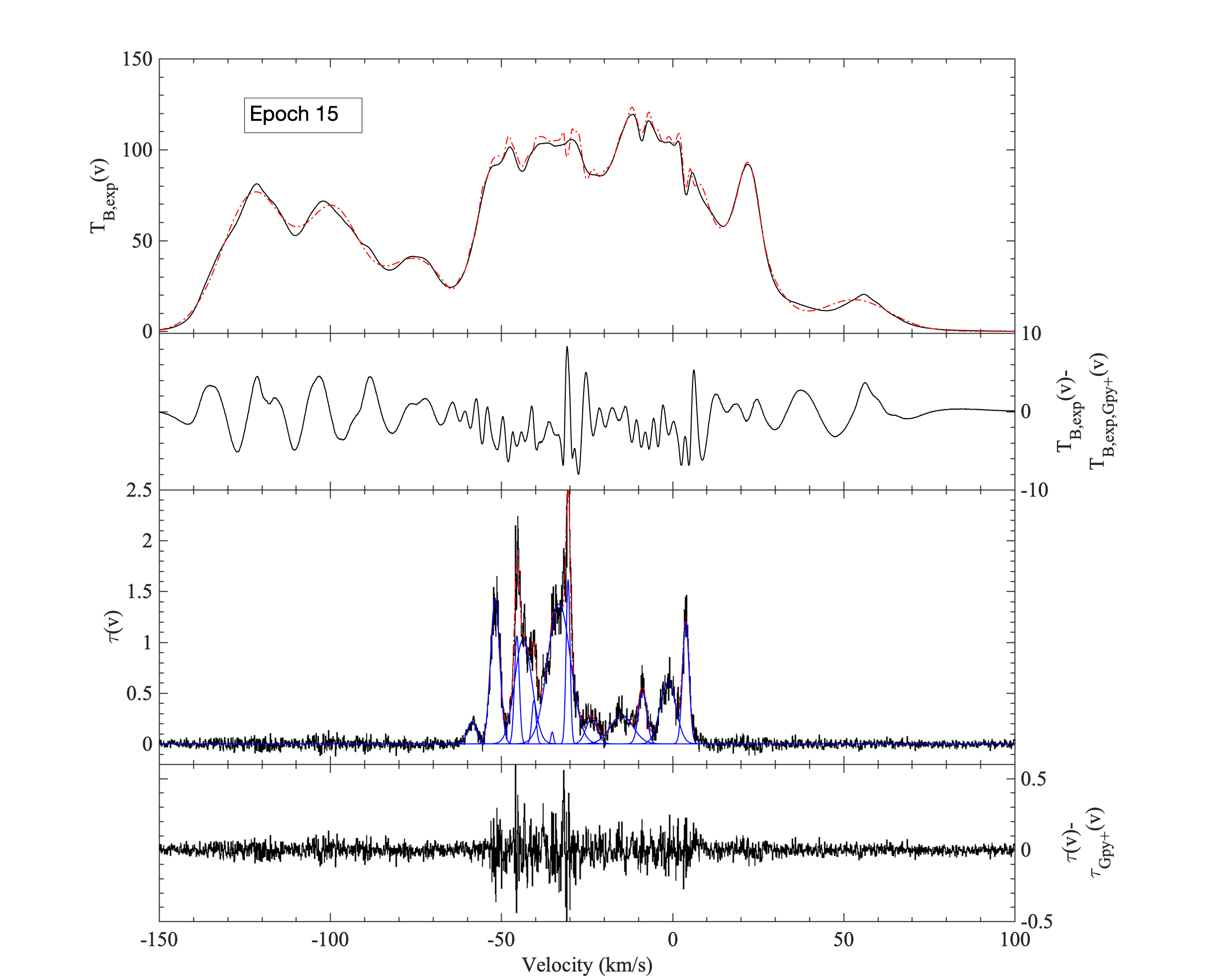}
\includegraphics[width=0.5\linewidth]{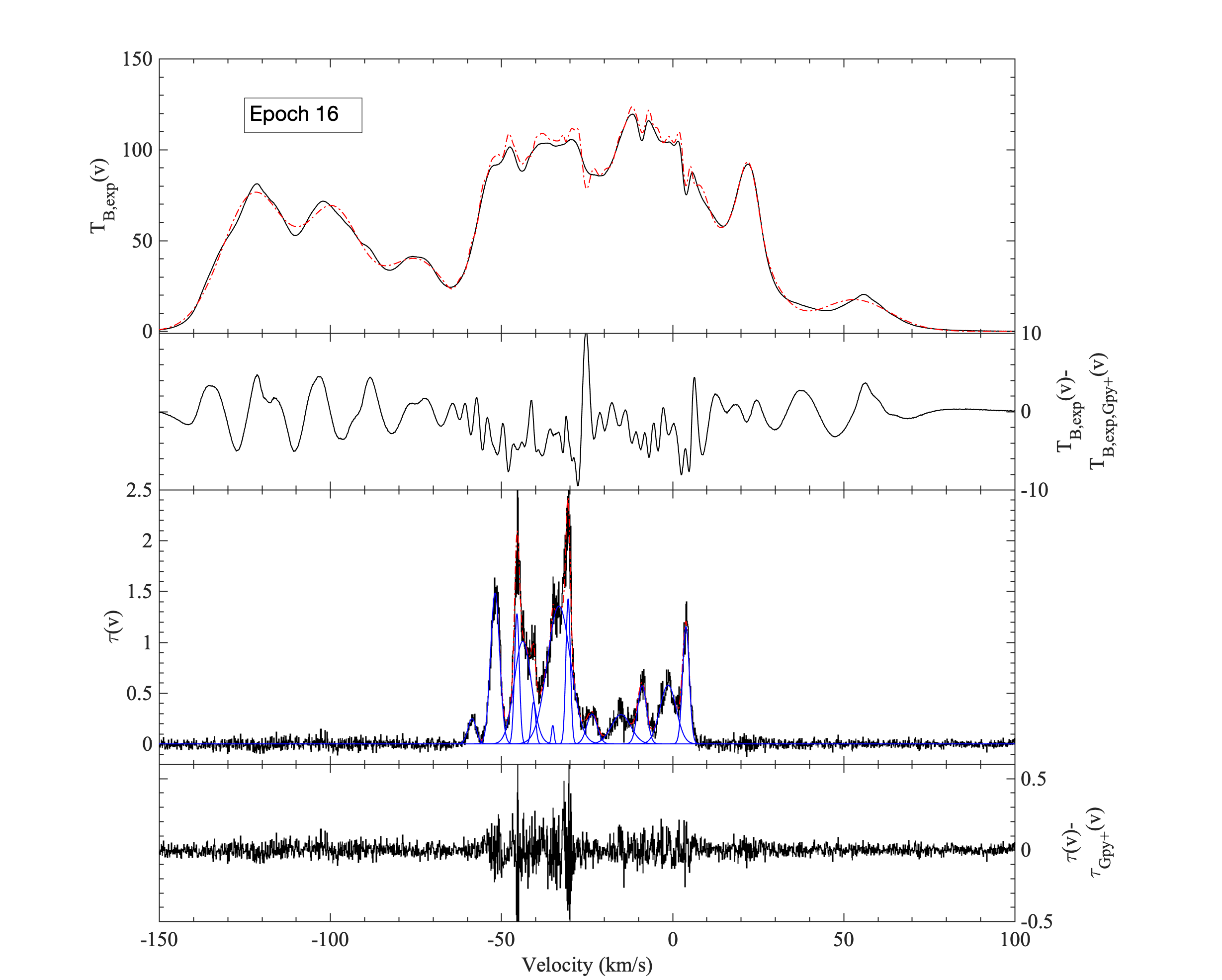}}
\gridline{\includegraphics[width=0.5\linewidth]{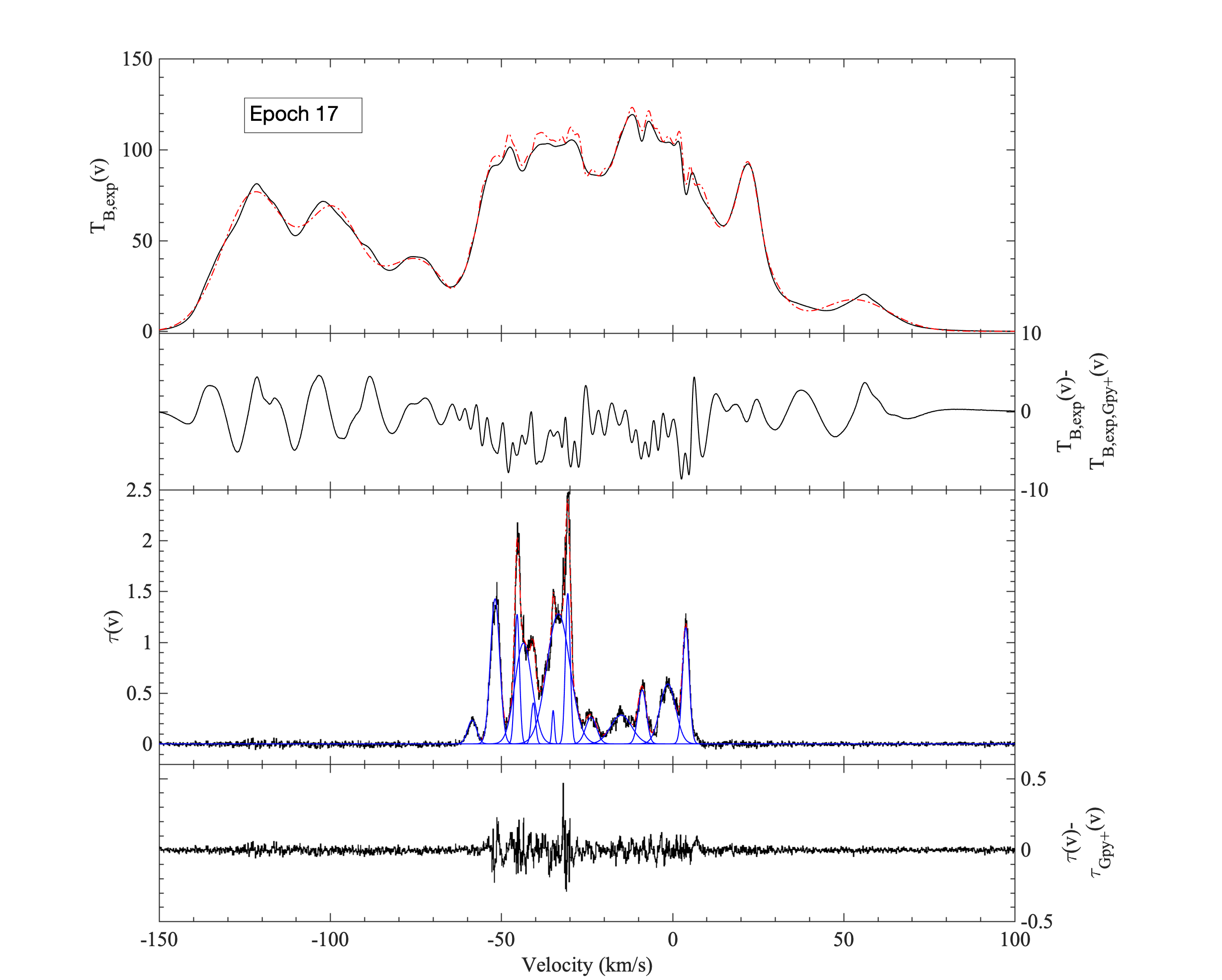}
\includegraphics[width=0.5\linewidth]{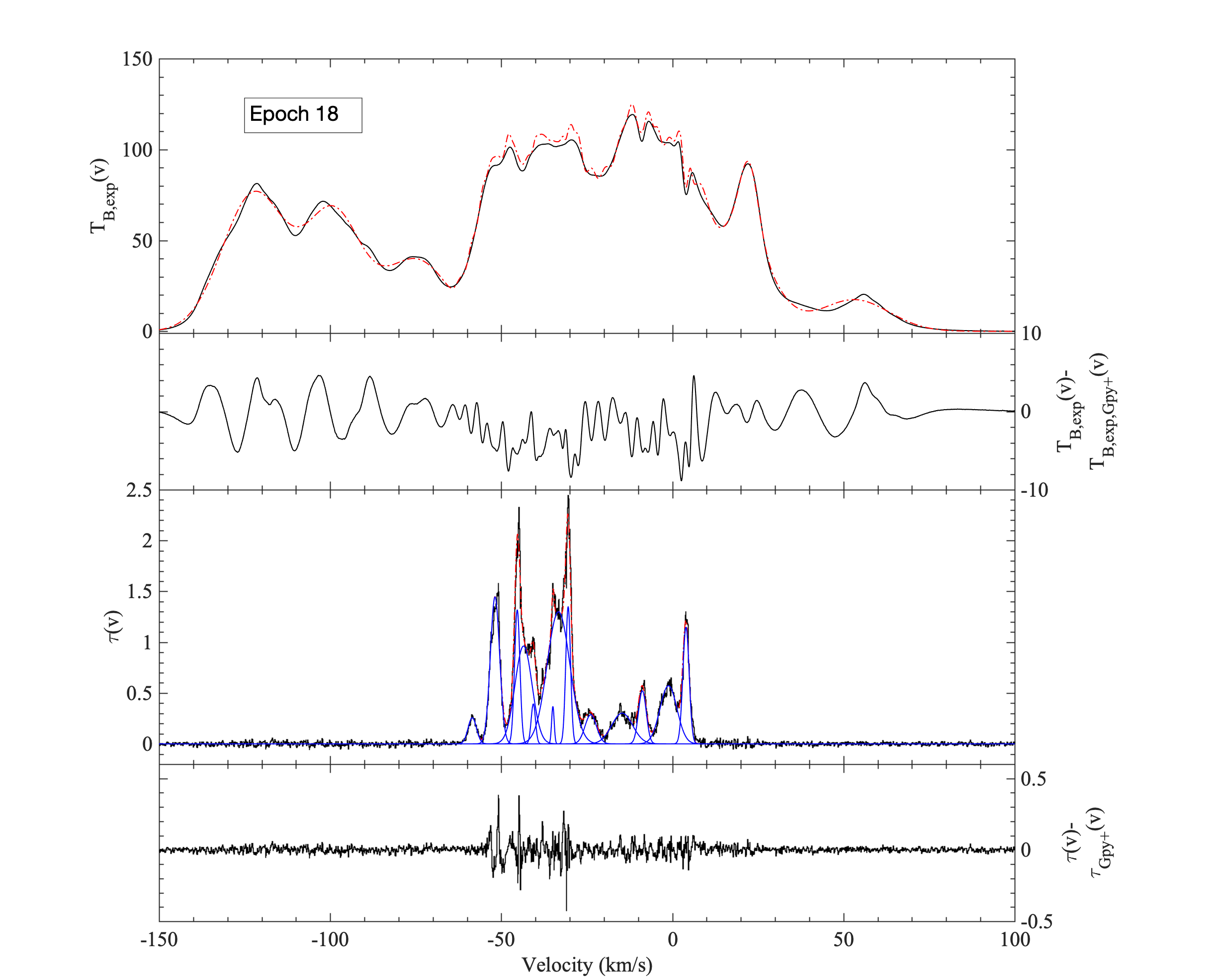}}
\caption{From epoch 13 to 18: similar to Figure~\ref{fig:compB_gaussian_Tex_1420_fit0}.} 
\vspace{0.2cm}
\label{fig:compB_gaussian_Tex_1420_fit2} 
\end{figure*}

\begin{figure*}
\centering
\includegraphics[width=0.5\linewidth]{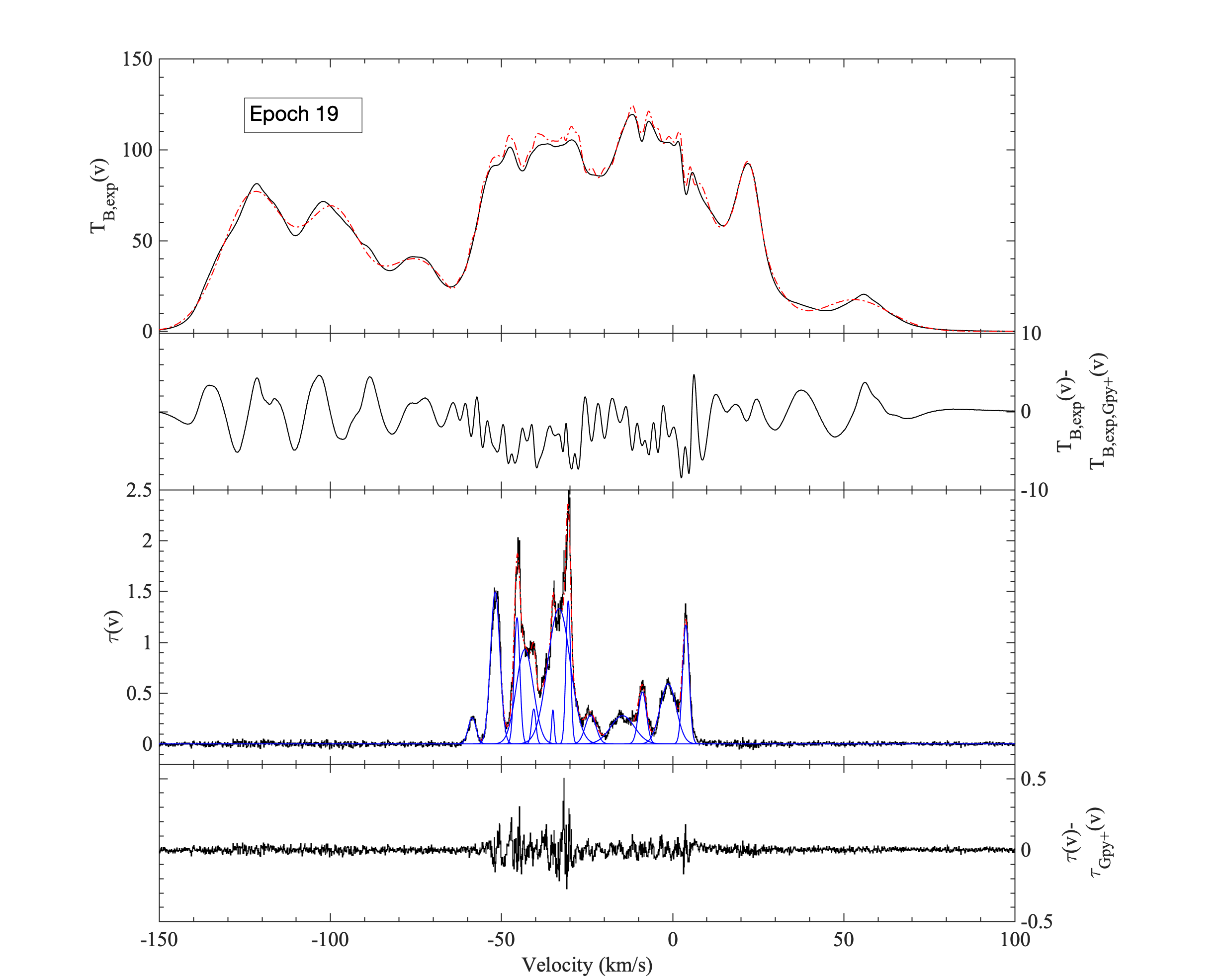}
\caption{For epoch 19: similar to Figure~\ref{fig:compB_gaussian_Tex_1420_fit0}.} 
\vspace{0.2cm}
\label{fig:compB_gaussian_Tex_1420_fit3} 
\end{figure*}

%%%updated 0926
\begin{rotatetable*}
\begin{deluxetable*}{c|ccccccccccccc}
\setlength{\tabcolsep}{0.04in} 
%\resizebox{\textwidth}
\tablecolumns{17} 
\tabletypesize{\scriptsize}
\tablewidth{0pt}
\tablecaption{Gaussian Decomposition Results of Amplitude for the \hi\ Optical Depth Spectra over 19 observing epochs}
\label{tab:fit_tau_1420_amp_compB}
\tablehead{\colhead{Epoch} & \colhead{$\tau_{0,1}$} & \colhead{$\tau_{0,2}$} & \colhead{$\tau_{0,3}$} & \colhead{$\tau_{0,4}$} & \colhead{$\tau_{0,5}$} & \colhead{$\tau_{0,6}$} & \colhead{$\tau_{0,7}$} & \colhead{$\tau_{0,8}$} & \colhead{$\tau_{0,9}$} & \colhead{$\tau_{0,10}$} & \colhead{$\tau_{0,11}$} & \colhead{$\tau_{0,12}$}& \colhead{$\tau_{0,13}$} } 
\startdata
No. &     &  &   &   & & &  & &  &  & & & \\
\hline
1 & 0.29 $\pm$ 0.02 & 1.42 $\pm$ 0.02 & 1.30 $\pm$ 0.04 & 0.96 $\pm$ 0.02 & 0.41 $\pm$ 0.04 & 0.26 $\pm$ 0.03 & 1.29 $\pm$ 0.02 & 1.42 $\pm$ 0.03 & 0.28 $\pm$ 0.01 & 0.27 $\pm$ 0.01 & 0.51 $\pm$ 0.02 & 0.60 $\pm$ 0.01 & 1.12 $\pm$ 0.02\\
2 & 0.27 $\pm$ 0.02 & 1.37 $\pm$ 0.01 & 1.28 $\pm$ 0.03 & 0.92 $\pm$ 0.02 & 0.40 $\pm$ 0.04 & 0.34 $\pm$ 0.03 & 1.33 $\pm$ 0.01 & 1.44 $\pm$ 0.02 & 0.28 $\pm$ 0.01 & 0.27 $\pm$ 0.01 & 0.51 $\pm$ 0.02 & 0.57 $\pm$ 0.01 & 1.15 $\pm$ 0.02\\
3 & 0.26 $\pm$ 0.02 & 1.38 $\pm$ 0.02 & 1.46 $\pm$ 0.03 & 0.96 $\pm$ 0.02 & 0.33 $\pm$ 0.04 & 0.42 $\pm$ 0.03 & 1.29 $\pm$ 0.01 & 1.52 $\pm$ 0.03 & 0.28 $\pm$ 0.01 & 0.27 $\pm$ 0.01 & 0.54 $\pm$ 0.02 & 0.58 $\pm$ 0.01 & 1.16 $\pm$ 0.02\\
4 & 0.28 $\pm$ 0.02 & 1.48 $\pm$ 0.02 & 1.33 $\pm$ 0.04 & 0.97 $\pm$ 0.02 & 0.41 $\pm$ 0.04 & 0.28 $\pm$ 0.03 & 1.38 $\pm$ 0.02 & 1.63 $\pm$ 0.03 & 0.29 $\pm$ 0.02 & 0.28 $\pm$ 0.01 & 0.52 $\pm$ 0.02 & 0.59 $\pm$ 0.01 & 1.17 $\pm$ 0.02\\
5 & 0.26 $\pm$ 0.02 & 1.39 $\pm$ 0.01 & 1.30 $\pm$ 0.04 & 0.95 $\pm$ 0.02 & 0.43 $\pm$ 0.04 & 0.26 $\pm$ 0.03 & 1.36 $\pm$ 0.01 & 1.32 $\pm$ 0.03 & 0.28 $\pm$ 0.01 & 0.29 $\pm$ 0.01 & 0.51 $\pm$ 0.02 & 0.59 $\pm$ 0.01 & 1.16 $\pm$ 0.02\\
6 & 0.26 $\pm$ 0.02 & 1.47 $\pm$ 0.02 & 1.29 $\pm$ 0.05 & 0.95 $\pm$ 0.03 & 0.41 $\pm$ 0.05 & 0.29 $\pm$ 0.04 & 1.35 $\pm$ 0.02 & 1.51 $\pm$ 0.03 & 0.28 $\pm$ 0.02 & 0.28 $\pm$ 0.01 & 0.51 $\pm$ 0.03 & 0.59 $\pm$ 0.02 & 1.18 $\pm$ 0.02\\
7 & 0.25 $\pm$ 0.01 & 1.47 $\pm$ 0.01 & 1.40 $\pm$ 0.03 & 0.97 $\pm$ 0.02 & 0.39 $\pm$ 0.04 & 0.25 $\pm$ 0.03 & 1.33 $\pm$ 0.01 & 1.31 $\pm$ 0.02 & 0.28 $\pm$ 0.01 & 0.28 $\pm$ 0.01 & 0.50 $\pm$ 0.02 & 0.59 $\pm$ 0.01 & 1.16 $\pm$ 0.02\\
8 & 0.23 $\pm$ 0.04 & 1.44 $\pm$ 0.04 & 1.40 $\pm$ 0.08 & 0.98 $\pm$ 0.04 & 0.39 $\pm$ 0.08 & 0.50 $\pm$ 0.07 & 1.50 $\pm$ 0.03 & 1.36 $\pm$ 0.06 & 0.29 $\pm$ 0.03 & 0.28 $\pm$ 0.03 & 0.56 $\pm$ 0.04 & 0.61 $\pm$ 0.03 & 1.19 $\pm$ 0.04\\
9 & 0.25 $\pm$ 0.02 & 1.44 $\pm$ 0.01 & 1.24 $\pm$ 0.03 & 0.96 $\pm$ 0.02 & 0.40 $\pm$ 0.04 & 0.23 $\pm$ 0.03 & 1.33 $\pm$ 0.02 & 1.51 $\pm$ 0.03 & 0.27 $\pm$ 0.01 & 0.28 $\pm$ 0.01 & 0.52 $\pm$ 0.02 & 0.58 $\pm$ 0.01 & 1.17 $\pm$ 0.02\\
10 & 0.25 $\pm$ 0.02 & 1.49 $\pm$ 0.02 & 1.23 $\pm$ 0.04 & 0.98 $\pm$ 0.03 & 0.42 $\pm$ 0.05 & 0.30 $\pm$ 0.04 & 1.26 $\pm$ 0.02 & 1.79 $\pm$ 0.03 & 0.27 $\pm$ 0.02 & 0.28 $\pm$ 0.01 & 0.50 $\pm$ 0.03 & 0.57 $\pm$ 0.02 & 1.17 $\pm$ 0.02\\
11 & 0.24 $\pm$ 0.03 & 1.49 $\pm$ 0.03 & 1.43 $\pm$ 0.06 & 0.98 $\pm$ 0.03 & 0.34 $\pm$ 0.06 & 0.41 $\pm$ 0.05 & 1.34 $\pm$ 0.02 & 1.38 $\pm$ 0.04 & 0.30 $\pm$ 0.02 & 0.28 $\pm$ 0.02 & 0.51 $\pm$ 0.04 & 0.63 $\pm$ 0.02 & 1.14 $\pm$ 0.03\\
12 & 0.24 $\pm$ 0.03 & 1.43 $\pm$ 0.03 & 1.49 $\pm$ 0.06 & 0.99 $\pm$ 0.04 & 0.41 $\pm$ 0.07 & 0.35 $\pm$ 0.06 & 1.32 $\pm$ 0.03 & 1.66 $\pm$ 0.05 & 0.28 $\pm$ 0.02 & 0.28 $\pm$ 0.02 & 0.49 $\pm$ 0.04 & 0.58 $\pm$ 0.02 & 1.12 $\pm$ 0.04\\
13 & 0.25 $\pm$ 0.03 & 1.46 $\pm$ 0.03 & 1.27 $\pm$ 0.06 & 1.01 $\pm$ 0.04 & 0.46 $\pm$ 0.06 & 0.29 $\pm$ 0.05 & 1.43 $\pm$ 0.02 & 1.66 $\pm$ 0.05 & 0.31 $\pm$ 0.03 & 0.26 $\pm$ 0.02 & 0.49 $\pm$ 0.04 & 0.60 $\pm$ 0.02 & 1.18 $\pm$ 0.03\\
14 & 0.23 $\pm$ 0.03 & 1.44 $\pm$ 0.03 & 1.35 $\pm$ 0.06 & 0.92 $\pm$ 0.03 & 0.37 $\pm$ 0.06 & 0.38 $\pm$ 0.05 & 1.34 $\pm$ 0.03 & 1.45 $\pm$ 0.04 & 0.25 $\pm$ 0.02 & 0.28 $\pm$ 0.02 & 0.48 $\pm$ 0.04 & 0.62 $\pm$ 0.02 & 1.15 $\pm$ 0.03\\
15 & 0.22 $\pm$ 0.03 & 1.42 $\pm$ 0.03 & 1.07 $\pm$ 0.06 & 1.03 $\pm$ 0.04 & 0.43 $\pm$ 0.07 & 0.12 $\pm$ 0.06 & 1.38 $\pm$ 0.03 & 1.62 $\pm$ 0.05 & 0.24 $\pm$ 0.03 & 0.28 $\pm$ 0.02 & 0.51 $\pm$ 0.04 & 0.61 $\pm$ 0.02 & 1.20 $\pm$ 0.04\\
16 & 0.25 $\pm$ 0.03 & 1.48 $\pm$ 0.03 & 1.28 $\pm$ 0.07 & 1.01 $\pm$ 0.05 & 0.41 $\pm$ 0.08 & 0.18 $\pm$ 0.06 & 1.36 $\pm$ 0.03 & 1.43 $\pm$ 0.05 & 0.29 $\pm$ 0.03 & 0.29 $\pm$ 0.02 & 0.57 $\pm$ 0.04 & 0.58 $\pm$ 0.02 & 1.15 $\pm$ 0.04\\
17 & 0.23 $\pm$ 0.02 & 1.43 $\pm$ 0.02 & 1.28 $\pm$ 0.04 & 0.99 $\pm$ 0.02 & 0.40 $\pm$ 0.04 & 0.33 $\pm$ 0.03 & 1.28 $\pm$ 0.02 & 1.48 $\pm$ 0.03 & 0.26 $\pm$ 0.01 & 0.28 $\pm$ 0.01 & 0.53 $\pm$ 0.02 & 0.59 $\pm$ 0.01 & 1.15 $\pm$ 0.02\\
18 & 0.26 $\pm$ 0.02 & 1.45 $\pm$ 0.02 & 1.32 $\pm$ 0.04 & 0.96 $\pm$ 0.03 & 0.39 $\pm$ 0.05 & 0.37 $\pm$ 0.04 & 1.30 $\pm$ 0.02 & 1.35 $\pm$ 0.03 & 0.29 $\pm$ 0.02 & 0.30 $\pm$ 0.01 & 0.52 $\pm$ 0.02 & 0.57 $\pm$ 0.01 & 1.15 $\pm$ 0.02\\
19 & 0.26 $\pm$ 0.02 & 1.49 $\pm$ 0.02 & 1.24 $\pm$ 0.04 & 0.93 $\pm$ 0.02 & 0.34 $\pm$ 0.04 & 0.34 $\pm$ 0.03 & 1.33 $\pm$ 0.02 & 1.41 $\pm$ 0.03 & 0.28 $\pm$ 0.02 & 0.28 $\pm$ 0.01 & 0.51 $\pm$ 0.03 & 0.59 $\pm$ 0.01 & 1.17 $\pm$ 0.02\\
\enddata
\tablecomments{Column (1): observing epoch. Columns (2-14): Gaussian amplitude fitting results to each \hi\ optical depth component. 
}
\end{deluxetable*} 
\end{rotatetable*}

%%%updated 0926
\begin{rotatetable*}
\begin{deluxetable*}{c|ccccccccccccccc}
\setlength{\tabcolsep}{0.01in} 
\tablecolumns{14} 
\tabletypesize{\scriptsize}
\tablewidth{0.1pt}
\tablecaption{Gaussian Decomposition Results of Central Velocity for the \hi\ Optical Depth Spectra over 19 observing epochs}
\label{tab:fit_tau_1420_v_compB}
\tablehead{\colhead{Epoch}& \colhead{$v_{0,1}$} & \colhead{$v_{0,2}$} & \colhead{$v_{0,3}$} & \colhead{$v_{0,4}$} & \colhead{$v_{0,5}$} & \colhead{$v_{0,6}$} & \colhead{$v_{0,7}$} & \colhead{$v_{0,8}$} & \colhead{$v_{0,9}$} & \colhead{$v_{0,10}$} & \colhead{$v_{0,11}$} & \colhead{$v_{0,12}$} & \colhead{$v_{0,13}$} } 
\startdata
No. & (\kms\ )& (\kms\ ) &  (\kms\ ) & (\kms\ )  &(\kms\ ) &(\kms\ ) & (\kms\ ) & (\kms\ ) &  (\kms\ )  & (\kms\ ) & (\kms\ )& (\kms\ ) & (\kms\ )\\
\hline
1 & -58.6 $\pm$ 0.1 & -51.9 $\pm$ 0.0 & -45.5 $\pm$ 0.0 & -43.7 $\pm$ 0.1 & -40.7 $\pm$ 0.0 & -34.9 $\pm$ 0.1 & -33.4 $\pm$ 0.1 & -30.6 $\pm$ 0.0 & -24.1 $\pm$ 0.1 & -15.1 $\pm$ 0.2 & -8.9 $\pm$ 0.1 & -1.5 $\pm$ 0.1 & 3.9 $\pm$ 0.0\\
2 & -58.6 $\pm$ 0.1 & -51.9 $\pm$ 0.0 & -45.4 $\pm$ 0.0 & -43.6 $\pm$ 0.1 & -40.8 $\pm$ 0.0 & -34.9 $\pm$ 0.0 & -33.4 $\pm$ 0.1 & -30.6 $\pm$ 0.0 & -24.0 $\pm$ 0.1 & -14.8 $\pm$ 0.2 & -8.9 $\pm$ 0.0 & -1.3 $\pm$ 0.1 & 3.9 $\pm$ 0.0\\
3 & -58.6 $\pm$ 0.1 & -51.9 $\pm$ 0.0 & -45.5 $\pm$ 0.0 & -43.4 $\pm$ 0.1 & -40.6 $\pm$ 0.1 & -35.0 $\pm$ 0.0 & -33.3 $\pm$ 0.1 & -30.6 $\pm$ 0.0 & -23.9 $\pm$ 0.1 & -14.9 $\pm$ 0.2 & -8.9 $\pm$ 0.1 & -1.4 $\pm$ 0.1 & 3.9 $\pm$ 0.0\\
4 & -58.4 $\pm$ 0.1 & -51.9 $\pm$ 0.0 & -45.4 $\pm$ 0.0 & -43.6 $\pm$ 0.1 & -40.7 $\pm$ 0.0 & -35.0 $\pm$ 0.1 & -33.2 $\pm$ 0.1 & -30.5 $\pm$ 0.0 & -23.8 $\pm$ 0.1 & -15.0 $\pm$ 0.2 & -8.9 $\pm$ 0.1 & -1.4 $\pm$ 0.1 & 3.9 $\pm$ 0.0\\
5 & -58.6 $\pm$ 0.1 & -51.9 $\pm$ 0.0 & -45.4 $\pm$ 0.0 & -43.5 $\pm$ 0.1 & -40.7 $\pm$ 0.0 & -35.0 $\pm$ 0.0 & -33.3 $\pm$ 0.1 & -30.6 $\pm$ 0.0 & -23.8 $\pm$ 0.1 & -15.0 $\pm$ 0.2 & -8.9 $\pm$ 0.1 & -1.3 $\pm$ 0.1 & 3.9 $\pm$ 0.0\\
6 & -58.6 $\pm$ 0.1 & -51.9 $\pm$ 0.0 & -45.5 $\pm$ 0.0 & -43.5 $\pm$ 0.2 & -40.6 $\pm$ 0.1 & -34.9 $\pm$ 0.1 & -33.4 $\pm$ 0.1 & -30.5 $\pm$ 0.0 & -24.0 $\pm$ 0.2 & -14.9 $\pm$ 0.2 & -8.9 $\pm$ 0.1 & -1.3 $\pm$ 0.1 & 3.9 $\pm$ 0.0\\
7 & -58.5 $\pm$ 0.1 & -51.9 $\pm$ 0.0 & -45.4 $\pm$ 0.0 & -43.6 $\pm$ 0.1 & -40.6 $\pm$ 0.0 & -35.0 $\pm$ 0.0 & -33.3 $\pm$ 0.1 & -30.6 $\pm$ 0.0 & -23.9 $\pm$ 0.1 & -14.9 $\pm$ 0.2 & -8.8 $\pm$ 0.0 & -1.3 $\pm$ 0.1 & 3.9 $\pm$ 0.0\\
8 & -58.6 $\pm$ 0.2 & -51.9 $\pm$ 0.0 & -45.6 $\pm$ 0.0 & -43.3 $\pm$ 0.3 & -40.5 $\pm$ 0.1 & -34.6 $\pm$ 0.1 & -33.2 $\pm$ 0.1 & -30.4 $\pm$ 0.0 & -23.8 $\pm$ 0.2 & -15.0 $\pm$ 0.4 & -9.2 $\pm$ 0.1 & -1.6 $\pm$ 0.1 & 3.9 $\pm$ 0.0\\
9 & -58.6 $\pm$ 0.1 & -51.9 $\pm$ 0.0 & -45.5 $\pm$ 0.0 & -43.8 $\pm$ 0.1 & -40.8 $\pm$ 0.0 & -34.9 $\pm$ 0.1 & -33.5 $\pm$ 0.1 & -30.7 $\pm$ 0.0 & -23.8 $\pm$ 0.1 & -14.9 $\pm$ 0.2 & -8.9 $\pm$ 0.1 & -1.4 $\pm$ 0.1 & 3.8 $\pm$ 0.0\\
10 & -58.5 $\pm$ 0.1 & -51.9 $\pm$ 0.0 & -45.5 $\pm$ 0.0 & -43.7 $\pm$ 0.2 & -40.8 $\pm$ 0.1 & -35.0 $\pm$ 0.1 & -33.5 $\pm$ 0.1 & -30.7 $\pm$ 0.0 & -24.0 $\pm$ 0.2 & -14.9 $\pm$ 0.2 & -8.8 $\pm$ 0.1 & -1.4 $\pm$ 0.1 & 3.9 $\pm$ 0.0\\
11 & -58.6 $\pm$ 0.2 & -51.8 $\pm$ 0.0 & -45.6 $\pm$ 0.0 & -43.3 $\pm$ 0.2 & -40.6 $\pm$ 0.1 & -35.0 $\pm$ 0.1 & -33.3 $\pm$ 0.1 & -30.7 $\pm$ 0.0 & -23.9 $\pm$ 0.2 & -14.8 $\pm$ 0.4 & -8.7 $\pm$ 0.1 & -1.5 $\pm$ 0.1 & 3.8 $\pm$ 0.0\\
12 & -58.6 $\pm$ 0.2 & -52.0 $\pm$ 0.0 & -45.5 $\pm$ 0.0 & -43.7 $\pm$ 0.2 & -40.9 $\pm$ 0.1 & -35.0 $\pm$ 0.1 & -33.5 $\pm$ 0.1 & -30.6 $\pm$ 0.0 & -23.8 $\pm$ 0.2 & -15.1 $\pm$ 0.4 & -9.0 $\pm$ 0.1 & -1.3 $\pm$ 0.1 & 3.9 $\pm$ 0.0\\
13 & -58.6 $\pm$ 0.2 & -51.9 $\pm$ 0.0 & -45.4 $\pm$ 0.0 & -43.7 $\pm$ 0.2 & -40.6 $\pm$ 0.1 & -35.1 $\pm$ 0.1 & -33.3 $\pm$ 0.1 & -30.6 $\pm$ 0.0 & -24.1 $\pm$ 0.2 & -14.9 $\pm$ 0.5 & -8.9 $\pm$ 0.1 & -1.5 $\pm$ 0.1 & 3.8 $\pm$ 0.0\\
14 & -58.6 $\pm$ 0.2 & -51.9 $\pm$ 0.0 & -45.5 $\pm$ 0.0 & -43.4 $\pm$ 0.2 & -40.8 $\pm$ 0.1 & -34.9 $\pm$ 0.1 & -33.4 $\pm$ 0.1 & -30.6 $\pm$ 0.0 & -24.0 $\pm$ 0.2 & -14.7 $\pm$ 0.4 & -8.9 $\pm$ 0.1 & -1.4 $\pm$ 0.1 & 3.9 $\pm$ 0.0\\
15 & -58.7 $\pm$ 0.2 & -51.9 $\pm$ 0.0 & -45.5 $\pm$ 0.0 & -43.7 $\pm$ 0.2 & -40.6 $\pm$ 0.1 & -35.2 $\pm$ 0.2 & -33.3 $\pm$ 0.1 & -30.6 $\pm$ 0.0 & -23.8 $\pm$ 0.3 & -14.9 $\pm$ 0.4 & -8.8 $\pm$ 0.1 & -1.4 $\pm$ 0.1 & 3.8 $\pm$ 0.0\\
16 & -58.6 $\pm$ 0.2 & -51.9 $\pm$ 0.0 & -45.5 $\pm$ 0.0 & -43.9 $\pm$ 0.2 & -40.7 $\pm$ 0.1 & -35.1 $\pm$ 0.1 & -33.3 $\pm$ 0.1 & -30.6 $\pm$ 0.0 & -23.7 $\pm$ 0.2 & -15.1 $\pm$ 0.3 & -8.9 $\pm$ 0.1 & -1.4 $\pm$ 0.1 & 3.9 $\pm$ 0.0\\
17 & -58.6 $\pm$ 0.1 & -51.9 $\pm$ 0.0 & -45.5 $\pm$ 0.0 & -43.7 $\pm$ 0.1 & -40.8 $\pm$ 0.0 & -35.0 $\pm$ 0.0 & -33.5 $\pm$ 0.1 & -30.7 $\pm$ 0.0 & -23.9 $\pm$ 0.1 & -15.1 $\pm$ 0.2 & -8.9 $\pm$ 0.1 & -1.5 $\pm$ 0.1 & 3.8 $\pm$ 0.0\\
18 & -58.6 $\pm$ 0.1 & -51.9 $\pm$ 0.0 & -45.4 $\pm$ 0.0 & -43.6 $\pm$ 0.2 & -40.8 $\pm$ 0.1 & -35.0 $\pm$ 0.0 & -33.4 $\pm$ 0.1 & -30.6 $\pm$ 0.0 & -24.0 $\pm$ 0.1 & -14.8 $\pm$ 0.2 & -8.9 $\pm$ 0.1 & -1.3 $\pm$ 0.1 & 3.9 $\pm$ 0.0\\
19 & -58.6 $\pm$ 0.1 & -51.9 $\pm$ 0.0 & -45.5 $\pm$ 0.0 & -43.3 $\pm$ 0.2 & -40.7 $\pm$ 0.1 & -35.0 $\pm$ 0.0 & -33.3 $\pm$ 0.1 & -30.6 $\pm$ 0.0 & -24.0 $\pm$ 0.1 & -14.7 $\pm$ 0.3 & -8.9 $\pm$ 0.1 & -1.4 $\pm$ 0.1 & 3.8 $\pm$ 0.0\\
\enddata
\tablecomments{Column (1): observing epoch. Columns (2-14): Gaussian central velocity fitting results to each \hi\ optical depth component.  
}
\end{deluxetable*} 
\end{rotatetable*}

%%updated 0926
%\begin{rotatetable*}
\begin{deluxetable*}{c|ccccccccccccc}
\setlength{\tabcolsep}{0.01in} 
%\resizebox{\textwidth}
\tablecolumns{14} 
\tabletypesize{\scriptsize}
\tablewidth{0pt}
\tablecaption{Gaussian Decomposition Results of the FWHM for the \hi\ optical Depth Spectra over 19 observing epochs}
\label{tab:fit_tau_1420_dv_compB}
\tablehead{\colhead{Epoch} & \colhead{$\Delta v_{0,1}$} & \colhead{$\Delta v_{0,2}$} & \colhead{$\Delta v_{0,3}$} & \colhead{$\Delta v_{0,4}$} & \colhead{$\Delta v_{0,5}$} & \colhead{$\Delta v_{0,6}$} & \colhead{$\Delta v_{0,7}$} & \colhead{$\Delta v_{0,8}$} & \colhead{$\Delta v_{0,9}$} & \colhead{$\Delta v_{0,10}$} & \colhead{$\Delta v_{0,11}$} & \colhead{$\Delta v_{0,12}$} & \colhead{$\Delta v_{0,13}$}} 
\startdata
No. & (\kms\ )& (\kms\ ) &  (\kms\ ) & (\kms\ )  &(\kms\ ) &(\kms\ ) & (\kms\ ) & (\kms\ )&(\kms\ )  & (\kms\ ) & (\kms\ )& (\kms\ )&(\kms\ )\\
\hline
1 & 2.5 $\pm$ 0.2 & 3.2 $\pm$ 0.0 & 1.7 $\pm$ 0.1 & 5.9 $\pm$ 0.2 & 1.5 $\pm$ 0.2 & 1.0 $\pm$ 0.1 & 7.8 $\pm$ 0.1 & 1.6 $\pm$ 0.0 & 4.0 $\pm$ 0.3 & 7.5 $\pm$ 0.6 & 2.8 $\pm$ 0.1 & 5.4 $\pm$ 0.2 & 2.3 $\pm$ 0.1\\
2 & 2.6 $\pm$ 0.2 & 3.3 $\pm$ 0.0 & 1.9 $\pm$ 0.1 & 5.9 $\pm$ 0.2 & 1.5 $\pm$ 0.1 & 0.9 $\pm$ 0.1 & 7.8 $\pm$ 0.1 & 1.6 $\pm$ 0.0 & 3.8 $\pm$ 0.3 & 8.0 $\pm$ 0.7 & 2.7 $\pm$ 0.1 & 5.8 $\pm$ 0.2 & 2.2 $\pm$ 0.0\\
3 & 2.7 $\pm$ 0.2 & 3.3 $\pm$ 0.0 & 1.7 $\pm$ 0.0 & 5.9 $\pm$ 0.2 & 1.5 $\pm$ 0.2 & 0.9 $\pm$ 0.1 & 7.8 $\pm$ 0.1 & 1.5 $\pm$ 0.0 & 3.8 $\pm$ 0.3 & 7.5 $\pm$ 0.6 & 2.8 $\pm$ 0.1 & 5.8 $\pm$ 0.2 & 2.2 $\pm$ 0.0\\
4 & 2.7 $\pm$ 0.2 & 3.1 $\pm$ 0.0 & 1.6 $\pm$ 0.1 & 5.9 $\pm$ 0.2 & 1.5 $\pm$ 0.2 & 0.9 $\pm$ 0.1 & 7.7 $\pm$ 0.1 & 1.4 $\pm$ 0.0 & 3.8 $\pm$ 0.3 & 7.5 $\pm$ 0.7 & 2.9 $\pm$ 0.2 & 5.5 $\pm$ 0.2 & 2.2 $\pm$ 0.1\\
5 & 2.8 $\pm$ 0.2 & 3.3 $\pm$ 0.0 & 1.8 $\pm$ 0.1 & 5.9 $\pm$ 0.2 & 1.5 $\pm$ 0.1 & 0.9 $\pm$ 0.1 & 7.6 $\pm$ 0.1 & 1.6 $\pm$ 0.0 & 4.1 $\pm$ 0.3 & 6.7 $\pm$ 0.5 & 3.0 $\pm$ 0.1 & 5.7 $\pm$ 0.2 & 2.2 $\pm$ 0.0\\
6 & 2.9 $\pm$ 0.3 & 3.2 $\pm$ 0.1 & 1.8 $\pm$ 0.1 & 5.9 $\pm$ 0.3 & 1.5 $\pm$ 0.2 & 1.0 $\pm$ 0.2 & 7.6 $\pm$ 0.2 & 1.6 $\pm$ 0.0 & 4.0 $\pm$ 0.4 & 7.5 $\pm$ 0.8 & 3.0 $\pm$ 0.2 & 5.5 $\pm$ 0.2 & 2.2 $\pm$ 0.1\\
7 & 2.9 $\pm$ 0.2 & 3.1 $\pm$ 0.0 & 1.7 $\pm$ 0.0 & 5.9 $\pm$ 0.2 & 1.5 $\pm$ 0.1 & 0.9 $\pm$ 0.1 & 7.7 $\pm$ 0.1 & 1.6 $\pm$ 0.0 & 3.8 $\pm$ 0.3 & 7.7 $\pm$ 0.6 & 2.8 $\pm$ 0.1 & 5.7 $\pm$ 0.2 & 2.2 $\pm$ 0.0\\
8 & 2.8 $\pm$ 0.5 & 3.2 $\pm$ 0.1 & 1.6 $\pm$ 0.1 & 5.9 $\pm$ 0.4 & 1.5 $\pm$ 0.4 & 0.9 $\pm$ 0.2 & 7.2 $\pm$ 0.2 & 1.3 $\pm$ 0.1 & 3.7 $\pm$ 0.6 & 6.7 $\pm$ 1.2 & 3.0 $\pm$ 0.3 & 5.2 $\pm$ 0.4 & 2.1 $\pm$ 0.1\\
9 & 2.9 $\pm$ 0.2 & 3.1 $\pm$ 0.0 & 1.6 $\pm$ 0.1 & 5.9 $\pm$ 0.2 & 1.5 $\pm$ 0.2 & 1.1 $\pm$ 0.2 & 7.8 $\pm$ 0.1 & 1.6 $\pm$ 0.0 & 4.0 $\pm$ 0.3 & 7.3 $\pm$ 0.6 & 2.9 $\pm$ 0.1 & 5.5 $\pm$ 0.2 & 2.2 $\pm$ 0.0\\
10 & 2.8 $\pm$ 0.3 & 3.1 $\pm$ 0.0 & 1.7 $\pm$ 0.1 & 5.9 $\pm$ 0.3 & 1.5 $\pm$ 0.2 & 1.0 $\pm$ 0.2 & 7.9 $\pm$ 0.2 & 1.6 $\pm$ 0.0 & 4.0 $\pm$ 0.4 & 7.5 $\pm$ 0.7 & 2.7 $\pm$ 0.2 & 5.5 $\pm$ 0.2 & 2.3 $\pm$ 0.1\\
11 & 2.9 $\pm$ 0.4 & 3.2 $\pm$ 0.1 & 1.7 $\pm$ 0.1 & 5.9 $\pm$ 0.3 & 1.2 $\pm$ 0.3 & 0.9 $\pm$ 0.1 & 7.7 $\pm$ 0.2 & 1.5 $\pm$ 0.1 & 3.8 $\pm$ 0.5 & 7.9 $\pm$ 1.2 & 2.9 $\pm$ 0.3 & 5.0 $\pm$ 0.3 & 2.4 $\pm$ 0.1\\
12 & 3.0 $\pm$ 0.4 & 3.2 $\pm$ 0.1 & 1.6 $\pm$ 0.1 & 5.9 $\pm$ 0.3 & 1.5 $\pm$ 0.3 & 0.9 $\pm$ 0.2 & 7.7 $\pm$ 0.2 & 1.5 $\pm$ 0.1 & 4.1 $\pm$ 0.6 & 7.1 $\pm$ 1.2 & 3.2 $\pm$ 0.3 & 5.7 $\pm$ 0.4 & 2.2 $\pm$ 0.1\\
13 & 2.8 $\pm$ 0.4 & 3.2 $\pm$ 0.1 & 1.6 $\pm$ 0.1 & 5.9 $\pm$ 0.3 & 1.2 $\pm$ 0.2 & 0.9 $\pm$ 0.2 & 7.4 $\pm$ 0.2 & 1.3 $\pm$ 0.1 & 3.8 $\pm$ 0.5 & 8.2 $\pm$ 1.5 & 3.0 $\pm$ 0.3 & 5.1 $\pm$ 0.3 & 2.2 $\pm$ 0.1\\
14 & 3.0 $\pm$ 0.4 & 3.3 $\pm$ 0.1 & 1.8 $\pm$ 0.1 & 5.9 $\pm$ 0.3 & 1.5 $\pm$ 0.3 & 1.0 $\pm$ 0.2 & 7.7 $\pm$ 0.2 & 1.5 $\pm$ 0.1 & 4.1 $\pm$ 0.6 & 8.2 $\pm$ 1.4 & 3.0 $\pm$ 0.3 & 5.2 $\pm$ 0.3 & 2.3 $\pm$ 0.1\\
15 & 3.1 $\pm$ 0.5 & 3.1 $\pm$ 0.1 & 1.6 $\pm$ 0.1 & 5.9 $\pm$ 0.4 & 1.5 $\pm$ 0.3 & 0.9 $\pm$ 0.5 & 7.8 $\pm$ 0.2 & 1.3 $\pm$ 0.1 & 3.9 $\pm$ 0.7 & 7.8 $\pm$ 1.3 & 2.8 $\pm$ 0.3 & 5.4 $\pm$ 0.3 & 2.2 $\pm$ 0.1\\
16 & 2.8 $\pm$ 0.4 & 3.1 $\pm$ 0.1 & 1.6 $\pm$ 0.1 & 5.9 $\pm$ 0.4 & 1.5 $\pm$ 0.3 & 0.9 $\pm$ 0.4 & 8.0 $\pm$ 0.2 & 1.5 $\pm$ 0.1 & 3.6 $\pm$ 0.5 & 6.7 $\pm$ 1.0 & 3.1 $\pm$ 0.3 & 5.6 $\pm$ 0.4 & 2.3 $\pm$ 0.1\\
17 & 3.1 $\pm$ 0.2 & 3.2 $\pm$ 0.0 & 1.6 $\pm$ 0.1 & 5.9 $\pm$ 0.2 & 1.5 $\pm$ 0.2 & 0.9 $\pm$ 0.1 & 7.9 $\pm$ 0.1 & 1.6 $\pm$ 0.0 & 3.8 $\pm$ 0.3 & 7.2 $\pm$ 0.6 & 2.8 $\pm$ 0.1 & 5.4 $\pm$ 0.2 & 2.3 $\pm$ 0.1\\
18 & 2.9 $\pm$ 0.2 & 3.2 $\pm$ 0.0 & 1.8 $\pm$ 0.1 & 5.9 $\pm$ 0.2 & 1.5 $\pm$ 0.2 & 0.9 $\pm$ 0.1 & 7.8 $\pm$ 0.2 & 1.6 $\pm$ 0.0 & 4.0 $\pm$ 0.3 & 7.2 $\pm$ 0.6 & 2.7 $\pm$ 0.1 & 5.8 $\pm$ 0.2 & 2.2 $\pm$ 0.1\\
19 & 2.7 $\pm$ 0.2 & 3.2 $\pm$ 0.0 & 1.9 $\pm$ 0.1 & 5.9 $\pm$ 0.2 & 1.5 $\pm$ 0.2 & 0.9 $\pm$ 0.1 & 7.7 $\pm$ 0.1 & 1.6 $\pm$ 0.0 & 3.7 $\pm$ 0.3 & 8.2 $\pm$ 0.8 & 2.7 $\pm$ 0.2 & 5.6 $\pm$ 0.2 & 2.2 $\pm$ 0.1\\
\enddata
\tablecomments{Column (1): observing epoch. Columns (2-14): Gaussian FWHM fitting results to each \hi\ optical depth component. 
}
\end{deluxetable*} 
%\end{rotatetable*}

%updated 0926
\begin{rotatetable*}
\begin{deluxetable*}{c|ccccccccccccc}
\setlength{\tabcolsep}{0.02in} 
%\resizebox{\textwidth}
\tablecolumns{14} 
\tabletypesize{\scriptsize}
\tablewidth{0.1pt}
\tablecaption{Spin Temperatures for the \hi\ Optical Depth Spectra over 19 observing epochs}
\label{tab:fit_tau_1420_Ts_compB}
\tablehead{\colhead{Epoch}&  \colhead{$T_{0,1}$} & \colhead{$T_{0,2}$} & \colhead{$T_{0,3}$} & \colhead{$T_{0,4}$} & \colhead{$T_{0,5}$} & \colhead{$T_{0,6}$} & \colhead{$T_{0,7}$} & \colhead{$T_{0,8}$} & \colhead{$T_{0,9}$} & \colhead{$T_{0,10}$} & \colhead{$T_{0,11}$} & \colhead{$T_{0,12}$}& \colhead{$T_{0,13}$}  } 
\startdata
No. & (K )& (K ) &  (K ) & (K)  &(K ) &(K ) & (K) & (K)&(K )  & (K ) & (K )& (K )&(K )\\
\hline
1 & 126.7 $\pm$ 7.0 & 109.6 $\pm$ 7.2 & 25.8 $\pm$ 4.9 & 136.4 $\pm$ 6.0 & 47.7 $\pm$ 2.9 & 7.8 $\pm$ 8.5 & 141.8 $\pm$ 5.8 & 52.7 $\pm$ 7.8 & 306.0 $\pm$ 10.5 & 443.0 $\pm$ 3.5 & 151.2 $\pm$ 11.1 & 218.5 $\pm$ 5.8 & 96.8 $\pm$ 14.5\\
2 & 132.4 $\pm$ 7.8 & 111.4 $\pm$ 6.8 & 28.6 $\pm$ 5.9 & 138.5 $\pm$ 5.9 & 47.7 $\pm$ 3.0 & 4.3 $\pm$ 5.9 & 139.6 $\pm$ 5.9 & 50.8 $\pm$ 5.7 & 300.4 $\pm$ 10.3 & 460.3 $\pm$ 2.7 & 159.9 $\pm$ 5.9 & 225.1 $\pm$ 3.8 & 88.1 $\pm$ 12.4\\
3 & 135.7 $\pm$ 8.7 & 111.4 $\pm$ 6.6 & 32.3 $\pm$ 8.9 & 134.7 $\pm$ 4.7 & 47.7 $\pm$ 3.2 & 4.2 $\pm$ 5.4 & 141.6 $\pm$ 6.1 & 49.7 $\pm$ 5.4 & 307.5 $\pm$ 7.2 & 445.8 $\pm$ 2.6 & 151.4 $\pm$ 9.5 & 223.0 $\pm$ 4.7 & 89.0 $\pm$ 12.3\\
4 & 128.6 $\pm$ 6.6 & 107.0 $\pm$ 7.3 & 28.5 $\pm$ 4.4 & 134.3 $\pm$ 6.3 & 47.7 $\pm$ 3.0 & 7.7 $\pm$ 7.8 & 137.8 $\pm$ 6.2 & 44.4 $\pm$ 5.2 & 303.7 $\pm$ 4.6 & 439.9 $\pm$ 2.7 & 150.8 $\pm$ 13.2 & 218.7 $\pm$ 4.7 & 90.1 $\pm$ 12.9\\
5 & 134.5 $\pm$ 7.0 & 110.2 $\pm$ 7.1 & 31.9 $\pm$ 5.5 & 135.9 $\pm$ 6.1 & 47.7 $\pm$ 2.9 & 5.1 $\pm$ 6.9 & 138.5 $\pm$ 5.9 & 51.9 $\pm$ 8.6 & 307.5 $\pm$ 8.2 & 419.6 $\pm$ 4.7 & 162.8 $\pm$ 20.0 & 220.8 $\pm$ 4.1 & 89.2 $\pm$ 13.6\\
6 & 140.4 $\pm$ 5.6 & 108.0 $\pm$ 7.5 & 32.4 $\pm$ 6.2 & 136.6 $\pm$ 5.7 & 47.7 $\pm$ 2.9 & 6.0 $\pm$ 7.3 & 138.2 $\pm$ 5.8 & 53.2 $\pm$ 7.1 & 308.6 $\pm$ 12.2 & 433.6 $\pm$ 3.1 & 159.6 $\pm$ 15.4 & 219.4 $\pm$ 4.9 & 90.2 $\pm$ 11.8\\
7 & 139.4 $\pm$ 6.3 & 108.0 $\pm$ 7.5 & 26.9 $\pm$ 4.4 & 135.2 $\pm$ 6.3 & 47.7 $\pm$ 3.0 & 8.3 $\pm$ 9.0 & 139.6 $\pm$ 6.0 & 52.2 $\pm$ 8.4 & 305.2 $\pm$ 7.7 & 437.0 $\pm$ 2.9 & 164.5 $\pm$ 5.4 & 221.8 $\pm$ 4.2 & 87.9 $\pm$ 12.1\\
8 & 148.6 $\pm$ 9.1 & 108.1 $\pm$ 6.4 & 32.1 $\pm$ 7.7 & 135.2 $\pm$ 5.3 & 47.7 $\pm$ 3.0 & 3.8 $\pm$ 5.4 & 132.0 $\pm$ 7.0 & 39.2 $\pm$ 5.0 & 289.4 $\pm$ 3.9 & 438.7 $\pm$ 6.1 & 164.2 $\pm$ 12.5 & 218.5 $\pm$ 6.5 & 89.1 $\pm$ 7.4\\
9 & 142.7 $\pm$ 5.8 & 108.5 $\pm$ 7.4 & 19.7 $\pm$ 3.6 & 137.7 $\pm$ 6.2 & 47.7 $\pm$ 3.1 & 14.5 $\pm$ 13.5 & 137.9 $\pm$ 6.6 & 55.4 $\pm$ 5.9 & 320.2 $\pm$ 8.9 & 436.2 $\pm$ 3.2 & 156.7 $\pm$ 12.4 & 222.6 $\pm$ 5.0 & 91.3 $\pm$ 11.3\\
10 & 139.4 $\pm$ 6.5 & 106.9 $\pm$ 7.4 & 25.5 $\pm$ 4.9 & 134.8 $\pm$ 5.6 & 47.7 $\pm$ 3.0 & 13.1 $\pm$ 11.7 & 141.7 $\pm$ 6.3 & 57.8 $\pm$ 5.1 & 319.3 $\pm$ 12.4 & 437.1 $\pm$ 2.9 & 161.5 $\pm$ 5.7 & 227.4 $\pm$ 4.9 & 94.8 $\pm$ 14.4\\
11 & 146.0 $\pm$ 5.9 & 108.1 $\pm$ 7.0 & 37.0 $\pm$ 8.5 & 133.1 $\pm$ 4.8 & 31.9 $\pm$ 3.0 & 6.5 $\pm$ 6.7 & 139.6 $\pm$ 6.0 & 48.0 $\pm$ 5.8 & 286.2 $\pm$ 10.4 & 445.1 $\pm$ 2.7 & 169.7 $\pm$ 9.2 & 211.6 $\pm$ 6.1 & 103.8 $\pm$ 20.1\\
12 & 146.3 $\pm$ 6.6 & 107.8 $\pm$ 7.0 & 25.9 $\pm$ 5.4 & 133.8 $\pm$ 5.6 & 46.6 $\pm$ 3.0 & 8.0 $\pm$ 7.7 & 138.8 $\pm$ 6.1 & 51.0 $\pm$ 5.0 & 314.2 $\pm$ 13.6 & 431.0 $\pm$ 6.1 & 178.9 $\pm$ 16.1 & 222.1 $\pm$ 4.1 & 88.7 $\pm$ 10.6\\
13 & 143.2 $\pm$ 8.4 & 107.2 $\pm$ 6.5 & 23.5 $\pm$ 3.1 & 132.3 $\pm$ 7.5 & 31.9 $\pm$ 2.8 & 10.2 $\pm$ 8.8 & 135.3 $\pm$ 6.5 & 39.2 $\pm$ 5.1 & 281.7 $\pm$ 11.8 & 464.0 $\pm$ 4.1 & 158.0 $\pm$ 13.5 & 220.2 $\pm$ 6.4 & 91.8 $\pm$ 9.1\\
14 & 152.8 $\pm$ 6.4 & 108.8 $\pm$ 7.2 & 31.7 $\pm$ 6.7 & 137.6 $\pm$ 5.6 & 47.7 $\pm$ 3.1 & 3.9 $\pm$ 5.5 & 138.9 $\pm$ 5.6 & 51.4 $\pm$ 5.7 & 338.1 $\pm$ 12.9 & 440.3 $\pm$ 3.2 & 163.1 $\pm$ 15.1 & 213.1 $\pm$ 5.6 & 96.6 $\pm$ 13.7\\
15 & 161.0 $\pm$ 6.0 & 108.5 $\pm$ 7.2 & 29.7 $\pm$ 3.9 & 130.3 $\pm$ 6.0 & 45.5 $\pm$ 2.9 & 11.0 $\pm$ 11.0 & 138.2 $\pm$ 6.1 & 39.2 $\pm$ 5.2 & 334.7 $\pm$ 4.2 & 433.4 $\pm$ 3.1 & 157.9 $\pm$ 9.2 & 215.5 $\pm$ 4.9 & 91.9 $\pm$ 10.8\\
16 & 143.7 $\pm$ 5.0 & 106.8 $\pm$ 7.2 & 21.0 $\pm$ 4.5 & 133.5 $\pm$ 5.9 & 46.0 $\pm$ 3.0 & 9.9 $\pm$ 10.4 & 138.5 $\pm$ 6.9 & 51.8 $\pm$ 5.6 & 279.2 $\pm$ 3.8 & 420.0 $\pm$ 5.9 & 164.0 $\pm$ 17.4 & 224.6 $\pm$ 4.9 & 96.0 $\pm$ 15.8\\
17 & 151.4 $\pm$ 4.9 & 109.1 $\pm$ 7.3 & 28.6 $\pm$ 5.7 & 133.4 $\pm$ 5.5 & 47.7 $\pm$ 3.0 & 6.7 $\pm$ 7.5 & 141.4 $\pm$ 6.2 & 53.6 $\pm$ 7.1 & 310.1 $\pm$ 4.2 & 427.2 $\pm$ 3.7 & 148.7 $\pm$ 12.0 & 221.5 $\pm$ 5.5 & 98.0 $\pm$ 15.5\\
18 & 138.9 $\pm$ 6.4 & 108.1 $\pm$ 7.5 & 28.5 $\pm$ 5.0 & 135.3 $\pm$ 5.9 & 47.7 $\pm$ 3.0 & 5.5 $\pm$ 6.6 & 140.3 $\pm$ 5.8 & 52.6 $\pm$ 8.1 & 302.6 $\pm$ 11.8 & 410.5 $\pm$ 2.8 & 152.8 $\pm$ 5.9 & 225.4 $\pm$ 3.7 & 87.0 $\pm$ 12.2\\
19 & 138.6 $\pm$ 7.0 & 107.7 $\pm$ 6.9 & 39.7 $\pm$ 8.0 & 137.3 $\pm$ 4.7 & 47.7 $\pm$ 3.1 & 4.4 $\pm$ 6.0 & 139.4 $\pm$ 5.9 & 50.9 $\pm$ 6.8 & 293.4 $\pm$ 8.0 & 447.8 $\pm$ 2.5 & 158.2 $\pm$ 7.1 & 220.2 $\pm$ 4.6 & 93.4 $\pm$ 13.2\\
\enddata
\tablecomments{Column (1): observing epoch. Columns (2-14): Spin temperatures of each \hi\ optical depth component. 
}
\end{deluxetable*} 
\end{rotatetable*}

%%%%%updated 0926
\begin{rotatetable*}
\begin{deluxetable*}{c|ccccccccccccc}
\setlength{\tabcolsep}{0.01in} 
%\resizebox{\textwidth}
\tablecolumns{14} 
\tabletypesize{\scriptsize}
\tablewidth{0.1pt}
\tablecaption{Column Density for the 13 Gaussian Features over 19 observing epochs}
\label{tab:fit_tau_1420_N_compB}
\tablehead{\colhead{Epoch}& \colhead{$N_{0,1}$} & \colhead{$N_{0,2}$} & \colhead{$N_{0,3}$} & \colhead{$N_{0,4}$} & \colhead{$N_{0,5}$} & \colhead{$N_{0,6}$} & \colhead{$N_{0,7}$} & \colhead{$N_{0,8}$} & \colhead{$N_{0,9}$} & \colhead{$N_{0,10}$} & \colhead{$N_{0,11}$} &\colhead{$N_{0,12}$}  & \colhead{$N_{0,13}$} } 
\startdata
No. & ($10^{20}\rm cm^{-2}$ )& ($10^{20}\rm cm^{-2}$ ) &  ($10^{20}\rm cm^{-2}$ ) & ($10^{20}\rm cm^{-2}$)  &($10^{20}\rm cm^{-2}$ ) &($10^{20}\rm cm^{-2}$ ) & ($10^{20}\rm cm^{-2}$) & ($10^{20}\rm cm^{-2}$)&($10^{20}\rm cm^{-2}$ )  & ($10^{20}\rm cm^{-2}$ ) & ($10^{20}\rm cm^{-2}$ )& ($10^{20}\rm cm^{-2}$ )&($10^{20}\rm cm^{-2}$ )\\
\hline
1 & 1.8 $\pm$ 0.2 & 9.6 $\pm$ 0.7 & 1.1 $\pm$ 0.2 & 14.9 $\pm$ 0.9 & 0.6 $\pm$ 0.1 & 0.0 $\pm$ 0.0 & 27.7 $\pm$ 1.3 & 2.4 $\pm$ 0.4 & 6.5 $\pm$ 0.7 & 17.4 $\pm$ 1.6 & 4.2 $\pm$ 0.4 & 13.8 $\pm$ 0.6 & 4.8 $\pm$ 0.7\\
2 & 1.8 $\pm$ 0.2 & 9.7 $\pm$ 0.6 & 1.3 $\pm$ 0.3 & 14.4 $\pm$ 0.8 & 0.5 $\pm$ 0.1 & 0.0 $\pm$ 0.0 & 27.9 $\pm$ 1.3 & 2.2 $\pm$ 0.3 & 6.1 $\pm$ 0.6 & 19.1 $\pm$ 1.7 & 4.3 $\pm$ 0.3 & 14.6 $\pm$ 0.6 & 4.3 $\pm$ 0.6\\
3 & 1.9 $\pm$ 0.2 & 9.9 $\pm$ 0.6 & 1.6 $\pm$ 0.4 & 14.7 $\pm$ 0.8 & 0.5 $\pm$ 0.1 & 0.0 $\pm$ 0.0 & 27.7 $\pm$ 1.3 & 2.2 $\pm$ 0.2 & 6.3 $\pm$ 0.6 & 17.7 $\pm$ 1.6 & 4.4 $\pm$ 0.4 & 14.4 $\pm$ 0.7 & 4.4 $\pm$ 0.6\\
4 & 1.8 $\pm$ 0.2 & 9.6 $\pm$ 0.7 & 1.2 $\pm$ 0.2 & 14.9 $\pm$ 0.9 & 0.6 $\pm$ 0.1 & 0.0 $\pm$ 0.0 & 28.4 $\pm$ 1.4 & 2.0 $\pm$ 0.2 & 6.4 $\pm$ 0.6 & 17.5 $\pm$ 1.8 & 4.4 $\pm$ 0.5 & 13.9 $\pm$ 0.7 & 4.5 $\pm$ 0.7\\
5 & 2.0 $\pm$ 0.2 & 9.7 $\pm$ 0.7 & 1.4 $\pm$ 0.3 & 14.7 $\pm$ 0.9 & 0.6 $\pm$ 0.1 & 0.0 $\pm$ 0.0 & 27.7 $\pm$ 1.3 & 2.2 $\pm$ 0.4 & 7.0 $\pm$ 0.6 & 15.8 $\pm$ 1.3 & 4.9 $\pm$ 0.7 & 14.3 $\pm$ 0.6 & 4.4 $\pm$ 0.7\\
6 & 2.0 $\pm$ 0.3 & 9.7 $\pm$ 0.7 & 1.5 $\pm$ 0.3 & 14.7 $\pm$ 1.0 & 0.6 $\pm$ 0.1 & 0.0 $\pm$ 0.0 & 27.6 $\pm$ 1.4 & 2.6 $\pm$ 0.4 & 6.6 $\pm$ 0.8 & 17.6 $\pm$ 2.1 & 4.7 $\pm$ 0.6 & 13.9 $\pm$ 0.7 & 4.5 $\pm$ 0.6\\
7 & 1.9 $\pm$ 0.2 & 9.7 $\pm$ 0.7 & 1.2 $\pm$ 0.2 & 14.9 $\pm$ 0.9 & 0.5 $\pm$ 0.1 & 0.0 $\pm$ 0.0 & 28.0 $\pm$ 1.3 & 2.2 $\pm$ 0.4 & 6.3 $\pm$ 0.6 & 18.2 $\pm$ 1.5 & 4.4 $\pm$ 0.3 & 14.3 $\pm$ 0.6 & 4.3 $\pm$ 0.6\\
8 & 1.9 $\pm$ 0.5 & 9.8 $\pm$ 0.7 & 1.4 $\pm$ 0.4 & 15.1 $\pm$ 1.3 & 0.5 $\pm$ 0.2 & 0.0 $\pm$ 0.0 & 27.6 $\pm$ 1.8 & 1.4 $\pm$ 0.2 & 6.0 $\pm$ 1.2 & 15.7 $\pm$ 3.3 & 5.3 $\pm$ 0.8 & 13.4 $\pm$ 1.2 & 4.4 $\pm$ 0.4\\
9 & 2.0 $\pm$ 0.2 & 9.5 $\pm$ 0.7 & 0.8 $\pm$ 0.1 & 15.1 $\pm$ 0.9 & 0.6 $\pm$ 0.1 & 0.1 $\pm$ 0.1 & 28.0 $\pm$ 1.4 & 2.6 $\pm$ 0.3 & 6.7 $\pm$ 0.7 & 17.3 $\pm$ 1.6 & 4.6 $\pm$ 0.5 & 13.8 $\pm$ 0.6 & 4.5 $\pm$ 0.6\\
10 & 1.9 $\pm$ 0.3 & 9.6 $\pm$ 0.7 & 1.0 $\pm$ 0.2 & 15.0 $\pm$ 1.0 & 0.6 $\pm$ 0.1 & 0.1 $\pm$ 0.1 & 27.2 $\pm$ 1.4 & 3.3 $\pm$ 0.3 & 6.7 $\pm$ 0.8 & 17.6 $\pm$ 1.9 & 4.3 $\pm$ 0.4 & 13.8 $\pm$ 0.7 & 4.9 $\pm$ 0.8\\
11 & 2.0 $\pm$ 0.4 & 10.0 $\pm$ 0.7 & 1.7 $\pm$ 0.4 & 14.9 $\pm$ 1.0 & 0.3 $\pm$ 0.1 & 0.0 $\pm$ 0.0 & 28.0 $\pm$ 1.5 & 1.9 $\pm$ 0.3 & 6.4 $\pm$ 1.0 & 18.7 $\pm$ 3.1 & 5.0 $\pm$ 0.6 & 13.0 $\pm$ 0.9 & 5.6 $\pm$ 1.1\\
12 & 2.0 $\pm$ 0.4 & 9.5 $\pm$ 0.7 & 1.2 $\pm$ 0.3 & 15.1 $\pm$ 1.2 & 0.5 $\pm$ 0.1 & 0.0 $\pm$ 0.0 & 27.3 $\pm$ 1.6 & 2.5 $\pm$ 0.3 & 7.1 $\pm$ 1.2 & 16.6 $\pm$ 3.1 & 5.4 $\pm$ 0.8 & 14.2 $\pm$ 1.1 & 4.2 $\pm$ 0.6\\
13 & 1.9 $\pm$ 0.4 & 9.8 $\pm$ 0.7 & 0.9 $\pm$ 0.1 & 15.2 $\pm$ 1.3 & 0.3 $\pm$ 0.1 & 0.1 $\pm$ 0.0 & 27.8 $\pm$ 1.6 & 1.7 $\pm$ 0.2 & 6.3 $\pm$ 1.0 & 19.1 $\pm$ 3.7 & 4.6 $\pm$ 0.7 & 13.0 $\pm$ 0.9 & 4.6 $\pm$ 0.5\\
14 & 2.0 $\pm$ 0.4 & 9.9 $\pm$ 0.7 & 1.5 $\pm$ 0.3 & 14.4 $\pm$ 1.1 & 0.5 $\pm$ 0.1 & 0.0 $\pm$ 0.0 & 27.8 $\pm$ 1.5 & 2.2 $\pm$ 0.3 & 6.7 $\pm$ 1.2 & 19.3 $\pm$ 3.5 & 4.6 $\pm$ 0.8 & 13.4 $\pm$ 0.9 & 4.9 $\pm$ 0.7\\
15 & 2.1 $\pm$ 0.4 & 9.4 $\pm$ 0.7 & 1.0 $\pm$ 0.2 & 15.3 $\pm$ 1.3 & 0.5 $\pm$ 0.1 & 0.0 $\pm$ 0.0 & 28.7 $\pm$ 1.6 & 1.7 $\pm$ 0.2 & 6.2 $\pm$ 1.2 & 18.3 $\pm$ 3.3 & 4.4 $\pm$ 0.6 & 13.7 $\pm$ 1.0 & 4.8 $\pm$ 0.6\\
16 & 2.0 $\pm$ 0.4 & 9.7 $\pm$ 0.7 & 0.9 $\pm$ 0.2 & 15.4 $\pm$ 1.5 & 0.5 $\pm$ 0.2 & 0.0 $\pm$ 0.0 & 29.0 $\pm$ 1.8 & 2.2 $\pm$ 0.3 & 5.7 $\pm$ 1.0 & 15.6 $\pm$ 2.7 & 5.7 $\pm$ 0.9 & 14.1 $\pm$ 1.2 & 5.0 $\pm$ 0.9\\
17 & 2.1 $\pm$ 0.2 & 9.6 $\pm$ 0.7 & 1.2 $\pm$ 0.2 & 15.0 $\pm$ 0.9 & 0.6 $\pm$ 0.1 & 0.0 $\pm$ 0.0 & 27.9 $\pm$ 1.4 & 2.5 $\pm$ 0.4 & 5.9 $\pm$ 0.6 & 16.9 $\pm$ 1.5 & 4.3 $\pm$ 0.4 & 13.6 $\pm$ 0.6 & 5.0 $\pm$ 0.8\\
18 & 2.0 $\pm$ 0.2 & 9.7 $\pm$ 0.7 & 1.3 $\pm$ 0.2 & 14.8 $\pm$ 1.0 & 0.5 $\pm$ 0.1 & 0.0 $\pm$ 0.0 & 27.6 $\pm$ 1.3 & 2.3 $\pm$ 0.4 & 6.7 $\pm$ 0.7 & 17.2 $\pm$ 1.5 & 4.1 $\pm$ 0.3 & 14.6 $\pm$ 0.7 & 4.2 $\pm$ 0.6\\
19 & 1.9 $\pm$ 0.2 & 10.1 $\pm$ 0.7 & 1.9 $\pm$ 0.4 & 14.5 $\pm$ 0.8 & 0.5 $\pm$ 0.1 & 0.0 $\pm$ 0.0 & 27.7 $\pm$ 1.3 & 2.2 $\pm$ 0.3 & 6.0 $\pm$ 0.6 & 19.6 $\pm$ 2.1 & 4.3 $\pm$ 0.4 & 14.0 $\pm$ 0.7 & 4.7 $\pm$ 0.7\\
\enddata
\tablecomments{Column (1): observing epoch. Columns (2-14): Column density to each \hi\ optical depth component.  
}
\end{deluxetable*} 
\end{rotatetable*}

\bibliography{current-ms_HI}{}
\bibliographystyle{aasjournal}



\end{document}